\documentclass[a4paper,11pt]{article}
\usepackage{bigfoot}
\usepackage{varioref}
\usepackage{amsmath,amsfonts,amsthm,mathrsfs}

\usepackage[normalem]{ulem}
\usepackage{multirow,color,graphics}
\usepackage{amsmath}
\usepackage{amsfonts}
\usepackage{amssymb}
\usepackage{jheppub}
\usepackage{enumerate}
\usepackage{fancyvrb}
\usepackage{verbatim}
\usepackage{wrapfig}
\usepackage{appendix}
\usepackage{amstext}
\usepackage{amssymb}
\usepackage{graphicx}
\usepackage{color}
\usepackage{varioref}
\usepackage{multirow,graphics}
\usepackage{epstopdf}
\usepackage{simplewick}
\usepackage[makeroom]{cancel}

\usepackage[dvipsnames]{xcolor}
 
\definecolor{Green}{RGB}{0, 142, 0}

\newcommand{\nn}{\nonumber}

\numberwithin{equation}{section}

\def\[{\left[}
\def\]{\right]}
\def\({\left(}
\def\){\right)}
\def\d{\partial}

    \newcommand{\beq}{\begin{equation}}
    \newcommand{\eeq}{\end{equation}}

    \newcommand\beqa{\begin{eqnarray}}
    \newcommand\eeqa{\end{eqnarray}}
\newcommand\bea{\begin{array}}
\newcommand\eea{\end{array}}

\newcommand{\bQ}{{\bf Q}}

\newcommand{\bP}{{\bf P}}

\newcommand{\cD}{{\mathbb D}}

\newcommand{\cQ}{{\cal Q}}

\newcommand{\la}[1]{\label{#1}}
\newcommand{\eq}[1]{(\ref{#1})}

\usepackage[dvipsnames]{xcolor}

    \def\bQ{{\bf Q}}
    \def\bP{{\bf P}}

\makeatletter
     \@ifundefined{usebibtex}{} {}
\makeatother

    \def\bQ{{\bf Q}}

        \def\bP{{\bf P}}

\newcommand{\ket}[1]{| #1 \rangle}
\newcommand{\cN}{\mathcal{N}}

\renewcommand{\d}{\partial}

\newcommand{\bh}{\left\{ \mathbf{h} \right\}}

\title{Separation of Variables in AdS/CFT: Functional Approach for the Fishnet CFT}

\author[a]{~~Andrea Cavagli\`a,}
\author[a,b]{~~Nikolay Gromov,}
\author[c,1]{~~Fedor Levkovich-Maslyuk\note{Also at Institute for Information Transmission Problems, Moscow 127994, Russia}}

\affiliation[a]{Mathematics Department, King's College London, The Strand, London WC2R 2LS, UK}
\affiliation[b]{St.Petersburg INP, Gatchina, 188 300, St.Petersburg,
  Russia}
\affiliation[c]{
Institut de Physique Th\'{e}orique, Universit\'{e} Paris Saclay, CEA, CNRS, F-91191 Gif-sur-Yvette, France
}

\emailAdd{andrea.cavaglia${\bullet}$kcl.ac.uk,  nikgromov${\bullet}$gmail.com, fedor.levkovich${\bullet}$gmail.com}

\abstract{
The major simplification in a number of  quantum integrable systems 
is the existence of special coordinates in  which the eigenstates take a factorised form. 
Despite many years of studies, the
basis realising the 
separation of variables (SoV) remains unknown in ${\mathcal{N}=4}$ SYM and similar models, even though it is widely believed they are integrable.
In this paper we initiate the SoV approach for observables with nontrivial coupling dependence in a close cousin of ${\cal N}=4$ SYM -- the fishnet 4D CFT.
We develop the functional SoV formalism in this theory, which allows us to compute non-perturbatively some nontrivial  observables  
in a form suitable for numerical evaluation. 
We present some applications of these methods. In particular, we
discuss the possible SoV structure of the one-point correlators in presence of a defect, and write down a  SoV-type   expression for diagonal OPE coefficients involving an arbitrary state and the Lagrangian density operator. 
We believe that many of the findings of this paper can be applied in the  
${\mathcal{N}=4}$ SYM case, as we speculate in the last part of the article.
}

\begin{document}

\maketitle

\newpage
\setcounter{page}{1}
\section{Introduction}
The Separation of Variables (SoV), which traces its origin to the Hamilton-Jacobi approach
and later to Sklyanin's works
 \cite{Sklyanin:1984sb, Sklyanin:1987ih,Sklyanin:1991ss,Sklyanin:1995bm}, is often regarded as the most powerful method to solve quantum integrable systems (for recent developments see \cite{Derkachov:2021wja,Gombor:2021uxz,Gromov:2020fwh,Pei:2020ljw,Derkachov:2020zvv,Maillet:2020ykb,Ryan:2020rfk,Derkachov:2019tzo,Gromov:2019wmz,Maillet:2019ayx,Cavaglia:2019pow,Ryan:2018fyo,Derkachov:2018ewi,Liashyk:2018qfc,Gromov:2018cvh,Gromov:2016itr}). 
It is based on the expected property that eigenfunctions of the integrals of motion  factorise when evaluated in  a special basis $\langle \mathbf{x}  |$ labelled by the ``separated variables'' $\mathbf{x} \equiv \left\{{x}_n \right\}_{n=1}^L$, where $L$ is the number of degrees of freedom.  Schematically, the eigenstates decompose as
\beq
\langle \mathbf{x} |\Psi \rangle = \prod_{n=1}^L Q_{(n)}( x_n )\; ,
\eeq
where the factors can be determined by solving
linear functional equations in one variable. Typically, they coincide with the so-called Q-functions (or simple combinations of them), and the functional  equations are known as Baxter TQ equations or, in certain contexts, Quantum Spectral Curve (QSC) equations, which also include nontrivial analyticity conditions. 
When it can be worked out, the SoV gives access not only to the spectrum and eigenfunctions of the integrals of motion (IM),  but   also usually leads to simple expressions for  correlation functions (see  e.g. \cite{Smirnov:1998kv,Lukyanov:1997bp} for examples in two-dimensional integrable field theories and recent work \cite{Gromov:2020fwh} for spin chains). 

The SoV paradigm is expected to hold true for quantum integrable systems carrying any representation of the symmetry algebra, in contrast e.g. with the Bethe Ansatz (BA), which can be applied only in special cases. 
In particular, there is a growing body of evidence  that this approach is applicable for such complicated systems as 4D $\mathcal{N}=4$ SYM.
In \cite{Cavaglia:2018lxi,Giombi:2018qox,Giombi:2018hsx,McGovern:2019sdd}, it was shown that 
 some correlation functions at finite coupling, obtained by various techniques such as supersymmetric localisation or the direct resummation of diagrams, can be written as an integral of a product of Q-functions, in agreement with the type of structures expected from SoV.
This is especially exciting as these examples appear in the non-perturbative regime of observables involving short operators, where the BA methods are not applicable.\footnote{
At weak coupling or for very long operators, one can get spectacular results with BA-inspired methods
\cite{Hexagons1,Hexagons2}, and even go beyond the planar limit \cite{Bargheer:2017nne}.
} 
This gives hopes that the complete non-perturbative solution of planar $\mathcal{N}=4$ SYM could be obtainable by means of SoV methods. 
The result for correlators should be given in terms of the Q-functions, which are,
luckily,  already under full control  at finite coupling: they are solutions of the QSC equations  \cite{Gromov:2013pga,Gromov:2014caa}\footnote{The QSC formulation is also known for the ABJM theory  \cite{Cavaglia:2014exa,Bombardelli:2017vhk,Bombardelli:2018bqz} and for some non-local operators \cite{Cusp}.  A similar construction with nontrivial analyticity conditions for the Q-functions also exists for the Hubbard model \cite{Cavaglia:2015nta}. },  obtained in one-to-one correspondence with the anomalous dimensions of primary operators. 
In addition, there are other examples of the applicability of SoV for studying various corners of the theory~ \cite{Jiang:2015lda,Derkachov:2018rot,Derkachov:2019tzo,Derkachov:2020zvv,Belitsky:2014rba,Belitsky:2016fce,Belitsky:2019ygi}, and these techniques could eventually lead to the first principles derivation of the BA-based Hexagon formalism~\cite{Hexagons1,Hexagons2}. 
 
Developing the SoV approach in $\cN=4$ SYM from first principles is very challenging. The main conceptual difficulty is that we do not even  have a precise \emph{definition} (in terms of a Hilbert space and Hamiltonian) of the integrable system which controls the spectrum of $\mathcal{N}=4$ SYM at finite coupling. It can be in principle obtained in AdS/CFT by quantising appropriately the worldsheet $\sigma$-model, which has not been done beyond a few quasi-classical orders.
The situation improves considerably if we consider a theory that is closely related to $\mathcal{N}=4$ SYM, the so-called fishnet CFT. This model was obtained as a double-scaling limit of the  $\gamma$-deformation of $\mathcal{N}$=4 SYM in \cite{Gurdogan:2015csr}. 
Remarkably, it is a non unitary, non supersymmetric, but exactly conformal theory (at least in the planar limit) defined by a very simple  Lagrangian:\footnote{We  are  omitting  three  double-trace  vertices  in the Lagrangian, which  are needed perturbatively for  quantum  conformal  invariance,  see  \cite{Grabner:2017pgm}. The double traces enter only into a limited number of diagrams and will not play a crucial role in this paper.} 
\beq
\mathcal{L} = N_c \text{Tr} \left( \partial_{\mu}\phi_1^{\dagger}\partial^{\mu} \phi_1 + \partial_{\mu}\phi_2^{\dagger}\partial^{\mu} \phi_2 + (4 \pi)^2 \xi^2 \phi_1^{\dagger} \phi_2^{\dagger} \phi_1 \phi_2 \right),
\eeq
where the complex scalar fields $\phi_i$, $i=1,2$ are $N_c\times N_c$ matrices. 
The model inherits integrability from the $\gamma$-deformed $\mathcal{N}=4$ SYM: in particular, the Q-functions and QSC equations can be understood as a limit of the ones in the ``parent theory''\cite{Gromov:2017cja}. 
However, in the fishnet model this integrable structure can be understood much more clearly~\cite{Zamolodchikov:1980mb,Gurdogan:2015csr,Gromov:2017cja,Gromov:2019jfh}, making it an ideal playground for developing the SoV program.

Integrability arises directly as a property of the fishnet Feynman diagrams, which were  already studied by Zamolodchikov in  \cite{Zamolodchikov:1980mb}. 
The resummation of fishnet graphs with the topology of a cylinder, and the related Dyson-Schwinger equations, define rigorously the integrable system associated with the spectral problem.  It consists in a spin chain carrying an infinite-dimensional representation of the 4D conformal group. 
The role of Hamiltonian is played by the graph-building operator which constructs the Feynman diagrams  \cite{Gromov:2017cja}, or more conveniently its inverse, which acts as a differential operator. In the following, we will refer to the conformal spin chain with this Hamiltonian as the \emph{fishchain}, following \cite{Gromov:2019aku}. 

\begin{figure}[t!]
\centering
\begin{minipage}{0.45\textwidth}
\includegraphics[scale=0.3]{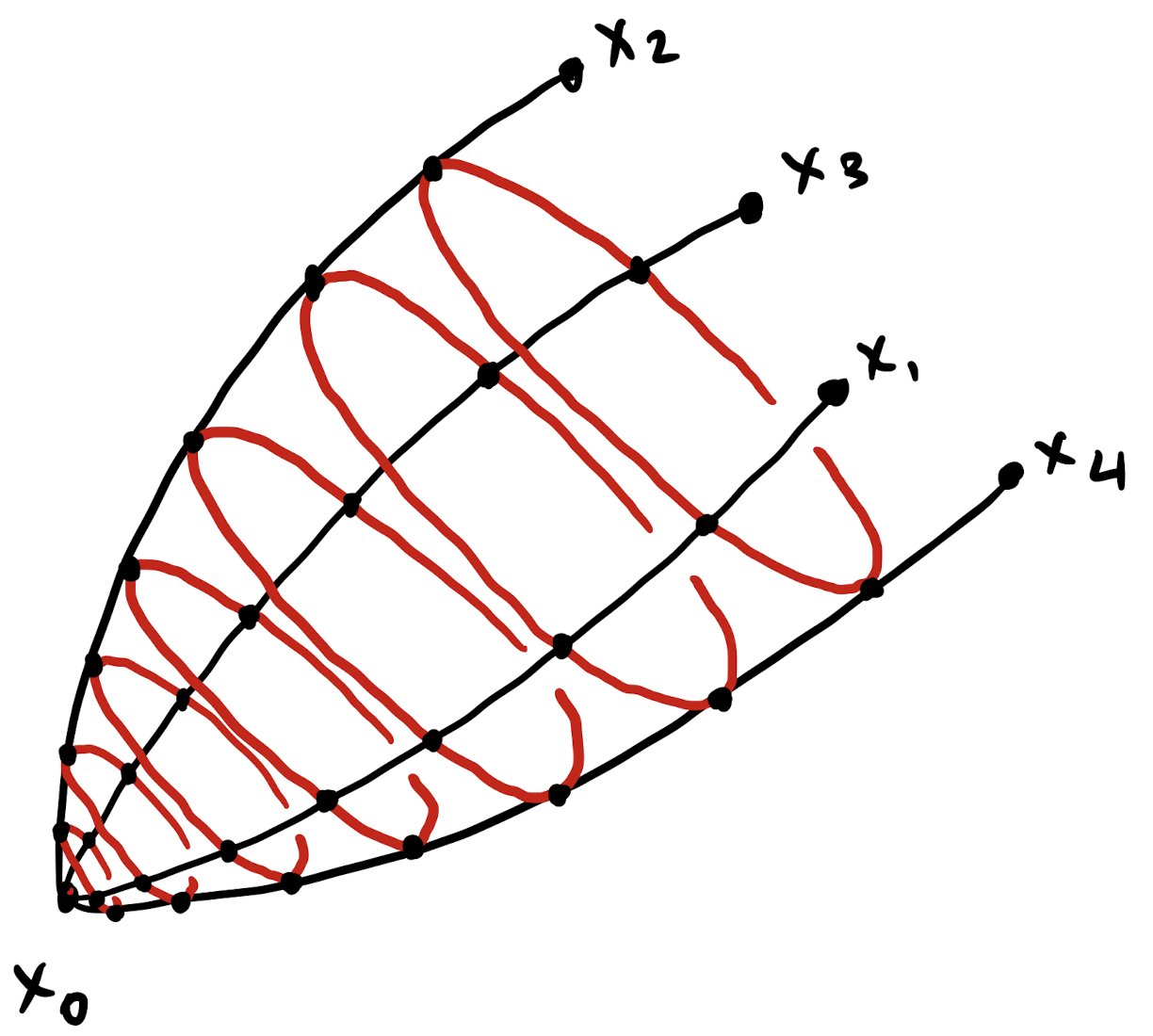}
\end{minipage}
\begin{minipage}{0.45\textwidth}
\centering
\includegraphics[scale=0.5]{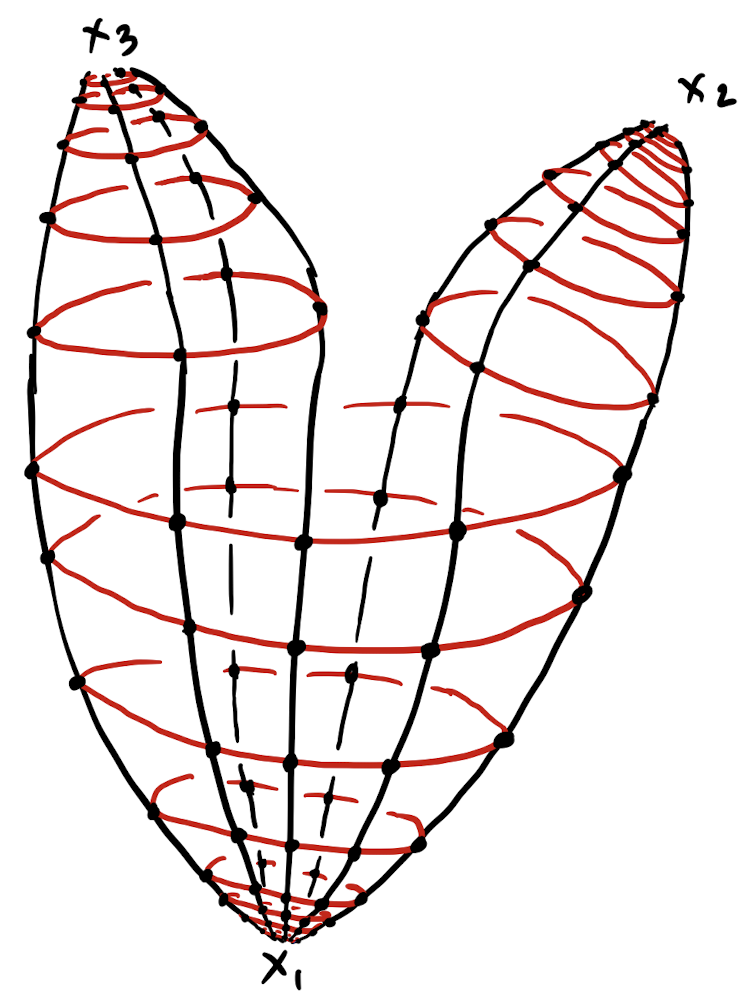}
\end{minipage}
\caption{
Left: Illustration of a CFT wave function. The point $x_0$ represents the insertion of the operator and implies the renormalised sum of an infinite ladder of wheel diagrams.  Right: 3 CFT wave functions glued together into a 3-point function at finite coupling. The wave functions contain all the nontrivial coupling dependence of the 3-point correlator. 
\label{fig:wf}}
\end{figure}

As we review in the main text of the article, the eigenstates of the integrable system can be interpreted as certain trace-trace correlators introduced in \cite{Gromov:2019bsj}, and called \emph{CFT wave functions}. The CFT wave functions encode all information on the single-trace operators at planar level, and moreover appear as natural building blocks in correlators, see Figure \ref{fig:wf}. 
One expects that, written in a basis of Separated Variables, they can be computed in terms of the Q-functions solving the QSC equations. The chain of relations between operators, wave functions, and Q-functions, is illustrated here:
\beq\label{eq:mapIntro}
\begin{array}{ccccc} & \texttt{{\small spin chain  }} & & \texttt{{\small SoV map}} & \\ \mathcal{O}(x_0) & \longrightarrow &  \varphi_{\mathcal{O}}(x_0 | x_1, \dots, x_J) & \longleftrightarrow & q_a(u) \\
\text{{\small single-trace operator}} & & \text{{\small CFT wave function}} &  & \text{{\small Q-functions}} 
\end{array}
\eeq 
For the simplest family of nontrivial operators, those with length one in the presence of twists~\cite{Cavaglia:2020hdb}, the map between wave functions and Q-functions was found by the present authors with A. Sever in  \cite{withAmitupcoming}. Generalising this result to all states is currently an important open problem, but the general methods of \cite{Sklyanin:1995bm,Gromov:2016itr,Gromov:2020fwh} give a clear indication of how to construct the SoV transformation at least formally. 

The explicit construction of 
such map is referred to as the \emph{operatorial approach} to SoV. In this paper, we do not address this problem yet. However, we take a \emph{functional approach}~\cite{Cavaglia:2019pow}, which is based only on the Baxter TQ equations, and allows us to derive a number of mathematical results related to the  structure of the SoV. 
 
 A key element of our construction is the introduction of  inhomogeneities, and arbitrary scaling weights, on the sites of the spin chain. Such parameters are typically very useful in the SoV program. Here we introduce them for the first time for the single-trace operators in the fishnet model at finite coupling. 
 Following \cite{Cavaglia:2019pow,Gromov:2020fwh}, we show how variations of these parameters can be used to generate nontrivial  operators acting on the spin chain, and to compute their expectation values in terms of Q-functions, as we discuss below.

 Another important technical ingredient used in this paper are quasi-periodic, or twisted, boundary conditions along the spin chain, which break the conformal symmetry of local operators \cite{Cavaglia:2020hdb}. The twists are crucial for the SoV, since they lift the degeneracies of the spectrum and guarantee that solutions of the QSC are identified one-to-one with the eigenstates of the spin chain. We use the  construction of the  colour-twist~\cite{Cavaglia:2020hdb}, which is particularly simple for the fishnet model, and introduces the twists directly at the level  of Feynman graphs.

\subsection{Summary of the main results}

In the first half of the paper we develop the functional SoV formalism for the fishchain. In the second part, we discuss some applications, which  broadly concern three topics: scalar products, spin chain form factors, and the g-function. We also speculate on the applicability of our methods to ${\cal N}=4$ SYM.

\paragraph{Integral orthogonality relations.}
It was shown in \cite{Cavaglia:2019pow} that one can  realise the orthogonality properties of different integrable eigenstates, in terms of an integral  over certain combinations of their Q-functions. 
This type of orthogonality relations in the context of the spin chains are very useful, 
e.g. made it possible \cite{Gromov:2019wmz,Gromov:2020fwh} to extract the SoV measure explicitly 
 for higher-rank models. 

In the previously studied cases,  the Q-functions were polynomials, but we demonstrate in this paper how the approach extends to the situations when all Q-functions are non-polynomial. The key building blocks for the integral orthogonality relations are integrals of the product of two Q-functions, which we call \emph{Q-bilinear forms}. We discuss in detail their properties. 
We show that some of these forms are finite precisely for ``on-shell" solutions of the TQ-relations, i.e. those satisfying the quantisation condition giving a discrete spectrum of conformal dimensions. This gives a new interpretation of the quantisation conditions, which were previously quite mysterious and could be justified mainly by invoking properties of the QSC in  $\mathcal{N}=4$ SYM.

\paragraph{Variation of parameters and spin chain form factors.}
We obtained an analytic formula  for the variation of any integral of motion of the fishchain with respect to a parameter. The expression, similar to those found in \cite{Gromov:2019wmz,Gromov:2020fwh} for HW spin chains,  consists in a ratio of determinants built with integrals over products of Q-functions. 
Such observables can be interpreted as spin chain form factors, i.e. expectation values of operators acting on the spin chain. Since we have $\sim 4 J$ integrals of motion, and we can use as a parameter any of the $2 J$ inhomogeneities + weights, a naive counting suggests that we have access to  $\sim (4 J) \times (2J)$ different observables, where $J$ is the length of the chain.  

The simplest example is the derivative of the scaling dimension $\Delta_A$ of an arbitrary single-trace operator $\mathcal{O}_A$ with respect to the coupling constant. It is well known from conformal perturbation theory that such variation can be  identified with the OPE coefficient~\cite{Costa:2010rz}
\beq\label{eq:LagrOPE}
\partial_{\xi^2} \Delta_A = C_{\mathcal{O}_A^{\dagger} \mathcal{O}_A \mathcal{L}_{\text{int}} },
\eeq
where $\mathcal{L}_{\text{int}} = \text{Tr}(\phi_1 \phi_2 \phi_1^{\dagger} \phi_2^{\dagger} )$ is the interaction vertex in the Lagrangian. 
Even though such correlators can also be extracted from the known spectrum, our result gives a closed answer in terms of the Q-functions at a given value of the coupling $\xi$ -- as a determinant of one-dimensional integrals of Q-functions. Based on the example of \cite{Cavaglia:2018lxi}, we expect this to provide rich structural information for generalisations to more complicated OPE coefficients. 
Correlators of the type (\ref{eq:LagrOPE}) were also recently studied in \cite{FishnetHexagon} with the hexagon decomposition approach. Understanding fully the relation with our result could help to advance  with the resummation of the Hexagon series.  We were also able to compute the expectation value of the variation of the Hamiltonian w.r.t. local weights, which is a natural observable and provides an exact result for a class of Feynman diagrams with nontrivial local insertions.

\paragraph{The structure of the g-function.}
We will also make some observations on the computation of the so-called g-functions inspired by the recent papers \cite{Caetano:2020dyp} and \cite{Gombor:2021uxz}. The $g$-functions capture the overlap between a spin chain eigenstate and an integrable boundary state (for a more detailed definition see section \ref{sec:gsec}), and have appeared in connection to several observables in $\mathcal{N}$=4 SYM \cite{Jiang:2019xdz,Komatsu:2020sup,Gombor:2020kgu,WLToappear}. The traditional integrable formalism for computing the g-function \cite{Dorey:2004xk} is based on the thermodynamic Bethe Ansatz, and organises the result as a product of two factors. One, simpler to compute, is boundary-dependent, while the other is a universal factor which depends only on the state. 
 In \cite{Caetano:2020dyp}, a conjecture was presented expressing the universal factor in terms of determinants built with the  Q-functions, for the case of the Sinh-Gordon model. This proposal -- which still  contains an undetermined proportionality constant -- passes an infinite number of nontrivial consistency checks, as it reproduces a selection rule for the $g$-functions. Moreover it matches the SoV structure of the overlaps which was later established rigorously for the Heisenberg spin chain in \cite{Gombor:2021uxz}. The arguments of \cite{Caetano:2020dyp} are in the spirit of the functional SoV method, and are based on the analysis of the Q-function representation for the scalar product. We show that a similar analysis can be extended to our case, and  present the analogous  conjecture for the  fishnet model. 
 
 \paragraph{Towards $\mathcal{N}=4$ SYM. }
 Finally, we point out that there are many aspects of the functional SoV setup that can be generalised directly to $\mathcal{N}$=4 SYM. In particular, it is quite clear how a notion of scalar product as a multiple integral of the Q-functions can be obtained in this theory, as we discuss in section \ref{sec:secYM}. In turn, this gives a good starting point to construct the g-function, which is expected to correspond to the  TBA expressions written down in  \cite{Jiang:2019xdz,Komatsu:2020sup}. 
 
 The key peculiarity of the $\mathcal{N}=4$  SYM case is that, even for  short operators, the underlying integrable system has an infinite number of degrees of freedom at finite coupling (which is also expected from the dual worldsheet description).
 So the closest analogy is \cite{Smirnov:1998kv,Lukyanov:1997bp}, where indeed the size of the determinant was infinite in the SoV-type integrals appearing in the description of correlators. Nevertheless, at weak and at finite coupling, it should be possible to truncate the determinants to a finite size, as the number of relevant parameters in the Q-functions can be taken to be finite with very high precision when studying the spectrum.

\vspace{5mm}
The paper is organised as follows: In section \ref{sec:basic}, we review the general spin chain formulation,  presenting a precise definition of the scalar product, and we summarise the Baxter equations and properties of the Q-functions, introducing the new ingredients of the inhomogeneities and weights.
In section \ref{sec:fn} we link the general spin chain formalism 
with the fishnet CFT. In section~\ref{sec:sovq} we describe technical details about an important notion of Q-bilinear forms, and discuss the new interpretation of the quantisation condition.
In section \ref{sec:orthosec}, we derive  the form of the variation of the integrals of motion. 
In section \ref{sec:sovscalar}, we discuss the orthogonality relations and the expectations for the scalar product in the  SoV basis.
In section \ref{sec:gsec}, we discuss the natural candidate for the universal factor of the g-function in terms of our construction. In the last section \ref{sec:secYM}, we sketch how the arguments of this paper can be generalised to $\mathcal{N}=4$ SYM. We close with a short summary, and some appendices where  technical details are unrolled.  

\section{Conformal spin chain}\label{sec:basic}
In this section we review in general the spin chains with 4D conformal symmetry at each site.
We start by discussing the conformal spin chain abstractly. We generalise previous constructions, e.g.~\cite{Gromov:2019jfh}, by considering arbitrary conformal weights on different sites and arbitrary inhomogeneities (similarly to~\cite{Derkachov:2021rrf}). We construct a natural notion of conformally invariant scalar product on the spin chain, and 
 we review the Baxter TQ equation,  which is the engine behind the integrability construction. 
 
At the end of the section we also review the quantisation condition for the wave-functions and its counterpart
in terms of the Q-functions, which singles out a discrete spectrum of anomalous dimensions as a function of one parameter, which will be identified with the `t Hooft coupling in the next section.

\subsection{Representation of the conformal group}\label{sec:conformal0}

In this section we introduce some notations for the representations of the Euclidean conformal group $SO(1,5)$ which we use below.
We found it beneficial to use the 6D formalism especially when it comes to fusion of transfer matrices~\cite{Gromov:2019jfh}. Some expressions in 4D can be found in e.g.~\cite{Chicherin:2012yn}.

For simplicity we consider only scalar representations on each site, with a generic scaling dimension $h$, realised as functions $f(x^\mu)$ of points $x \in \mathbb{R}^4$. The space of such fields will be denoted as $\mathcal{F}_h$. 
We will treat $\mathcal{F}_h$ as a real vector space, so that we do not have to make any assumption on the reality of $h$. We will often focus on the $h=1$ or $h=2$ cases, which are connected to the fishnet theory, and correspond (in the Euclidean signature in which we work) to a representation of the conformal group without a highest nor lowest weight vector.

In 6D space, the conformal group generators take the standard $SO(1,5)$ form
\beq\la{q6}
\hat q^{MN}=-{i}\(X^N \frac{\partial}{\partial X_{M}}-
X^M \frac{\partial}{\partial X_{N}}\)\;\;,\ N,M=-1,0,\dots,4 \ ,
\eeq
which satisfy the standard commutation relation
\begin{equation}\la{CMqq}
\[\hat q^{MN},\hat q^{KL}\]={i}\(-\eta^{MK}\hat q^{NL}+\eta^{NK}\hat q^{ML}+\eta^{ML}\hat q^{NK}-\eta^{NL}\hat q^{MK}\).
\end{equation}
Here $\eta$ is the 6D Minkowsky metric with signature $-+++++$.
The action on the functions of $4$ variables is defined by acting on the expression
\beq\la{f6f4}
\frac{1}{(X^{-1}+X^{0})^h}
f\(\frac{X^1}{X^{-1}+X^{0}},\dots,\frac{X^4}{X^{-1}+X^{0}}\)\;.
\eeq
As $X_M X^M$ is invariant under $SO(1,5)$
we can set it to zero and thus exclude $X^-\equiv X^{-1}-X^0$. As furthermore the action by $\hat q^{MN}$
commutes with the re-scaling $X^M\to \alpha X^M$
and its eigenvalue is measured by $h$, 
we see that the form \eq{f6f4} must be preserved under the $SO(1,5)$ action. In this way we induce the action of \eq{q6} on functions $f(x^\mu)$ of $4$ variables.
The generators $\hat q^{MN}$
can be related to the dilatation ${\mathbb D}$,
translation ${\mathbb P}_\mu$ and special conformal transformations ${\mathbb K}_\mu$ as follows
\beqa\label{4Dgen}
&& \hat{q}^{-1 , 0}  =   {\mathbb D} \; , \\
&&\hat{q}^{0 , \mu}+\hat{q}^{-1 , \mu}   = - {\mathbb P}_{\mu}  \;,
\\ &&
\hat{q}^{0 , \mu}-\hat{q}^{-1 , \mu}  = - {\mathbb K}_{\mu}  \;,
\\ && \hat{q}^{\mu , \nu} =  {\mathbb S}_{\mu \nu} 
\;.
\eeqa
On can deduce from \eq{q6} and \eq{f6f4} the explicit action on the 4D fields
 \beqa\label{eq:diffoperators}
&& \mathbb{P}_{\mu} \equiv -i \partial_{\mu}, \;\;\;\;\; \mathbb{K}_{\mu} \equiv i \left( x^2 \partial_{\mu} - 2 x_{\mu} (x \cdot \partial )  - 2 h x_\mu \right) , \\
&& \mathbb{S}_{\mu\nu} \equiv i x_{\mu} \partial_{\nu} - i x_{\nu} \partial_{\mu}, \;\;\;\;\; \mathbb{D} \equiv -i x^{\mu} \partial_{\mu} - i h \;.\nn
\eeqa 
It is also useful to note that for a finite conformal transformation  $g \in SO(1,5)$, the representation acts as
\beq\label{eq:group}
g \circ f(x) \equiv  \left| \frac{\partial g^{-1}(x) }{\partial x} \right|^{\frac{h}{d} } \, f( g^{-1}(x) ) , \;\;\;\;f \in \mathcal{F}_h\; ,
\eeq
where $g(x)$ denotes the action of the conformal transformation on a point $x \in \mathbb{R}^4$.

The Hilbert space of the spin chain is a space of functions of $J$ variables transforming in the tensor product representation
$$
\mathcal{F}_{ \mathbf{h}} \equiv \mathcal{F}_{h_1} \otimes \mathcal{F}_{h_{2}} \otimes \dots \otimes \mathcal{F}_{h_{J-1}}   \otimes \mathcal{F}_{h_J} \;, 
$$
where for generality we  consider different weights at every site and we denoted 
$\left\{\mathbf{h} \right\} \equiv \left\{h_1, \dots, h_J \right\}$. 
On this space we can define an action of the conformal group at each site separately, and denote it by $g_\alpha$
\beq
g_\alpha\circ F(x_1, \dots, x_J ) \equiv   \left| \frac{\partial g_\alpha^{-1}(x_\alpha) }{\partial x_\alpha} \right|^{\frac{h_\alpha}{4} } \, F(x_1,\dots, g^{-1}(x_\alpha), \dots, x_J )\;.
\eeq
The global action we denote as $g$ is given by the product $\prod_{\alpha=1}^J g_\alpha$ as usual. We will also use the notation for the sum of all weights,
\beq
    D_0\equiv \sum_{\alpha=1}^Jh_\alpha \ .
\eeq

\subsection{The conformally invariant scalar product }\la{sec:scaladef}
Now we introduce a scalar product, invariant under the conformal group:
\beq
f_1, f_2 \in \mathcal{F}_{h} \longrightarrow \langle \langle f_1\, , \, f_2 \rangle\rangle_h\;.
\eeq
The requirement that
\beq\label{eq:requirement}
\langle \langle f_1\, ,  f_2 \rangle \rangle_h = \langle \langle  g \circ  f_1\, , \, g\circ f_2 \rangle \rangle_h\;\;, \;\;  g \in SO(1,5)\;,
\eeq
fixes completely the form of the scalar product up to an arbitrary overall constant factor, see \cite{Dobrev:1977qv} and Appendix \ref{app:scalar}. 
The result is given by:
 \beq\label{eq:scalar1s}
 \langle \langle f_1,  \, f_2 \rangle \rangle_h = \int d^D x\, f_1(x)\;\;  \Box^{ D/2 - h } f_2(x )\;,
 \eeq
where $D=4$ and in general we define the fractional
power of the d'Alambert operator by (see e.g. \cite{Kazakov:2018qbr}):
 \beq\label{eq:boxfrac}
\Box^{\beta} f(x) \equiv \frac{ (-4)^{\beta} \Gamma(2 + \beta )}{\pi^{2} \Gamma(-\beta) } \, \int d^4 y \frac{f(y) }{|x-y|^{4 + 2 \beta } }\;.
\eeq
For the Hilbert space of the spin chain the scalar product is defined by
\beq\label{eq:scalar}
\langle \langle F_1\, , \, F_2 \rangle \rangle = \int \prod_{\alpha=1}^J d^D x_\alpha\, F_1(x_1, \dots, x_J ) \prod_{\alpha=1}^J \Box_{x_\alpha}^{ D/2 - h_\alpha } \circ F_2(x_1, \dots, x_J )\;.
\eeq
We will see in section~\ref{sec:quantwf} that the above scalar product needs regularisation in some important cases.

\subsection{Integrable fishchain with inhomogeneities}\label{sec:fishchain}
Here we introduce notations for the spin chain and build the integrals of motion via transfer matrices. 

Following \cite{Gromov:2019bsj,Gromov:2019jfh}, we define an integrable system on the spin chain introduced above. At this stage, we will keep the weights $\left\{h_1, \dots, h_J \right\}$ completely generic, even though, in applications to the fishnet CFT, the relevant weights are either $h_\alpha=1$ (non-magnon) or $h_\alpha=2$ (magnon). 

To define the integrable system, we introduce a family of nontrivially commuting transfer matrices following the construction of \cite{Zamolodchikov:1980mb,Chicherin:2012yn,Gromov:2017cja,Gromov:2019bsj,Gromov:2019jfh}.
Their eigenvalues provide us with a complete family of integrals of motion. 
We will also introduce inhomogeneities in the definition of the transfer matrices, which will be labelled by $\left\{ \vartheta_{\alpha} \right\}$, $1\leq \alpha\leq J$, and twisted boundary conditions.

\subsubsection{Commuting transfer  matrices}
 The transfer matrices will be denoted as $\hat{ \mathbb{T} }^{{\bf r}}$ with ${\bf r} \in \left\{ \mathbf{1},\mathbf{4},\mathbf{6},   \bar{\mathbf{4}} , \bar{\mathbf{1}} \right\}$ labelling the representation in the auxiliary space, corresponding to antisymmetric tensors with $0,\;1,\;2,\;3,\;4$ indices. To build them, we start from Lax matrices, which are differential operators acting on a single site of the spin chain, and evaluated as matrices in the auxiliary space.  
The Lax matrices defined in \cite{Gromov:2019bsj,Gromov:2019jfh} are:\footnote{We use a redefinition of the spectral parameter as compared to the original papers, to remove the explicit dependence on the Planck constant which was natural in the holographic setting of \cite{Gromov:2019bsj,Gromov:2019jfh}.} 
\beqa
\mathbb{L}_\alpha^{\mathbf{1}}(u) &=& 1 , \label{eq:Lax1}\\ \mathbb{L}_\alpha^{\mathbf{4}}(u) &=& u - \frac{i}{2} \hat q_\alpha^{MN} \Sigma_{MN} , \label{eq:Lax4}\\
\label{eq:Lax6}
\mathbb{L}_\alpha^{\mathbf{6}}(u) &=& u^2 + (u-i) \hat q_\alpha + \frac{\hat q_\alpha^2 }{2} + \frac{(1 + \mathcal{C}_{h_\alpha})}{4}, \\
\mathbb{L}_\alpha^{\bar{\mathbf{4}}}(u) &=& (u^2 + \frac{\mathcal{C}_{h_\alpha}}{4} + 1 )\left[ - \mathbb{L}_\alpha^{{\mathbf{4}}}(-u) \right]^{T}\\
\mathbb{L}_\alpha^{\bar{\mathbf{1}}}(u) &=& 
\left( u^2 +\frac{\mathcal{C}_{h_\alpha} }{4} + \frac{5}{4} \right)^2 - \frac{\mathcal{C}_{h_\alpha} }{4} -1 \;
,\label{eq:Lax4b}
\eeqa
where $\mathcal{C}_h \equiv  h (h-4)$ is the Casimir invariant for the representation at the site. 
The Sigma matrices $\Sigma^{MN}$ are  $4 \times 4$  for any choice of the indices $M,N$, satisfy $\Sigma^{MN} = -\Sigma^{NM}$, and obey the same commutation relations (\ref{CMqq}).  
An explicit realisation of such  matrices, which we will assume in the following, can be found in Appendix \ref{app:sigmas}.

The transfer matrices are defined as
\beq
\hat{\mathbb{T} }^{{\bf r}}(u) = \text{Tr}_{\bf r}\left[  \hat{\mathbb{L}}^{{\bf r}}_{J}(u - \vartheta_{J}  ) \hat{\mathbb{L}}^{{\bf r}}_{J-1}(u - \vartheta_{J-1}  )\dots\hat{\mathbb{L}}^{{\bf r}}_1(u - \vartheta_{1}  ) \, G^{{\bf r}} \right] ,
\eeq
where $G^{{\bf r}}$ is an element of the conformal group in the representation ${{\bf r}}$, introducing quasi-periodic boundary conditions, and $\vartheta_i$'s are the \textit{inhomogeneities}\footnote{Later we also use the notation $\theta_\alpha$ (rather than $\vartheta_\alpha$) for inhomogeneities with some extra shifts, see \eq{eq:shiftedtheta}. This is done so that the limit $\theta_\alpha\to 0$ corresponds to the actual fishnet CFT.}. 

Since the Lax operators satisfy the Yang-Baxter equation as shown in \cite{Gromov:2019bsj,Gromov:2019jfh}, the transfer matrices commute for different values of the spectral parameter and for any choice of the auxiliary space representation:
\beq
\left[ \hat{\mathbb{T} }^{{\bf r}}(u)  , \hat{\mathbb{T} }^{{\bf r}'}(u') \right] = 0\; . 
\eeq
Therefore, they can be diagonalised simultaneously. In the following, we often restrict the discussion to the eigenvalues which are denoted as ${\mathbb{T} }^{{\bf r}}(u)$. 

\subsubsection{Parametrisation of the twist}
We will consider twist transformations $G$ which admit two  (real) fixed points in $4$D. 
 Such transformations are diagonalisable in the $\mathbf{4}$ representation $G^{\bf 4}$. We will consider the case where the eigenvalues  $\lambda_1,\lambda_2,\lambda_3$ and $
\lambda_4=\frac{1}{\lambda_1\lambda_2\lambda_3}$ are distinct.  
 Accordingly, in the vector representation $G^{\bf 6}$
 will have $6$ eigenvectors $Y^M_{a}$,  which correspond to $6$ fixed points $y^\mu_a=\frac{Y^\mu_a}{Y_a^+}$ after projection to $4$D. Due to the basic properties of $SO(1,5)$ orthogonal matrices, 
for generic eigenvalues we find that  $Y_a.Y_b\propto (y_a-y_b)^2=0$ for all $a,b$, except for  $(a,b)=(1,2)$ or $(3,4)$ or $(5,6)$. 
Since in Euclidean space real vectors cannot be null-separated, this means that one can only have at most two fixed points which are simultaneously real.  
There are indeed precisely two real fixed points for
\beq
 e^{-\alpha_0}=\lambda_1\lambda_2 , \;\;,\;\;e^{i\alpha_1}=\lambda_1\lambda_3
 ,\;\;e^{i\alpha_2}=\lambda_2\lambda_4\;,\;\;\; \alpha_i \in \mathbb{R}\; .
\eeq
We will denote them as  $x_0$ and $x_{\bar{0}}$.  To get an explicit parametrisation of the map, one can send $x_0\to 0$ and $x_{\bar{0}}\to \infty$ by a similarity transformation
\beq
G=K R K^{-1} ,
\eeq
where $K$ is a special conformal transformation acting on coordinates in 4D as
\beq\label{eq:Kdef}
x^{\mu} \rightarrow x_0^{\mu} + \frac{ x^{\mu} - b^{\mu} x^2}{1 - 2 (b \cdot x) + b^2 \, x^2 }\;\; , \;\; b^{\nu} \equiv \frac{x_0^{\nu} - x_{\bar{0}}^{\nu}}{(x_0 - x_{\bar{0}} )^2 } ,
\eeq
and correspondingly,  on functions $f\in \mathcal{F}_h$ as
\beq
K \circ f(x) = \frac{(x_0-x_{\bar{0}})^{2h}}{
(x-x_0+ x_{\bar{0}} )^{2h}
}
f\(x_0^{\mu} + \frac{ x^{\mu} + b^{\mu} x^2}{1 + 2 (b \cdot x) + b^2 \, x^2 } \right),
\eeq
and $R$ is of the form 
\beq
R^{\bf 6}  =\left[\begin{array}{c|c}
\bea{cc}
\cosh\alpha_0&-\sinh\alpha_0\\
-\sinh\alpha_0&\cosh\alpha_0
\eea
 & 0 \\ \hline 
 0 &
 R_{4\times 4}
 \end{array}\right]\;\;,
\eeq
with $R_{4\times 4}$ being a $4$D rotation matrix.
A $4$D rotation normally has two invariant orthogonal planes. In a conventional standard position, we can take those planes to be $(1,2)$ and $(3,4)$. In this case we get
\beq\label{eq:R4x4}
R_{4\times 4}=\left[\bea{c|c}
\bea{cc}
\cos\alpha_1&-\sin\alpha_1\\
\sin\alpha_1&\cos\alpha_1
\eea & 0 \\ \hline
0 & 
\bea{cc}
\cos\alpha_2&-\sin\alpha_2\\
\sin\alpha_2&\cos\alpha_2
\eea
 \eea\right] . 
\eeq
We will frequently assume that in the standard frame the twist matrix $G$ is of this ``diagonal" form
\beq
\Lambda = (\lambda_1 \lambda_2)^{-i {\mathbb{ D}} } \, (\lambda_1 \lambda_3)^{ {\mathbb{ S}}_{1,2}} \, (\lambda_2 \lambda_4)^{{\mathbb{ S}}_{3,4}}\;  = e^{i \alpha_0 \mathbb{D} + i \alpha_1 \mathbb{S}_{1,2} + i \alpha_2  \mathbb{S}_{3,4} }\;,
\eeq
where ${ \mathbb{D}},\;{ \mathbb{S}}_{1,2}$ and ${ \mathbb{S}}_{3,4}$ are defined in \eq{4Dgen}.  
In the representation $\bf 4$ the twist matrix in the standard frame becomes 
$$
\Lambda^{\bf 4}={\rm diag}(\lambda_1,\lambda_2,
\lambda_3,
\lambda_4)=
{\rm diag}\(e^{-\frac{\alpha _0}{2}+\frac{i \alpha _1}{2}+\frac{i \alpha _2}{2}},e^{-\frac{\alpha
   _0}{2}-\frac{i \alpha _1}{2}-\frac{i \alpha _2}{2}},e^{\frac{\alpha _0}{2}+\frac{i \alpha
   _1}{2}-\frac{i \alpha _2}{2}},e^{\frac{\alpha _0}{2}-\frac{i \alpha _1}{2}+\frac{i \alpha
   _2}{2}}\)\;.
$$
The eigenvalues $\lambda_i$'s, satisfying $\prod_{a=1}^4 \lambda_a = 1$, are called  \textit{twist parameters}, and will appear throughout the paper. 

\subsubsection{Global conformal charges}
Under a conformal transformation of the wave functions connected to the identity\footnote{For the spinorial representations $\mathbf{4}$ and $\bar{\mathbf{4}}$, the relation (\ref{eq:brokenconf}) can be applied only for those conformal transformations $g$ which are in the connected component of the identity in the conformal group.  In particular, it cannot be applied to the transformation $F$ introduced in \eq{eq:defF2}. }, the transfer matrices are invariant, up to a redefinition of the twist map:
\beq\label{eq:brokenconf}
g\circ \hat{\mathbb{T}}^{{\bf r}} \circ g^{-1} = \left. \hat{\mathbb{T}}^{{\bf r}} \right|_{G \rightarrow g  G g^{-1} } .
\eeq
Therefore, they commute with the Cartan generators of the form 
\beq\label{eq:confres}
\left\{ 
\hat{\mathbb{Q}}_0, 
\hat{\mathbb{Q}}_1, 
\hat{\mathbb{Q}}_2 \right\} \equiv  K \circ \left\{ { \mathbb{D} } , \, {\mathbb{S}}_{1,2} , \, {\mathbb{S}}_{3,4} \right\} \circ K^{-1}\;.
\eeq
We will consider a basis of common eigenstates of the transfer matrices as well as the Cartan charges, such that 
\beq\label{eq:chargesQ}
\left( \hat{\mathbb{Q}}_a  - \mathbb{Q}_a \right) \circ \Psi( x_1, \dots, x_J ) =  0 ,\;\;\;\;\; \left( \hat{\mathbb{T}}^{{\bf r}}(u)- {\mathbb{T}}^{{\bf r}}(u)\right) \circ \Psi( x_1, \dots, x_J ) =  0 \;,
\eeq
and we will denote the conformal charges as 
\beq
\label{eq:Qrel}
\mathbb{Q}_0 \equiv i \Delta, \;\;\  \mathbb{Q}_n \equiv S_n, \ \ \;n=1,2\;.
\eeq
Single-valuedness of the wave functions implies that $S_n \in \mathbb{Z}$, while $\Delta$ can be in principle a generic complex number. 

In the next section, we will see how these parameters are naturally identified with the scaling dimensions and spins of twisted  operators in the fishnet model. An appropriate quantisation condition and the introduction of the coupling constant will restrict the conformal dimensions $\Delta$ to the  physical values in the spectrum of the CFT.

\subsubsection{Integrals of motion}
The transfer matrices have a polynomial dependence on the spectral parameter, therefore they are parametrised in terms of a finite number of commuting operators, the integrals of motion. 

There is also a trivial polynomial dependence in some of the transfer matrices, that does not depend on the state. To separate this part, it is convenient to introduce the fixed polynomials depending on the inhomogeneities:
\beq
\mathcal{Q}_{+}(u) \equiv \prod_{\alpha=1}^J \( u + i \frac{h_\alpha-2}{2} - \vartheta_\alpha\) , \;\;\;\; \mathcal{Q}_{-}(u) \equiv \prod_{\alpha=1}^J \( u - i \frac{h_\alpha-2}{2} - \vartheta_\alpha \)\;.
\eeq
After that the eigenvalues of the transfer matrices can then be written as
\beqa\la{TTs}
{ \mathbb{T}  }^{\mathbf{1}}(u) &=&1\\
{ \mathbb{T}  }^{\mathbf{4}}(u) &=& P_J^{\mathbf{4}}(u)\\
\mathbb{T}^{\mathbf{6}}(u) &=& P_{2J}^{\mathbf{6}}(u)\\
{ \mathbb{T} }^{\bar{\mathbf{4}}}(u) &=&    \mathcal{Q}_+(u) \, \mathcal{Q}_-(u) \, P_J^{\bar{\mathbf{4}} }(u)\\
{ \mathbb{T}  }^{\bar{\mathbf{1}}}(u) &=& \mathcal{Q}_+^{[+1]}(u) \, \mathcal{Q}_+^{[-1]}(u) \, \mathcal{Q}_-^{[+1]}(u) \, \mathcal{Q}_-^{[-1]}(u)
\eeqa
where $P_n^{{\bf r}}(u)$ is a  polynomial of degree $n$ and we use the notation
\beq
    f^\pm\equiv f(u\pm i/2) \ , \ f^{[+a]}\equiv f(u+ia/2) \ .
\eeq
The highest degree coefficients in such polynomials are fixed in terms of the twist eigenvalues: 
\beqa
&& P_{J}^{\mathbf{4}}(u ) =  u^J \chi_{\mathbf{4}} + \sum_{\alpha=1}^{J} I_{(2,\alpha)} u^{\alpha-1}  ,  \nn \\
&&P_{2J}^{{\mathbf{6}}}(u) =  u^{2 J} \chi_{{\mathbf{6}}} + \sum_{\alpha=1}^{J} \left( I_{(0,\alpha)}  +   I_{(\bar 0,\alpha)} u^{J} \right)u^{\alpha-1}  ,\\
&&P_{J}^{\bar{\mathbf{4}}}(u ) =  u^J \chi_{\bar{\mathbf{4}}}+ \sum_{\alpha=1}^{J} I_{(-2,\alpha)} u^{\alpha-1}\nn ,
\eeqa
with 
\beq
\label{defchi}
    \chi_{\mathbf{4}} = \sum_{a=1}^4 \lambda_a\ , \ \  \chi_{\mathbf{6}}= \sum_{1\leq a<b \leq 4} \lambda_a \lambda_b \ , \ \  \chi_{\bar{\mathbf{4}}} = \sum_{a=1}^4 1/\lambda_a\ .
\eeq
The remaining $4J$ coefficients, denoted as $I_{(a,\alpha)}$, depend nontrivially on the state. 
The conformal charges appear in the next-to-highest order terms:
 \beqa\la{III}
I_{(+2,J)} &=& \frac{1}{2 i}\left[ -   2 i \vartheta \lambda_{++++} -\Delta \,\lambda_{++--} 
+ S_1 \lambda_{+-+-}
+ S_2 \lambda_{+--+} 
\right], \\
I_{(-2,J)} &=& \frac{1}{2 i}\left[ - 2 i\vartheta  \bar\lambda_{++++} +\Delta \bar\lambda_{++--}
- S_1 \bar\lambda_{+-+-}
- S_2 \bar\lambda_{+--+}
\right],  \nn  \\
I_{(\bar 0,J)} &=& -2\vartheta \sum_{i<j} \lambda_i \lambda_j  + i \left[  \Delta ( \lambda_1 \lambda_2 - \lambda_3 \lambda_4 ) 
- S_1 ( \lambda_1 \lambda_3 - \lambda_2 \lambda_4 ) 
+ S_2 ( \lambda_2 \lambda_3 - \lambda_1 \lambda_4 ) 
\right] , \nn
\eeqa
where we introduced\footnote{to be clear, in this notation we have e.g. $\bar\lambda_{++--}=\bar\lambda_1+\bar\lambda_2-\bar\lambda_3-\bar\lambda_4$.}
\beq
\label{deflpm}
\lambda_{s_1s_2s_3s_4}\equiv\sum_{i=1}^4s_i\lambda_i \ , \ \ \  \bar{\lambda}_i \equiv 1/\lambda_i\ , \ \ \ \ \vartheta \equiv  \sum_{\alpha=1}^J \vartheta_{\alpha} \ .
\eeq
The remaining $4J - 3$ coefficients are the nontrivial eigenvalues of the commuting family of  integrals of motion.

\subsection{The quantisation condition for the wave functions}\label{sec:quantwf}
To define a well posed diagonalisation problem for the transfer matrices, we have to specify an appropriate function space for the eigenvectors, or, in other words, a quantisation condition. 
The condition we are interested in, for applications to the fishnet CFT, is that wave functions are single-valued, and do not have singularities, except when all the coordinates approach the fixed points $x_0$ and $x_{\bar{0}}$ simultaneously.\footnote{In principle we can allow sub-leading singularities when only a subset of the coordinates approaches the fixed point. In  the examples we analysed explicitly for wave functions in the fishnet theory, we have not found such sub-leading singularities. }

The approach to these points is obtained by evolving the coordinates with the transformation $e^{-i \rho \hat{Q}_0}$, for $\rho \rightarrow -\infty$ and $+\infty$, respectively. Using the fact that the wave function is an eigenvector under such transformation, we find, in the limit,
\beqa\label{eq:singularities}
\Psi( x_1, x_2, \dots, x_J) &\sim& \epsilon^{-\Delta - D_0 } \times O(1), \;\;\;\;\; |x_\alpha - x_0| \sim \epsilon \rightarrow 0\;, \\
\nn\Psi( x_1, x_2, \dots, x_J ) &\sim & \epsilon^{ +\Delta - D_0 } \times O(1), \;\;\;\;\; |x_\alpha - x_{\bar{0}}| \sim \epsilon \rightarrow 0\;,
\eeqa
where $\epsilon \propto e^{- |\rho|}$, ($\rho \rightarrow \pm \infty$) is a cutoff measuring the distance from the fixed points.
Notice that (\ref{eq:singularities}) distinguishes between the points $x_0$ and $x_{\bar{0}}$ by the sign of $\Delta$.
The interpretation of these equations will become clear when we relate the wave functions to a particular class of correlators in the fishnet CFT where $x_0$ is the location of a non-protected operator.

We expect that the regularity of the  wave function 
everywhere except for $x_0$ and $x_{\bar 0}$,  together with \eq{eq:singularities}  at these points, are sufficient conditions to
 restrict the spectrum of integrals of motion to a discrete set, for any choice of the Cartan charges $\left\{ \Delta, S_1, S_2 \right\}$.  Alternatively, we can say that the quantisation condition restricts the  $4 J$ integrals of motion to a set of curves, parametrised by a continuous variable which can be identified with $\Delta$. In applications to the fishnet CFT, we will  alternatively identify the continuous parameter   on which the spectrum depends with the coupling constant.

Note that \eq{eq:singularities} implies that there is certain relation between a state with dimension $\Delta$ and a state with dimension  $-\Delta$, which is obtained by interchanging the fixed points $x_0\leftrightarrow x_{\bar 0}$
with a conformal map. For definiteness we fix this map to be
\beq
    \label{eq:defF2}
F \equiv K \circ \tilde{\mathcal{I}} \circ K^{-1} 
\eeq
where $K$ is the special conformal transformation  appearing in the decomposition of the twist (\ref{eq:Kdef}), and  $\tilde{\mathcal{I}}$ is the ``holomorphic inversion'' defined as 
\beq\label{eq:holoI2}
\tilde{\mathcal{I}} \circ \left(x^1,x^2,x^3, x^4\right) \equiv \frac{1}{x\cdot x}\left(x^1,-x^2,-x^3,+x^4\right)\;,
\eeq
or, in 6D representation, as $\tilde{\mathcal{I}^{\bf 6}}={\rm diag}\{1, -1, 1, -1, -1, 1\}$. Note, however,
that whereas this transformation keeps the diagonal form of the twist matrix, it flips the parameters  $\alpha_a\to -\alpha_a,\;a=0,1,2$. 
Therefore this transformation inverts the signs of the spins in addition to $\Delta\to-\Delta$ (see (\ref{eq:CartanF})), while exchanging the fixed points $x_0$ and $x_{\bar{0}}$.  As discussed in section \ref{sec:orthogstart} and appendix \ref{app:Ts}, this map  appears in the relation between left and right eigenvectors of the transfer matrices. 

\subsubsection{Convergence of the scalar product}\la{sec:div}
Due to the singular behaviour of the eigenfunctions of the transfer matrices \eq{eq:singularities} the convergence of the scalar product is not obvious.
The general prescription, which may not be particularly practical, is to make an analytic continuation in parameter
space to reach the values of $\Delta$ where the convergence is manifest. This is essentially a $\zeta$-function type of regularisation.
This prescription allows to regularise most of the cases, except for the special case of the scalar product of two wave functions with opposite $\Delta$'s 
in the same frame (i.e. with the same singular points), for example the scalar product of a state and its conjugate. In this case, near the singular point $x_0$ 
we get an integral of the type
\beq
\int |x_\alpha-x_0|^{+\Delta-D_0}\prod_\beta \Box_\beta^{2-h_\beta}
|x_\beta-x_0|^{-\Delta-D_0}\prod_\beta d^4 x_\beta\sim
\int 
|x_\alpha-x_0|^{-4J}\prod_\beta d^4 x_\beta
\eeq
which is log-divergent and thus cannot be regularised in this way. The only option in this case is to introduce the cut-off. In this case the nontrivial object is the coefficient in front of the log, which is likely to be regularisation independent and will play an important role 
\beq\la{logdiv}
\langle\langle
\Psi,\Phi
\rangle\rangle \equiv \log \frac{\epsilon_{x_0}\epsilon_{x_{\bar 0}}}{(x_0-x_{\bar 0 })^2}
\langle\langle
\Psi,\Phi
\rangle\rangle_{fin}+{\text{ sub-leading}}\;.
\eeq

Without analytic continuation in parameters, 
 for generic states  $\langle\langle
\Psi,\Phi
\rangle\rangle$ is divergent near the singularities,
however, we expect the expansion in the small regulator $\epsilon$
to be power-like. In this case we define 
$\langle\langle
\Psi,\Phi
\rangle\rangle$ as the finite part in the small $\epsilon$
expansion, which is a non-ambiguous quantity in the absence of log type of divergences.

\subsection{The Baxter TQ relations}\la{sec:bax}
The Baxter TQ equation gives a complete reformulation for the problem of  diagonalising the transfer matrices, as an equation in just one variable. 
The SoV basis should be possible to obtain by diagonalising ${\bf B}$ and ${\bf C}$ along the lines of \cite{Gromov:2020fwh}.
Instead of doing this, in this paper we will bypass the explicit SoV construction by studying the TQ relation,
which for integrable models with $SL(n)$ global symmetry takes a known form \cite{Krichever:1996qd,Chervov:2006xk}. For the rank-3 case relevant for the fishnet theory, it reduces to
\beq
\mathbb{T}^{\mathbf{1} [+2]} Q^{[+4]} - \mathbb{T}^{\mathbf{4}+} Q^{[+2]}  + \mathbb{T}^{\mathbf{6}} Q -\mathbb{T}^{\bar{\mathbf{4}} [-1]} Q^{[-2]} + \mathbb{T}^{\bar{\mathbf{1}} [-2]} Q^{[-4]} =0\; ,
\eeq
where $\mathbb{T}^{\mathbf{r}}$
are the transfer matrix eigenvalues \eq{TTs}.
After the ``gauge" transformation 
\beq
Q(u) = q(u) e^{\pi u J/2} \prod_{\alpha=1}^J \Gamma\left(-i \left(u-\vartheta_{\alpha}  + i \frac{h_{\alpha}-1}{2}\right) \right),
\eeq
using \eq{TTs} we find that it takes the form
\beq\label{BaxIn}
0=q^{[+4]} \, \mathcal{Q}_+^{[+3]} \mathcal{Q}_+^{[+1]}
-
q^{[+2]} \, \mathcal{Q}_+^{[+1]} \,P^{{\bf 4}+}_{J}
+
q\,P^{\bf 6}_{2J}
-
q^{[-2]}
\mathcal{Q}_-^{[-1]}\, P^{\bar{\bf 4}-}_{J}
+
q^{[-4]} 
\, \mathcal{Q}_-^{[-3]} \mathcal{Q}_-^{[-1]} \ .
\eeq
Being a fourth order difference equation, it has four linearly independent  solutions that we denote  as $q_a(u)$ with $a=1,\dots,4$. For what follows, it is also important to introduce the dual Q-functions $q^a(u)$ which are built as $3\times 3$ determinants times an overall factor,
\beq\label{eq:qupdef0}
q^a(u) \equiv 
\frac{\mathcal{Q}_+^{[+1]}}{\mathcal{Q}_-^{[-1]}}\prod_{\alpha=1}^J\frac{\Gamma^2\left(-i\( u  - \vartheta_\alpha \)+ \frac{h_\alpha}{2} -\frac{1}{2}\right) }{\Gamma^2\left(-i\( u - \vartheta_\alpha\)- \frac{h_\alpha}{2}+\frac{1}{2}  \right)} \,  \,\epsilon^{abcd} \, q_b(u+i) q_c(u) q_d(u-i) \ .
\eeq

Using the results from Appendix \ref{app:dualBproof} we find that these functions $q^a$ satisfy a `dual' Baxter equation, which 
 reads
 \beq\label{eq:dualBax0}
0=q^{[+4]} \, \mathcal{Q}_-^{[+3]} \mathcal{Q}_-^{[+1]}
-
q^{[+2]} \, \mathcal{Q}_-^{[+1]} \,P^{\bar{\bf 4}+}_{J}
+
q\,P^{\bf 6}_{2J}
-
q^{[-2]}
\mathcal{Q}_+^{[-1]}\, P^{{\bf 4}-}_{J}
+
q^{[-4]}
\, \mathcal{Q}_+^{[-3]} \mathcal{Q}_+^{[-1]}
\; .
\eeq
We see that it is related to the Baxter equation \eq{BaxIn} for $q_a$ by simply exchanging $\cQ_+\leftrightarrow\cQ_-,\;P_J^{\bf 4}\leftrightarrow P_J^{\bf \bar4}$.

The four functions $q_a$ are distinguished according to their  asymptotic behaviour for large $u$:
\begin{equation}\label{eq:Qasy}
q_a(u) \simeq \lambda_a^{-i u} \, u^{\hat M_a}, 
\end{equation}
where $\lambda_a$ are the twist eigenvalues, and $\hat M_a$ are charges defined as
\beq\label{eq:asyQMnew}
\hat M_a = \left\{ \tfrac{+\Delta - D_0 -S_1 - S_2}{2}  , \, \tfrac{+\Delta - D_0 + S_1 + S_2}{2}  ,\, \tfrac{-\Delta - D_0 -S_1 + S_2}{2}   , \,\tfrac{-\Delta - D_0 + S_1 - S_2}{2} \right\}. 
\eeq
To fix the basis completely, one should furthermore impose analyticity conditions, as we discuss in section \ref{sec:analyt}. 

The large $u$ asymptotics of $q^a$ reads, similarly to \eq{eq:Qasy} 
\beq
    q^a(u)\simeq C_a\lambda_a^{iu}u^{-\hat M_a-D_0} \; ,
\eeq
with the  normalisation coefficients fixed by \eq{eq:Qasy} as follows
\beq
C_a={(-1)^a}{\lambda_a}{\prod\limits_{b,c\neq a,\;b<c}
(\lambda_c-\lambda_b)
}\;.
\eeq

\subsection{Quantisation condition for the Q-functions}\label{sec:refo}

To complete the solution for the spectrum, one has to further constrain the analytical properties of the  Q-functions, which will lead to the \textit{quantisation condition}. This condition is the Q-function counterpart of the requirement  that the wave functions have no singularities except for the ones in (\ref{eq:singularities}) we discussed above.  The quantisation condition for Q-functions is known at least in the case of weights $h_\alpha \in \left\{1,2\right\}$ which are relevant for applications to the fishnet model \cite{Cavaglia:2020hdb,GrabnerToAppear}, and in this subsection 
we describe how it works. Here we also verified it for generic $h_\alpha$. Once it is imposed, the values of the integrals of motion are fixed to a discrete set, for any given assignment of the twists, inhomogeneities and conformal charges $(\Delta, S_1, S_2)$.

\subsubsection{Analytic properties of Q-functions}\label{sec:analyt}

\paragraph{Solutions of maximal analyticity. } 
Let us define bases of solutions to the Baxter equation \eq{BaxIn}
 with the largest possible region of analyticity.

One possibility is choosing the Q-functions to be analytic for very large positive ${\rm Im}\;u$. Such solutions will be denoted as $q^{\downarrow}$ (the direction of the arrow indicates that one starts from large positive $\text{Im}(u)$, and iterates the Baxter equation to move down). The form of the Baxter equation implies that they unavoidably have infinitely many single poles extending into the lower half plane, located at $u \in \mathcal{P}_{\downarrow}$, with
 \beq\la{Pdn}
 \mathcal{P}_{\downarrow} =\left\{\vartheta_{\alpha} + i\frac{(h_{\alpha} -3)}{2}-i m \right\}_{\alpha=1,\dots,J} \ , \ \ m\in {\mathbb Z}_{\geq 0}\;,
 \eeq
which implies that
as we decrease ${\rm Im}\;u$ from a large positive value, the first poles will occur at  $u=\vartheta_\alpha - i\frac{h_\alpha-3}{2}$, $1\leq \alpha \leq J$, and the other poles will be reached by shifts of $-i$.
 
An alternative basis of solutions, denoted as $q_a^{\uparrow}$, can be chosen by requiring analyticity for large negative $\text{Im}\;u$. The singularities of these functions are  single poles for $u \in \mathcal{P}_{\uparrow}$,
\beq\la{Puup}
 \mathcal{P}_{\uparrow} =\left\{\vartheta_{\alpha} - i\frac{(h_{\alpha} -3)}{2} + i m \right\}_{\alpha=1,\dots,J} \ , \ \ m\in {\mathbb Z}_{ \geq 0}\; .
 \eeq
 These two bases are uniquely defined once their asymptotics -- in the respective domains of analyticity -- are specified as follows
\beqa
q^{\downarrow}_a(u)&\simeq&  \lambda_a^{-i u} u^{\hat M_a} \, \(1+\sum_{n=1} \frac{B_{a, n}}{u^n}\) \;\;,\;\; u \to + i \infty\;,\label{eq:asyQB}\\
q^{\uparrow}_a(u)&\simeq&  \lambda_a^{-i u} u^{\hat M_a} \, \(1+\sum_{n=1} \frac{B_{a, n}}{u^n}\) \;\;,\;\; u \to - i \infty\;.\label{eq:asyQB2}
\eeqa
The coefficients $B_{a,n}$ in the above asymptotic expansions are the same for both sets and can be found systematically from the TQ-relations. The powers
$\hat M_a$ are defined in (\ref{eq:asyQMnew}). 

A useful property is that the Wronskians built with the full set of solutions are explicitly fixed functions of the spectral parameter, generalising the case of differential equations where they are constants.   In the present case, defining 
\beq\label{eq:quantumdet}
q_{1234}(u)  \equiv  \det\limits_{n=-\frac{3}{2},-\frac{1}{2}, \frac{1}{2},\frac{3}{2}}\left\{ q_{1}(u+i n),\; q_{2}(u+i n),\; q_{3}(u+i n),\;q_{4}(u+i n) \right\}
\eeq
we have as a consequence of the Baxter equation~\eq{BaxIn}:
\beq
    \frac{q_{1234}(u-i/2)}{q_{1234}(u+i/2)}=\prod_{\alpha=1}^J\frac{\(u-\vartheta_\alpha+\frac{ih_\alpha}{2}+\frac{i}{2}\)\(u-\vartheta_\alpha+\frac{ih_\alpha}{2}-\frac{i}{2}\)}{\(u-\vartheta_\alpha-\frac{ih_\alpha}{2}-\frac{i}{2}\)\(u-\vartheta_\alpha-\frac{ih_\alpha}{2}+\frac{i}{2}\)} \ .
\eeq
Finding the solution with appropriate analytic properties, we find
\beqa
    q^{\downarrow}_{1234}&\propto&\prod_{\alpha=1}^J\frac{\Gamma \left(-\frac{h}{2}-i (u-\vartheta_\alpha)+1\right) \Gamma \left(-\frac{h}{2}-i
   (u-\vartheta_\alpha)\right)}{\Gamma \left(+\frac{h}{2}-i (u-\vartheta_\alpha)+1\right) \Gamma
   \left(+\frac{h}{2}-i (u-\vartheta_\alpha)\right)}\;,\label{eq:q1234dn}\\
    q^{\uparrow}_{1234}&\propto&\prod_{\alpha=1}^J\frac{\Gamma \left(-\frac{h}{2}+i (u-\vartheta_\alpha)+1\right) \Gamma \left(-\frac{h}{2}+i
   (u-\vartheta_\alpha)\right)}{\Gamma \left(+\frac{h}{2}+i (u-\vartheta_\alpha)+1\right) \Gamma
   \left(+\frac{h}{2}+i (u-\vartheta_\alpha)\right)}\;\label{eq:q1234up}
\eeqa
(with the same constant proportionality coefficient), so we see that the Wronskian is a state-independent function of $u$.

For the solutions of the dual Baxter equation we similarly define two bases
\beq
    {q^a}^{\;\downarrow/\uparrow}(u)\simeq\lambda_a^{iu}u^{-\hat M_a-D_0}\(1+\sum_{n=1}\frac{B^a_{ n}}{u^n}\)\;\;,\;\; u \to \pm i \infty \ .
\eeq
These functions have poles in the sets $\mathcal{P}^{\text{dual}}_{\downarrow}$, $\mathcal{P}^{\text{dual}}_{\uparrow}$, respectively, 
\beqa\label{eq:defPdual}
 \mathcal{P}_{\downarrow}^{\text{dual}} &=&\left\{\vartheta_{\alpha} - i\frac{(h_{\alpha} -1)}{2} - i m \right\}_{\alpha=1,\dots,J} \ , \ \ m\in {\mathbb Z}_{\geq 0} \;, \\
  \mathcal{P}_{\uparrow}^{\text{dual}} &=&\left\{\vartheta_{\alpha} + i\frac{(h_{\alpha} -1)}{2} + i m \right\}_{\alpha=1,\dots,J} \ , \ \ m\in {\mathbb Z}_{\geq 0} \;.
 \eeqa
For any generic choice of the parameters entering the Baxter equation (including the values of the integrals of motion),  $q_a^{\downarrow}$ and $q_a^{\uparrow}$ can be computed numerically with arbitrary precision with the method of \cite{Gromov:2015wca} (see \cite{Gromov:2019jfh} for the fishnet case), which we review in section \ref{sec:numericsreview}. To fix the values of the integrals of motion to a discrete set (depending on the continuous parameter $\Delta$) we need to impose the quantisation condition described below.

\paragraph{Relating the two bases. }
 Since $q_a^{\downarrow}$ and $q_a^{\uparrow}$ are both bases of solutions of the same difference equation, they must be related by a matrix of $i$-periodic functions:
\beq\label{eq:Om1}
q^{\uparrow}_a(u) = \Omega_a^{\;b}(u) \, q^{\downarrow}_b(u) , \;\;\;\; \Omega_a^{\;b}(u) = \Omega_a^{\;b}(u + i)\; .
\eeq
This matrix of coefficients can be {constructed explicitly} in terms of the Q-functions, as a solution of the linear system \eq{eq:Om1}: 
\beq\la{Omegadef}
\Omega_a^{\;b}(u) = \frac{\epsilon^{b b_1 b_2 b_3} \text{det}_{n=-1,\dots,2}\left\{ q_a^{\uparrow}(u-i n) q_{b_1}^{\downarrow}(u-i n) q_{b_2}^{\downarrow}(u-i n)q_{b_3}^{\downarrow}(u-i n) \right\}}{\epsilon^{b_1 b_2 b_3 b_4}\text{det}_{n=-1,\dots,2}\left\{ q_{b_1}^{\downarrow}(u-i n) q_{b_2}^{\downarrow}(u-i n) q_{b_3}^{\downarrow}(u-i n)q_{b_4}^{\downarrow}(u-i n) \right\}}\;.
\eeq

Notice that the denominator in (\ref{Omegadef}) is related by an argument shift to the quantum determinant $q_{1234}$ given by (\ref{eq:quantumdet}). 
From (\ref{Omegadef}), it is easy to see that, for generic positions of the inhomogeneities, $\Omega$ has first order poles at all points of the form $\frac{-i(h_\alpha-1)}{2}+\vartheta_{\alpha} + i \mathbb{Z}$, and no other singularities. Since we have the periodicity $\Omega(u) = \Omega(u + i)$, we have the pole decomposition
\beq\label{eq:poledec}
\Omega_{a}^{\;b}(u)=
\Omega_{0,a}^{\;\;\;\;\;b}+ 2 \pi \,
\sum_{\alpha=1}^J\frac{\Omega_{\alpha,a}^{\;\;\;\;\;b}}{1-e^{2\pi (u-\vartheta_\alpha+i(h_\alpha-1)/2)}} \ .
\eeq
It is also useful to analyse $\text{det}\,\Omega$. For that we can  repackage the equations (\ref{eq:Om1}) as
\beq
\left(\begin{array}{cccc}
q_1^{\uparrow[+2]} & q_1^{\uparrow} & q_1^{\uparrow[-2]} & q_1^{\uparrow[-4]} \\
q_2^{\uparrow[+2]} & q_2^{\uparrow} & q_2^{\uparrow[-2]} & q_2^{\uparrow[-4]}\\
q_3^{\uparrow[+2]} & q_3^{\uparrow} & q_3^{\uparrow[-2]} & q_3^{\uparrow[-4]}\\
q_4^{\uparrow[+2]} & q_4^{\uparrow} & q_4^{\uparrow[-2]} & q_4^{\uparrow[-4]}
\end{array} \right) = \Omega \cdot
\left(\begin{array}{cccc}
q_1^{\downarrow[+2]} & q_1^{\downarrow} & q_1^{\downarrow[-2]} & q_1^{\downarrow[-4]} \\
q_2^{\downarrow[+2]} & q_2^{\downarrow} & q_2^{\downarrow[-2]} & q_2^{\downarrow[-4]}\\
q_3^{\downarrow[+2]} & q_3^{\downarrow} & q_3^{\downarrow[-2]} & q_3^{\downarrow[-4]}\\
q_4^{\downarrow[+2]} & q_4^{\downarrow} & q_4^{\downarrow[-2]} & q_4^{\downarrow[-4]}
\end{array} \right)\; ,
\eeq
and take the determinant of the two sides 
to get 
$q_{1234}^{\uparrow [-1]}=\det\Omega\, q_{1234}^{\downarrow [-1]}$, so that $\det\Omega$  can be fixed explicitly from (\ref{eq:q1234dn}), (\ref{eq:q1234up}) and reads
\beq
\frac{q^{\uparrow[-1]}_{1234}}{q^{\downarrow[-1]}_{1234}}
=\frac{
\sin^2\pi\left(+\frac{h}{2}+\frac{1}{2}+i (u-\vartheta_\alpha)\right)
}
{
\sin^2\pi\left(-\frac{h}{2}+\frac{1}{2}+i (u-\vartheta_\alpha)\right)
}=\det\Omega\;.
\eeq
For integer weights $h_{\alpha} \in \mathbb{Z}$ (which includes the case of the fishnet model when $h_\alpha$ are 1 or 2), this implies that $\Omega$ has unit determinant.

Since the dual Q-functions are defined as determinants, we also have the following relations
\beq\la{qOq}
q^{a\uparrow}(u) = \bar{\Omega}^{\;a}_{b}(u) \, q^{b\downarrow}(u)
\;\;,\;\;
q^{\downarrow}_a(u) = \bar{\Omega}^{\;b}_{a}(u) \, q_b^{\uparrow}(u)\;,
\eeq
where
\beq
\bar {\Omega}^{\;a}_{b}=
\det\Omega\;\;({\Omega}^{-1})^{\;a}_{b}\;.
\eeq

\subsubsection{Quantisation condition}\label{sec:quantQ}
The quantisation condition found in \cite{Cavaglia:2020hdb,GrabnerToAppear}  simply reads:
\beq\label{eq:quant}
\Omega_{1}^{\;2}(u)=\Omega_2^{\;1}(u) = \Omega_3^{\;4}(u) = \Omega_4^{\;3}(u) = 0 \;.
\eeq
It has been previously verified for the cases when $h_\alpha$ take values $1$ or $2$, and we have also tested it numerically for non-integer values of $h_\alpha$ in the vicinity of $1$, so we expect it to be applicable for rather generic $h$ as well.
 We believe that (\ref{eq:quant}) is completely equivalent to the quantisation conditions for wave functions described in section \ref{sec:quantwf}. Indeed, this can be established for the $J=1$ case where we constructed the SoV map explicitly in \cite{Cavaglia:2020hdb}. It should be possible to prove it in general using the operatorial SoV  formalism. 

The conditions (\ref{eq:quant}) correspond to setting to zero $4 \times J$ coefficients entering the pole decompositon (\ref{eq:poledec})\footnote{In principle, every matrix element of $\Omega$ is parametrised by $J+1$ constants as in (\ref{eq:poledec}). However, it is possible to show that the vanishing of residues is sufficient for the full matrix element to vanish.}. It can be satisfied by tuning appropriately the values of the integrals of motion entering the Baxter TQ relation as parameters. 
More precisely, we find that, for fixed twists and inhomogeneities, (\ref{eq:quant}) admits several families of solutions, each family described by a one-parameter flow in the space of the integrals of motion.  
 As already mentioned, in the case of the fishnet theory the continuous parameter can be identified with the coupling constant.  

\paragraph{The gluing matrix. }
When (\ref{eq:quant}) are satisfied, the matrix $\Omega$ has a further  enhanced  symmetry: there exists a constant matrix (called the \emph{gluing matrix}) of the form
\beq\label{eq:glue}
\Gamma^{ab}=\left(
\begin{array}{cccc}
 0 & -\gamma  & 0 & 0 \\
 \gamma  & 0 & 0 & 0 \\
 0 & 0 & 0 & -1/\gamma \\
 0 & 0 & 1/\gamma & 0 \\
\end{array}
\right)
\eeq
(where the constant $\gamma$ is nontrivial and depends on the coupling), such that 
\beq\label{eq:symom}
\Gamma\Omega=\Omega^T \Gamma\; .
\eeq
The existence of this matrix reflects  the structure of the Quantum Spectral Curve of the parent theory $\mathcal{N}$=4 SYM, as  discussed in \cite{Cavaglia:2020hdb}, namely
the antisymmetric $i$-periodic matrix  $\omega=(\Gamma\Omega)^{-1}$ also appears in QSC. 
We checked numerically that this symmetry holds also for some generic values of $h$, beyond the fishnet model. 
It would be interesting to find a direct proof for the existence of the gluing matrix, using only the conformal spin chain  setup. Some nontrivial consequences of (\ref{eq:symom}), relevant for this purpose, were found  in \cite{GrabnerToAppear} and are reviewed in appendix \ref{app:Qplus}. 

\paragraph{Useful identities.}
We notice that one can use the gluing matrix $\Gamma^{ab}$ to raise indices
and $\Gamma^{-1}_{ab}$ to lower them.
For what follows it will be  convenient to introduce the notation\footnote{In this sub-section we restrict to the case when $h_\alpha$ are integer (such as in the fishnet CFT) so that ${\rm det}\;\Omega=1$.}
\beq\label{eq:pdef}
p_a(u) \equiv ( \Gamma^{-1})_{ab} q^b(u)\;\;,\;\;
p^a(u) \equiv ( \Gamma)^{ab} q_b(u)\;.
\eeq
These newly introduced functions  satisfy the same Baxter equations as 
$q^a$ and $q_a$ respectively.
They only differ by the way they transform 
between the UHP and LHP bases
$
p^{\downarrow}_a(u) = \Omega_a^{\;b}(u) \, p_b^{\uparrow}(u)\;,
$
i.e. they transform in the same way as $q_a$ in
\eq{eq:Om1} but with arrows reversed. It is convenient to introduce
\beq\la{omeg}
\omega_{ab}\equiv [(\Gamma\Omega)^{-1}]_{ab}=
\omega_{0,ab}+ {2 \pi }
\sum_{\alpha=1}^J\frac{\omega_{\alpha,ab}}{e^{2\pi (u-\vartheta_\alpha-i(h_\alpha-1)/2)}-1} \ ,
\eeq
so that we have
\beq\label{eq:Omw}
q^{\downarrow}_a(u) =\omega_{ab}(u) \, p^{\uparrow b}(u)\;\;,\;\;
p^{\uparrow}_a(u) =
\omega_{ab}(u) \, q^{\downarrow b}(u)\;,
\eeq
where as a consequence of \eq{eq:symom}, $\omega$
is anti-symmetric when the quantisation condition is satisfied.

\subsection{Orthogonality properties}\label{sec:orthogstart}
In this section, we discuss orthogonality properties for the eigenstates of the integrable system and relations between left and right eigenvectors of monodromy matrices.

Given the scalar product (\ref{eq:scalar}), we can construct the ``transpose" of the transfer matrices $( \hat{\mathbb{T}}^{{\bf r}} )^{T} $ with respect to this quadratic form, such that
\beq\label{eq:deftranspose}
\langle \langle f \,,\, \hat{\mathbb{T}}^{{\bf r}} \circ g \rangle \rangle = \langle \langle ( \hat{\mathbb{T}}^{{\bf r}} )^{T} \circ f \,,\,  g  \rangle \rangle ,
\eeq
for any $f,g \in \mathcal{F}_{\bh}$. 
As we know, all conformal generators are antisymmetric w.r.t. this bilinear form, so that 
$(\hat{\mathbb{T}}^{{\bf r}} )^{T}$ differ from
$\hat{\mathbb{T}}^{{\bf r}} $ by the replacement $q^{MN}\to-q^{MN}$ or, as we show in appendix~\ref{app:Ts}, this is equivalent to first exchanging $x_0$ and $x_{\bar 0}$ with the holomorphic inversion \eq{eq:defF2}, and then also reversing the order of particles in the chain.
As the TQ relations are not sensitive to the order of particles this shows, as expected, that the right and left eigenvalues of all the integrals of motion are the same (as would be in the finite-dimensional case), but the explicit relation between the left and right eigenvectors in the case of general $\vartheta_\alpha$'s and $h_\alpha$'s could be complicated (see an example of the map interchanging particles in some particular cases in \cite{Gromov:2019jfh}).
In the simplest case when all inhomogeneities and the weights are equal, the left and right eigenvectors of the integrals of motion are simply related by the holomorphic inversion \eq{eq:defF2} and relabelling coordinates $x_\alpha\to x_{J-\alpha+1}$.

With this insight in mind, we should have that the left and right eigenstates corresponding to different eigenvalues of the transfer matrices are orthogonal, just like in the finite-dimensional case.
At the same time, if the eigenvalues are equal then we are in the situation discussed around \eq{logdiv} where we concluded that the scalar product is log-divergent and the meaningful object is the finite coefficient in front of the logarithm. Furthermore, in the generic situation the spectrum is expected to be non-degenerate.

\section{Fishnet model as a spin chain}\la{sec:fn}
The goal of this section is to relate the abstract spin chain formulation with the field theory observables.
More precisely, we relate it to the planar fishnet CFT  desribed in the Introduction and defined by
\beq
\mathcal{L} = N_c \text{Tr} \left( \partial_{\mu}\phi_1^{\dagger}\partial^{\mu} \phi_1 + \partial_{\mu}\phi_2^{\dagger}\partial^{\mu} \phi_2 + (4 \pi)^2 \xi^2 \phi_1^{\dagger} \phi_2^{\dagger} \phi_1 \phi_2 \right)\;.
\eeq
\subsection{Operators and wave functions}\label{sec:wavefs}
The setup we are using in this paper includes the so-called colour-twist operators, which generalise the  traditional local single-trace operators. 
For the details of the construction we refer to~\cite{Cavaglia:2020hdb}. Below we briefly review the construction. 
\paragraph{Colour-twist operators.}
We consider the \emph{colour-twisted} version  of local single-trace operators, introduced in \cite{Cavaglia:2020hdb}. This gives a deformation of the concept of a local operator, depending on a symmetry transformation $G$, which in our case will be a generic conformal transformation with two fixed points. 
The deformation removes degeneracies in the spectrum and as a result makes the SoV construction more regular and uniform.  The presence of the twists is also technically convenient and will make the form of our results much more transparent.

Perturbatively, twisted single-trace operators are constructed  as 
\beq
\mathcal{O}(x_0) = \text{Tr}\left(\bar\partial_{1}^n\partial_{1}^{S_1-n}
\bar\partial_{2}^n\partial_{2}^{S_2-n}
\, \phi_1^J(x) \, \phi_2^M(x) \, (\phi_1(x) \phi_1^{\dagger}(x) )^{m_1} \, (\phi_2(x) \phi_2^{\dagger}(x) )^{m_2}   \mathcal{T}_{G} \right) + \dots ,
\eeq
where the dots stand for possible mixing with similar operators
and $\d_{1}=\d_{x_1}-i\d_{x_2},\;\d_{2}=\d_{x_3}-i\d_{x_4}$.
The marker $\mathcal{T}_G$ indicates the starting point of a ``twist-cut'' on the worldsheet of the planar Feynman graph.
Each propagator crossing this line gets deformed in accordance with the conformal group element $G$. For details
and examples of the construction see~\cite{Cavaglia:2020hdb}. 

\paragraph{Symmetries. }
The correlators involving twisted operators transform in a more complicated way under the conformal symmetry,
 since one should also keep track of the transformation of the colour-twist. Under the conformal transformation $C \in SO(1,5)$, the twist map transforms as $G \rightarrow \tilde G \equiv C G C^{-1}$, see Appendix \ref{app:maps}. 
This in particular means that all the usual degeneracies of the usual local operators are lifted -- in particular the operators transform nontrivially under the translation symmetry removing the necessity of considering conformal primaries and descendants separately.
Since the twist $G$ remains invariant under
$3$ generators in the Cartan subalgebra $\hat{\mathbb{Q}}_0 , \hat{\mathbb{Q}}_1 , \hat{\mathbb{Q}}_2$ defined in (\ref{eq:confres}) the twisted operators can be classified by the corresponding $3$ quantum numbers $\mathbb{Q}_a$, which are two spins and the scaling dimension. 

As we will see the scaling dimensions $\Delta$ of such operators can be computed non-perturbatively using integrability and are nontrivial functions of the coupling constant $\xi$, as well as the twist angles $\lambda_a$. We can recover the value of the scaling dimension for a standard local  operator by taking the untwisting limit $\lambda_i \rightarrow 1$. 

In addition, composite operators of the fishnet theory carry quantum numbers $J, M \in \mathbb{Z}$, associated to the $U(1)\times U(1)$ charges of the two scalar fields of the fishnet model. 

The fishnet Lagrangian admits a $\mathbb{Z}_4$ discrete symmetry generated by the elementary move\footnote{We thank D. Anninos for   pointing this out to us, and for interesting discussions, see also~\cite{Gromov:2018hut}.} $\left\{ \phi_1 , \phi_2, \phi^{\dagger}_1 , \phi_2^{\dagger} \right\}\rightarrow \left\{ \phi_2, \phi^{\dagger}_1 , \phi_2^{\dagger} , \phi_1\right\}$. This symmetry relates a quadruplet of states with charges $(J, M)$, $(-M , J )$, $(-J, -M )$, $(M, -J)$, leaving invariant their spectrum and wave functions.  Thanks to this symmetry, we can restrict our analysis to the case  $0\leq M \leq J$ without loss of generality.

In accordance with the connection with the conformal spin chain (reviewed below), we will  refer to $J$ as the \emph{length of the operator} (i.e. the number of $\phi_1$ fields\footnote{More precisely, $J$ is the number of $\phi_1$ fields minus the number of $\phi_1^{\dagger}$ fields. However, we are interested in the case of non-protected operators, which requires only $\phi_1$ to be present, as otherwise no Feynman diagrams are possible beyond tree level.}), and $M$ the \emph{number of magnons} (i.e. the number of $\phi_2$ fields minus the number of $\phi_2^\dagger$ fields).  
The case $M=J$ needs some special treatment (see appendix \ref{app:Qplus}), and for simplicity in this paper we restrict the analysis to the choice of quantum numbers $|M| < J$. 

It is known that there also exist infinitely many  non-dynamical operators in the fishnet theory, which have zero anomalous dimension at the planar level. 
In this paper, we will mostly focus on the operators ${\cal O}$ with  non-zero anomalous dimensions. 

Finally, notice that, since the fishnet theory is non-unitary,  the dilatation operator is strictly speaking not diagonalisable. Indeed, examples of  Jordan blocks of the anomalous dimension matrix have been studied in \cite{Caetano:2016ydc,Gromov:2018hut,Ipsen:2018fmu}. In all known  examples it was found that the generalised eigenvalues of the non-diagonalisable   Jordan blocks are zero,  which can be verified at all loops in some cases by diagrammatic arguments \cite{Gromov:2018hut}. 
It is our working assumption that this is true in general (at least at the planar level), so that the only operators acquiring anomalous dimensions are proper eigenvectors of the dilatation operator, which do not mix with the non-diagonalisable Jordan blocks. We believe that all such nontrivial operators are captured by the integrability methods discussed in this paper. 

\paragraph{The CFT wave function.}
All information on a single-trace operator can be encoded into a special kind of correlator called the \emph{CFT wave function}~\cite{Gromov:2019bsj}, which will play the role of the the spin chain wave function. It is defined as the renormalised correlator
\beq\label{eq:wfdef}
\varphi_{\mathcal{O}}(x_1, x_2, \dots, x_J ) 
=\langle \mathcal{O}(x_0) \; \text{Tr}\left( \chi_{{\bf I}_1}(x_1) \chi_{{\bf I}_2}(x_2)  \dots \chi_{{\bf I}_J}(x_J) \mathcal{T}_{G^{-1} } \right) \rangle\; ,
\eeq
where the indices ${\bf I}_\alpha \in \left\{0,1,-1,\bar 0\right\}$, and the fields in the second trace are
\beqa
\chi_0(x) &\equiv& \phi_1^{\dagger}(x)\;,\\
\chi_1(x) &\equiv& \phi_2^{\dagger}(x) \phi_1^{\dagger}(x)\;,\\
\chi_{-1}(x) &\equiv& \phi_1^{\dagger}(x) \phi_2(x)\;,\\
\chi_{\bar 0}(x) &\equiv& \phi_2^{\dagger}(x)\phi_1^{\dagger}(x)\phi_2(x)\;.
\eeqa
In order for the correlator \eq{eq:wfdef}
to be non-zero, these indices have to satisfy 
$M = \sum_{\alpha=1}^J {\bf I}_\alpha$, where $M$ is the number of magnons in the operator $\mathcal{O}$. 
We see that there is a weight $h_\alpha=1+|{\bf I}_\alpha|$ associated with each $x_\alpha$.

The conformal charges of the operator ${\cal O}$
can be extracted from the CFT wave-function by acting only on the  $x_1,\dots,x_J$ coordinates. Namely, the charges defined as in  (\ref{eq:Qrel}):
\beq
 \hat{\mathbb{Q}}_{a} \circ   \varphi_{\mathcal{O}}( x_1, \dots, x_J ) =\mathbb{Q}_a\varphi_{\mathcal{O}}( x_1, \dots, x_J ) , \;\;\; a=0,1,2
\label{eq:chargeswf}\; ,
\eeq
are related to the scaling dimension and spins of the operator:\beq
\mathbb{Q}_0 = i \Delta^{\mathcal{O}} ,
\;\;\; \mathbb{Q}_i =  S^{\mathcal{O}}_i , \;\;i=1,2. 
\eeq
This follows from the conformal invariance of the correlator, which involves transformation of all $J+1$ local operators in the correlator defining the CFT wavefunction. So by transforming only the $J$ protected legs,  we indirectly measure the quantum numbers of ${\cal O}$.

We see that there is a clear analogy between the CFT-wave function and the wave function of the conformal spin chain (or fishchain) which we introduced in the previous section. In order to complete the embedding, one needs to show that the CFT wavefunction diagonalises the transfer matrix for some values of inhomogeneities $\vartheta_\alpha$. This was done in~\cite{Gromov:2019jfh}, and in the next section we will briefly review the main elements of the construction.

\begin{figure}[t!]
\begin{center}
\includegraphics[width=10cm]{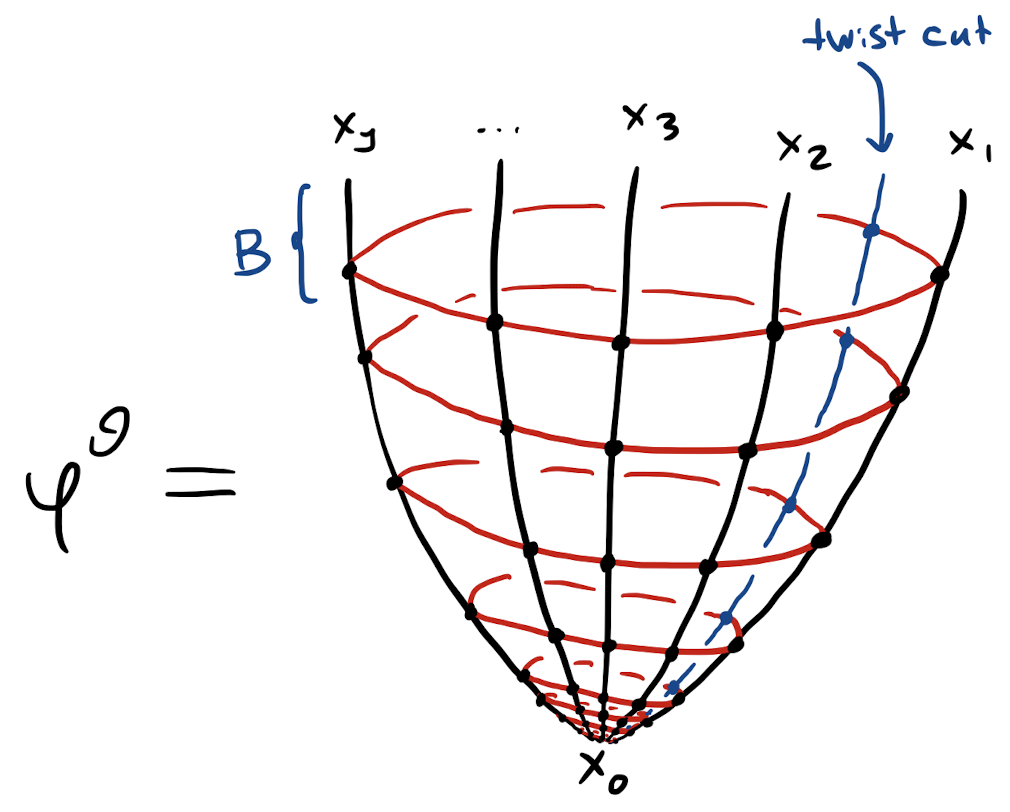}    
\end{center}
\caption{Feynman graphs contributing to the  CFT wave function
has an iterate structure.
The blue line crossing propagators indicates that these propagators are twisted with the global symmetry $G$.\la{CFTwave}}
\end{figure}

\subsection{CFT wave functions as integrable fishchain eigenstates}\label{sec:wfspinchain}
In this section we review the construction of~\cite{Gromov:2019jfh} which shows that the CFT wave functions are also eigenfunctions 
of the integrable spin chain defined in section \ref{sec:fishchain}. The basis for this consideration is the Dyson-Schwinger equation satisfied by the wave functions.
For simplicity we review the case $M=0$ and then present the general result.

\paragraph{The Dyson-Schwinger evolution equation.}
Based on the Feynman rules of the fishnet theory, the wave functions for physical operators are built out of Feynman diagrams which form an infinite ladder. Each rung of the ladder is an iteration of a  \emph{graph-building operator} $\hat B$. This is an integral operator acting on a function of $J$ variables  as\footnote{The first Jacobian under the integral comes from the rules in \cite{Cavaglia:2020hdb}, where the twisted propagator is given by $\frac{\left|\frac{\partial G }{\partial y}\right|^{\frac{1}{4} } }{(x - G(y) )^2 }$. It ensures that the graph-building operator is  covariant under conformal transformations, namely it satisfies the same equation (\ref{eq:brokenconf}) as the transfer matrices.
} \cite{Gurdogan:2015csr,Gromov:2018hut}:
\beq\label{eq:graphB}
\hat B \circ f(x_1, \dots, x_J) = \xi^{2 J} \, \int \, \left| \frac{\partial G(y_J) }{\partial y_J} \right|^{\frac{1}{4} } \, \prod_{\alpha=1}^J \frac{d^4 y_\alpha}{\pi^2}
\frac{1}{(x-y_\alpha)^2(y_\alpha-y_{\alpha-1})^2}
 \; f(y_1 , \dots, y_J) \;,
\eeq
Twisted periodic boundary conditions imply that $y_{0} \equiv G \circ y_J$, where $G$ is the twist map, corresponding to the convention illustrated in figure~\ref{CFTwave} for the twisted trace.

The ladder structure shown in 
figure~\ref{CFTwave}
implies that the CFT wave function satisfies a Dyson-Schwinger evolution equation: 
\beq\label{eq:DysSchw}
\hat B \cdot \varphi_{\mathcal{O}}( x_1, x_2, \dots, x_J ) = \varphi_{\mathcal{O}}( x_1, x_2, \dots, x_J ) \;.
\eeq
The integral operator $\hat B$ can be shown to commute with the Cartan generators 
$\hat {\mathbb Q}_a$.
Furthermore, we will see that the integral operator $\hat B$ belongs to a family of mutually commuting operators.

A useful property found in \cite{Gromov:2019bsj,Gromov:2019jfh} is that
even for the case with magnons, for $|M| < |J|$, the inverse of the graph-building operator is a differential operator. To demonstrate this property for $M=0$ case we notice that
\beq\label{eq:invB}
\xi^{2 J} \,\hat B^{-1} = \hat H \equiv \frac{1}{(-4)^J} 
\left|\frac{\partial G(x_J) }{\partial x_J} \right|^{-\frac{1}{4} }
\,  \prod_{\alpha=1}^J  x_{\alpha, \alpha-1}^2  \prod_{\alpha=1}^J \Box_{x_\alpha}\; ,
\eeq
(with boundary conditions $x_{\alpha-1} = G(x_J)$ for $\alpha=1$) 
and the Dyson-Schwinger equation can also be written as
\beq\label{eq:Schrod}
\hat H_{\vec x}\cdot \varphi_{\mathcal{O}}( x_1, x_2, \dots, x_J ) = \xi^{2 J} \, \varphi_{\mathcal{O}}( x_1, x_2, \dots, x_J ) \;,
\eeq 
which is a partial differential equation. 
Similar relation holds for the case with magnons too. 
Note that \eq{eq:Schrod} is satisfied everywhere except for the singular points $x_0$ and $x_{\bar 0}$,
where the behaviour of the wave function is dictated by the conformal dimension~\eq{eq:singularities}.

\paragraph{Integrability. }
At the heart of the integrability of the fishnet CFT is the fact that the Hamiltonian $\hat H$ (and its generalisation for the case with magnons) is embedded in the family of transfer matrices introduced in section \ref{sec:fishchain}. 
 It was proved in \cite{Gromov:2019jfh} that
\beq\label{eq:inverseB}
\hat H = \lim_{u\rightarrow 0} \hat{ \mathbb{T} }^{\mathbf{6}}(u)  = \hat I_{(0,1)}\;.
\eeq
For \eq{eq:inverseB} to be true one should  properly adjust the values of $h_\alpha$ and $\vartheta_\alpha$. The correct values of
$h_\alpha$ and $\vartheta_\alpha$ depend on the field content of the given site~\cite{Gromov:2019jfh}:
\beq
\bea{c|l|c|c}
{\bf I}_\alpha& \text{fields} & h_\alpha & \vartheta_\alpha \\ \hline
0 & \phi_1^{\dagger}(x_\alpha) & 1 & 0\\
+1 & \phi_2^{\dagger}(x_\alpha) \phi_1^{\dagger}(x_\alpha) & 2 & +\tfrac{i}{2}\\
-1 & \phi_1^{\dagger}(x_\alpha) \phi_2(x_\alpha) & 2 & -\tfrac{i}{2}\\
\bar 0 & \phi_2^{\dagger}(x_\alpha)\phi_1^{\dagger}(x_\alpha)\phi_2(x_\alpha) & 3 & 0
\eea\la{inhom}
\eeq
As a consequence of \eq{eq:inverseB}
the Hamiltonian (i.e., the inverse graph building operator) commutes with all the integrals of motion for the fishchain, and we can use all the power of integrability to diagonalise it. Even though  (\ref{eq:inverseB}) requires a special tuning of the inhomogeneities \eq{inhom}, we will still consider the general case in order to have more integrable parameters. 
In the following we introduce a convenient notation for the shifted inhomogeneities:
\beq\label{eq:shiftedtheta}
{\theta}_{\alpha} \equiv \vartheta_{\alpha} - i\frac{{\bf I}_{\alpha}}{2}\;, 
\eeq
so that the undeformed Hamiltonian is recovered when all $\theta_{\alpha} \rightarrow 0$. 

In the light of (\ref{eq:inverseB}), we see that the Hamiltonian is identified with the integral of motion $\hat{I}_{(0,1)}$.  For physical states solving the Schr\"odinger equation (\ref{eq:Schrod}), we can relate its eigenvalue to the coupling constant
\beq\label{eq:I31}
I_{(0,1)} \equiv \xi^{2 J} \;.
\eeq
This is the key equation for introducing the 't Hooft coupling  into the integrability formulation \cite{Gromov:2019bsj,Gromov:2019jfh,GrabnerToAppear}. 

The spectrum of this spin chain is somewhat unusual in comparison to more standard representations with highest weight. Usually 
the spectrum is discrete -- only some particular values of the integrals of motion can appear in the spectrum.
In the present case for any fixed $\Delta$ and integer $S_1,S_2$ we find an eigenfunction satisfying the correct quantisation condition \eq{eq:singularities} so that the eigenvalues of the integrals of motion $I_{(n,\alpha)}(\Delta,S_1,S_2)$ are in general multi-valued functions of $\Delta$. The physical discrete spectrum 
$\Delta_n(\xi)$ is then found by imposing the condition \eq{eq:I31}.

The numerical method described in~\cite{Gromov:2019jfh}, based on the TQ relation and the quantisation condition from section~\ref{sec:quantQ}, allows one to find the values of $\Delta_n(\xi^2)$ with high precision.

\subsection{The meaning of the spin chain norm in CFT}\label{sec:normmeaning}
\begin{figure}[t!]
\centering
\includegraphics[scale=0.5]{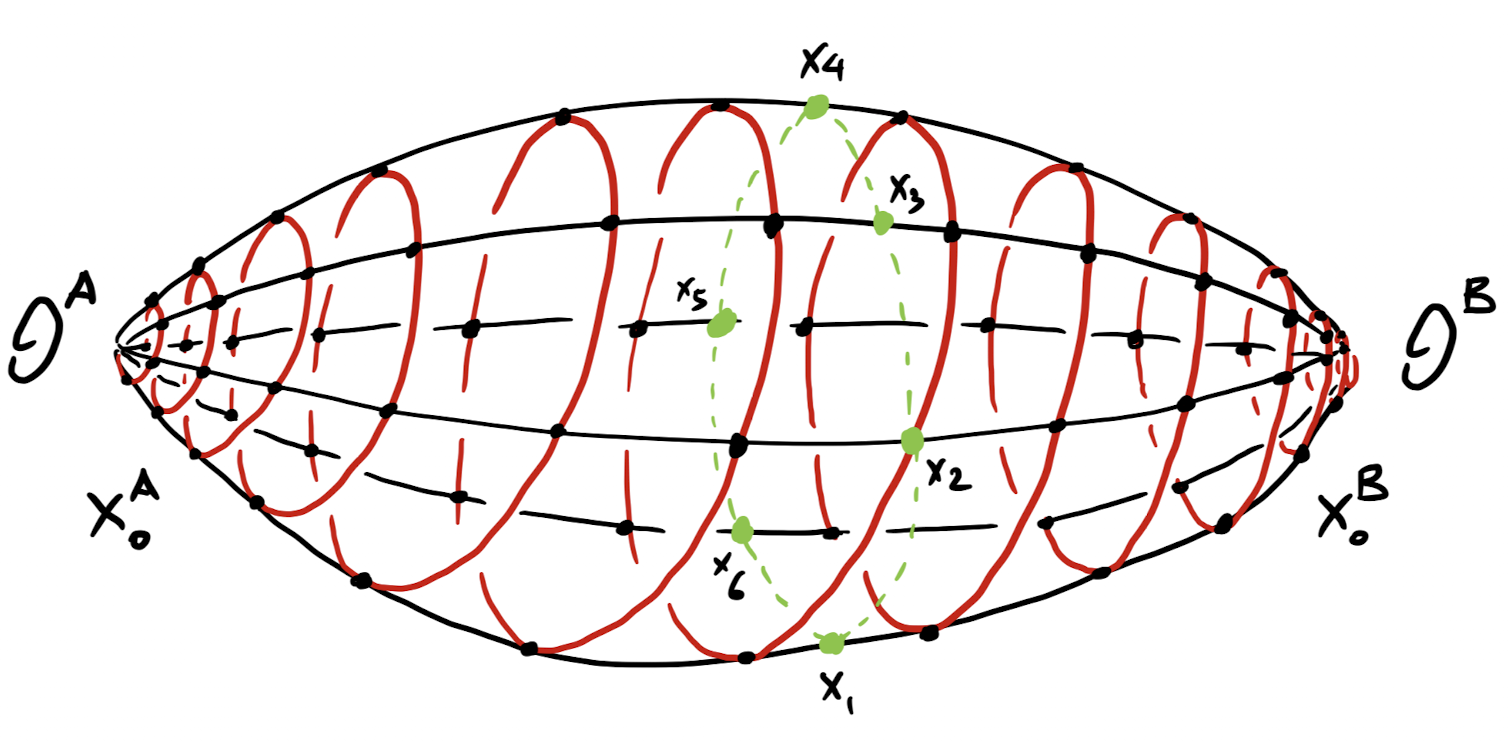}
\caption{\label{fig:scalar} The scalar product between two wave functions. The $\Box$ factors included in the definition of the scalar product remove some propagators, so that the two wave functions are joined seamlessly without altering the structure of the vertices.}
\end{figure}

In order to complete the embedding of the correlators of local operators into the general fishchain framework, 
we have to give an interpretation of the scalar product of two CFT wave functions, according to \eq{eq:scalar}.
To understand its meaning let us consider a particular case
of two operators of length $J=6$ with one magnon spiralling around $M=1$ as in fig.~\ref{fig:scalar}.
The scalar product~\eq{eq:scalar} is given by the product of these two wave functions with powers of d'Alembertian in between integrated over $J$ variables. In the case of
figure~\ref{fig:scalar}, all weights $h_\alpha$ are equal to $1$ except for $h_2=2$, which means that for all points we get d'Alembertian to the first power except the site with the magnon, where no d'Alembertian is added. Note that this is perfect from the Feynman diagram perspective, as we get one of two black propagators annihilated by the d'Alembertian at non-magnon sites, while where the magnons are sitting we glue them into the fishnet interaction $4$-vertex. Thus the scalar product gives a correlation function of two twisted operators.

The case when the operator $B$ is related to the operator $A$ by the holomorphic inversion 
\eq{eq:defF2} is particularly important.
Naively we get a two-point correlator in this case, however, as was discussed in section~\ref{sec:div} the scalar product is log-divergent, unlike the two point correlator which should be finite for two normalised operators. The reason for this discrepancy is that the 
combinatorics of the diagrams is a bit different: for example there are two diagrams with one wheel in the scalar product and only one in the two point correlator.
As was shown in~\cite{Gromov:2019bsj} and we review in appendix~\ref{app:proofdDelta}, the relation is as follows:
\beqa\label{eq:dDelta}
\langle \langle \tilde\varphi \; , \;{\varphi}  \rangle\rangle &=&  \log \frac{
\epsilon_0\epsilon_{\bar 0}}{(x_0-x_{\bar 0})^2}
 \times
 \frac{2}{J} \xi^2 \,\frac{\partial \Delta}{\partial \xi^2 } \times   \langle 
 \mathcal{O}(x_0)
 \tilde{\mathcal{O}}(x_{\bar{0}})
 \rangle
 \ ,
\eeqa
where in the r.h.s. we get the two-point function:
\beq
 \langle 
 \mathcal{O}(x_0)
 \tilde{\mathcal{O}}(x_{\bar{0}})
 \rangle
  = 
 \frac{\mathcal{N} }{(x_0 - x_{\bar{0}})^{ 2\Delta }}\;,
\eeq
which has the same kinematical dependence as in any CFT, even in the presence of twists \cite{Cavaglia:2020hdb} (see Appendix \ref{app:maps}).  
 The coefficient of the two-point function itself can be understood -- at least formally -- as a scalar product between a nontrivial state, and the same state extrapolated to zero coupling.\footnote{This quantity would have a power-type divergence and require multiplicative renormalisation,  for example by changing slightly the twist of one of the states so that they are not orthogonal.}

\subsection{Baxter TQ relations in the fishnet theory}\label{sec:refoquant}
In this section we speciallise the results of section~\ref{sec:bax} to the fishnet case, where $h_\alpha$ is either $1$ or $2$. 

Assuming $M\geq0$, we can always get rid of the anti-magnons by redefining the Q-functions as was discussed in \cite{Gromov:2019jfh}. 
For the sake of clarity, we will pick  a particular order for the magnons, and take ${\bf I}_i=1$ for $1\leq i\leq M$ and ${\bf I}_i=0$ for $M+1\leq i\leq J$ (different orderings are  related by a similarity transformation~\cite{Gromov:2019jfh}).
In this case, the Baxter equation~\eq{BaxIn} can be written explicitly as
\beqa\label{eq:Bax}
\nn
0&=&q^{[+4]}\prod_{\alpha=1}^M(u-\theta_\alpha)
\prod_{\beta=1}^J(u-\theta_\beta+i)
-
q^{[+2]}\prod_{\alpha=1}^M(u-\theta_\alpha)P^{\bf 4}_{J}(u+\tfrac{i}{2})
+
q(u)\frac{P^{\bf 6}_{2J}(u)}{\prod\limits_{\alpha=M+1}^J(u-\theta_\alpha)}
\\ 
&-&
q^{[-2]}
\prod_{\alpha=1}^M(u-\theta_\alpha-i)P^{\bar{\bf 4}}_{J}(u-\tfrac{i}{2})
+
q^{[-4]}
\prod_{\alpha=1}^M(u-\theta_\alpha-2i)
\prod_{\beta=1}^J(u-\theta_\beta-i)\;,
\eeqa
containing the shifted inhomogeneities (\ref{eq:shiftedtheta}). We denote the $4$ solutions of the above equation as $q_a$ (with lower index).
When $\theta_\alpha\to0$
we get the actual fishnet Baxter equation of~\cite{Gromov:2019jfh}. We will keep $\theta_\alpha$ arbitrary as an important regulator. In the simplest case of no magnons, and with $\theta_\alpha$ set to zero we get
\beqa\label{eq:Baxzero}
\nn
0&=&q^{[+4]}
(u+i)^J
-
q^{[+2]}P^{\bf 4}_{J}(u+\tfrac{i}{2})
+
q(u)\frac{P^{\bf 6}_{2J}(u)}{u^J} 
-
q^{[-2]}
P^{\bar{\bf 4}}_{J}(u-\tfrac{i}{2})
+
q^{[-4]}
(u-i)^J\;.
\eeqa
The dual Q-function,  constructed from $q_a$'s as a $3\times 3$ Wronskian.
In the case of the fishnet, it is convenient to introduce an extra multiplier w.r.t. \eq{eq:qupdef0},  $\prod_{\alpha=M+1}^{J}(u-\theta_\alpha)$, which simplifies the form of the dual Baxter equation for this particular choice of weights:
\beq
\label{qdnew}
 q^a(u) \equiv \left( \prod_{l=1}^J (u-\theta_l) \right) \left( \prod_{m=1}^M (u- \theta_m -i ) \right)\,\epsilon^{abcd} \, q_b(u+i) q_c(u) q_d(u-i)\; .
\eeq
 The dual Baxter equation has a form simply related to (\ref{eq:Bax}) by the interchange of the polynomials $P^{\bar{\bf 4}}_{J}(u) \leftrightarrow P^{{\bf 4}}_{J}(u)$.

Like before in section~\ref{sec:refo}, there are two distinguished bases for the Q-functions: $q_a^\uparrow$ that are lower-half-plane (LHP)  analytic and $q_a^\downarrow$ that are upper-half-plane (UHP) analytic, and similarly for the dual $q^a$'s.
The positions of the poles \eq{Pdn} and \eq{Puup}
simplify to
\beqa
\mathcal{P}_{\downarrow} &=& \left\{ \theta_{\alpha} - i m\; | \; m\in \mathbb{Z}_{>0} \right\}_{\alpha=1}^J \cup \left\{ \theta_{\alpha} \right\}_{\alpha=1}^M\; ,\\
\mathcal{P}_{\uparrow} &=& \left\{ \theta_{\alpha} + i m\; | \; m\in \mathbb{Z}_{>0} \right\}_{\alpha=1}^J\;,
\eeqa
and the same sets describe the poles for the dual  $q^{a \downarrow}$ and $q^{a \uparrow}$. This is indeed consistent with (\ref{eq:defPdual}),  taking into account that $q^a$ in this section are defined with an extra factor $\prod_{\alpha=M+1}^{J}(u-\theta_\alpha) $, which cancels some of the poles in $\mathcal{P}^{\text{dual}}$.  Notice that, for real inhomogeneities, the poles of $q^{\downarrow}$ are in the lower half plane or at most on the real axis, and the poles of $q^{\uparrow}$ are all in the upper half plane.

\section{SoV scalar product cookbook}
\label{sec:sovq}

Even though in this paper we have not attempted building the SoV basis, along the lines of recent papers~\cite{Gromov:2020fwh,Gromov:2019wmz,Cavaglia:2019pow} a lot of structure of the result can be predicted based on simple observations such as analyticity of Q-functions and TQ relations.
In this section we describe one of the main 
results of the paper -- the bilinear form built from two Q-functions, which should be a building block for the scalar product in the SoV basis. We will also see how the quantisation condition
described in section~\ref{sec:refo} naturally appears from the arguments of this section.

\subsection{Bilinear forms of Q-functions}
Here we introduce a set of bilinear forms pairing two Q-functions.
The construction is inspired by the findings in~\cite{Cavaglia:2018lxi}. We introduce the following notation for the bilinear form of two 
Q-functions, in general for different two states $A$ and $B$:
\beq\label{eq:Ka}
 \langle p^B_a\;\hat O\circ q^A_b \rangle_{\mu}\equiv \frac{1}{2\pi i}\int_{|} du \,\mu(u)\,  p_a^{B\uparrow}(u)
\;\hat O\circ q_b^{A\downarrow}(u)\;
,
\eeq
where $p_a$'s are defined in (\ref{eq:pdef}), $\mu(u)$ is a meromorphic $i$-periodic function (without poles on the integration contour), $\hat O$ is a finite difference operator 
of the form
$$
\hat{O} = \sum_{i=-K}^L o_i(u) \hat{D}^{2 i} , \;\;\;\;\; \hat{D}\circ f(u) \equiv f(u + \tfrac{i}{2} )\;, 
$$
with polynomial coefficients $o_i(u)$,
 and the integration contour is defined as follows
\beq\la{cont}
\int_{|} = \int_{c  -i \infty}^{c + i \infty} - \int_{-c  -i \infty}^{-c + i \infty}, \;\;\; c > \text{max}_{\alpha} \left\{ | \text{Re}(\theta_{\alpha}) | \right\}_{ \alpha = 1 }^J\; ,
\eeq
so that it contains all poles of the Q-functions. 
Furthermore, in section~\ref{sec:num2} below 
we show that, among all $\mu(u)$'s so defined, there are only $J$ distinguishable periodic functions $\mu_\alpha$, meaning that $\langle p^B_a\hat O \circ q^A_b \rangle_{\mu}$
for any $\mu(u)$ can be expressed as a linear combination (with state-independent coefficients) of $\langle p^B_a\hat O\circ q^A_b \rangle_{\mu_\beta}$. It is convenient to take the following basis
\beq\label{eq:defmu}
\mu_\beta(u) \equiv \prod_{\alpha \neq \beta}  \frac{1 - e^{2 \pi (u - \theta_{\alpha} )} }{ 1 - e^{2 \pi (\theta_\beta - \theta_{\alpha} )} } , \;\;\;\; \beta=1,\dots, J\;,
\eeq
which has zeroes at all $u=\theta_\alpha+i {\mathbb Z}$ except for $\alpha=\beta$ where it is equal to $1$.
For simplicity we denote 
\beq
K_{ab,\beta}^{AB}\equiv \langle p^B_a\hat O\circ q^A_b \rangle_{\mu_\beta}\; .
\eeq

In general the integrals \eq{cont} are not convergent, and have to be regularised by a $\zeta$-function type of regularisation. We will give an explicit analytic prescription on how to do that in the next section, 
and also explain how to compute this type of integrals numerically.

\subsubsection{Regularisation}
A convenient way of computing, or more precisely defining, the integrals of the type \eq{eq:Ka} is by re-expressing them as a sum over residues.
Assuming for simplicity that the finite difference operator $\hat O$
does not contain the shift $D^{-2n}$ with $n>1$, the integrand of $\langle p^{B\uparrow}_a\hat O\circ q^{A\downarrow}_b \rangle_{\mu_\beta}$ has simple poles at $\theta_\beta+i {\mathbb Z}$ due to the poles in the q- and p-functions.\footnote{In presence of such negative shifts, a finite number of double poles may be present. This is not a major complication, as we discuss in section \ref{sec:numericsreview}.} In the UHP the poles are coming from $p^{B\uparrow}$, to make them manifest one can use \eq{eq:Omw}
\beq
\underset{u=\theta_\beta+i n}{\rm res}\;p^{\uparrow}_a(u) = \omega_{\beta,ac} \; q^{c\downarrow}(\theta_\beta+i n)\;\;,\;\;n\in \mathbb{Z}_{>0}\;,
\eeq
and similarly $q^{\downarrow A}$ produces poles in the LHP, whose residues are obtained as
\beq
\underset{u=\theta_\beta-i n}{\rm res}\;q^{\downarrow}_a(u) = {\omega}_{\beta,ac} \; p^{c\uparrow}(\theta_\beta-i n)\;\;,\;\;n>0\; ,
\eeq
so that we can write formally for $\hat O = u^k \hat D^{2 m}$,
\beqa\la{asSum}
\langle p^{B\uparrow}_a\hat O q^{A\downarrow}_b \rangle_{\mu_\beta} &=& 
\sum_{n=1 + m}^\infty
\left.\omega_{\beta,ac}^B \; (q^B)^{c\downarrow}(u)
\;u^k\; q_b^{A\downarrow}(u+im)\right|_{u=\theta_\beta+i n}\\
&+&
\;\;\;\sum_{n=0}^\infty\;\;
\left.\omega_{\beta,bc}^A \; p_a^{B\uparrow}(u)
\;u^k\; (p^A)^{c\uparrow}(u+im)\right|_{u=\theta_\beta-i n}\;. \nn
\eeqa
For the cases with shift $m<0$ and for the states with magnons, there could be a finite number of double poles contributing. In the observables we consider below, the relevant case is $m=-1$, which gives a double pole contribution at the origin -- in this case one should add
\beq\la{extra}
\underset{u=\vartheta_{\beta}  }{\rm res}\left( p_a^{\uparrow B}(u+i) \, u^k\, q_b^{A \downarrow}(u) \right)  \;
\eeq
to the r.h.s.  of \eq{asSum}. We will discuss such terms in more detail in the next section. 
 Depending on the parameters of the states $A$ and $B$, the UHP and LHP sums in the first and second lines of  (\ref{asSum}) may or may not converge in the usual sense.
Indeed, at large $n$ we can use the asymptotic expansion \eq{eq:asyQB}, so that the combination we need reads
\beq\la{general}
(q^B)^{c\,\downarrow}(u)
\;u^k\; q_b^{A\downarrow}(u+im)\simeq 
(\lambda^A_b)^{-i u}(\lambda^B_c)^{+i u}
u^{+\hat M^A_a-\hat M^B_c-D_0+k}\(1+{\cal O}\(\frac{1}{u}\)\)
\eeq
using that the formal sum
\beq\label{eq:dilog}
\sum_{n=1}^\infty \frac{\lambda^n}{ n^\alpha} =  \text{Li}_{\alpha }(\lambda )
\eeq
is well defined for all $\alpha$ and $\lambda$, except for $\lambda=1$, where it reduces to the $\zeta$-function which has a pole at $\alpha=1$.\footnote{In general it also has a branch-cut $\lambda\in(1,+\infty)$.}
This means that for generic $\lambda_b^A$ and $\beta_c^B$
the sum is well defined in $\zeta$-regularisation.
In section~\ref{sec:numericsreview} we also describe how this sum can be computed very efficiently numerically.

We see that for two generic states $A$ and $B$ all the Q-bilinear 
forms  are well defined. However, in some particular situations, and especially when $A$ coincides with $B$ there is a danger to hit the singularity of (\ref{eq:dilog}), when one of the terms behaves as $1/n$ at large $n$'s.
In the next section we will be considering the diagonal form factors for which one has to face this problem.

Note that in fact the scalar product of two CFT wavefunctions is also log-divergent when $\Delta^A=\Delta^B$
as we discussed in section \ref{sec:quantwf}.
So the divergence we found in the Q-bilinear forms is consistent with the divergence in the coordinate representation of the scalar product. However, it is still meaningful to define the finite part in front of the log-divergence described in section \ref{sec:quantwf} in terms of the Q-functions as well.

In the next section we will show that there is still a sufficient amount of the bilinear combinations of Q-functions which are finite even when the states  $A$ and $B$ are identical.
We will then show in section~\ref{sec:orthosec} that those combinations can be used to express a vast class of diagonal form factors as a determinant of the finite Q-bilinear forms.
Intriguingly, we will see that those combinations are finite only when the Q-functions obey the correct quantisation condition from section~\ref{sec:quantQ} (in addition to the Baxter TQ relations).

\subsubsection{Conjugation under the Q-bilinear form}\la{sec:conjugationdef}
An important property of the Q-bilinear form is that it allows 
to define conjugate finite-difference operators $O^\dagger$,
independently of the measure $\mu(u)$.
For example, for the elementary operator $\hat O$ we define its conjugate $\hat O^\dagger$ by 
\beq
\hat O=\hat{D}^{m}u^k \hat{D}^{m}\;\;\Leftrightarrow\;\;
\hat O^\dagger=\frac{1}{\hat{D}^{m}}u^k \frac{1}{\hat{D}^{m}} \ .
\eeq
As the operators of this type, which can be also written as $\hat O=(u+\tfrac{im}{2})^k \hat{D}^m$, constitute a basis we thus have defined  (by linearity) the conjugation for all finite-difference operators with polynomial coefficients.

The key property of the Q-bilinear forms, which we will utilise in the next section, is the following:
\beq
\langle p_a \;\hat O \circ q_b\rangle_\mu = 
\langle (\hat O^\dagger \circ p_a)\;  q_b\rangle_\mu
\eeq
which can be seen by shifting the integration contour by $-i m$ 
along itself.

\subsubsection{Finite diagonal combinations}

We will now show that the quantisation
condition (\ref{eq:quant}) can be reinterpreted as the requirement that some Q-bilinear forms are finite.
As we explained in the previous section the problem
arises whenever the twist parameters get cancelled and also the powers combine so that we get a $1/n$ term at large $n$'s.
For identical states $A=B$, the dangerous terms are when in 
\eq{general} we get $b=c$, then both twists and the non-integer parts of the powers $\hat{M}_a$ get cancelled. Then the convergence will really depend on the value $k-D_0$, but we will see that we will need the values of $k$ such that $k-D_0\geq-1$, meaning that  divergence will occur in the sub-leading $1/u$ terms. So the best way to ensure convergence is to avoid $b=c$.

Let us show that the following combinations are finite:
\beq\label{eq:Kformdef}
K_{a,\beta}\equiv\langle p^{\uparrow}_a\hat O q^{\downarrow}_a \rangle_{\mu_\beta}\;\;,\;\;a=1,2,3,4
\eeq
(with no summation over $a$).
Indeed, on-shell the combination $\omega=\Omega^{-1}\Gamma^{-1}$ is anti-symmetric, as was discussed in section~\ref{sec:quantQ}, and thus $\omega_{\beta,ab}$ should be anti-symmetric in $a\leftrightarrow b$ as well.
From that observation, it immediately follows that in the pole expansions on the r.h.s. of
\eq{asSum} we never get both $q$'s  (or $p$'s) with the same indices, ensuring that the sum is well defined with our regularisation.

\subsection{Numerical evaluation of the Q-bilinear forms}\label{sec:numericsreview}
Here we explain how the Q-bilinear forms can be evaluated numerically in practice. It would be useful to recall first how we solve the TQ-relations and impose the quantisation condition. First, we find $q$ at large $u$ by plugging the asymptotic expansion \eq{eq:asyQB} into the Baxter equation \eq{eq:Bax}. We usually keep around $20$ coefficients of $B_{a,n}, \;B^{a}_{n}$
to get a very good (around $50$ digits) precision for $q$ at 
the values $|u|>100$. After that we use \eq{eq:Bax} to descend from large $u$'s along the imaginary axis to a finite value of $u$. This way we can compute both $q_a^\downarrow$
and $q_a^\uparrow$, with the difference that for
the former we use \eq{eq:Bax} to move down from large positive ${\rm Im}(u)$ values, whereas for $q_a^\uparrow$
we use \eq{eq:Bax} to move up from the asymptotic domain far below the real axis. Then we compute $\Omega$ using \eq{Omegadef} at $J+1$ different points, which then allows us to deduce $\Omega_{\beta},\;\beta=0,\dots,J$ from \eq{eq:poledec}. 
Then we have to adjust the coefficients in the polynomials $P$, which are the eigenvalues of the integrals of motion, to impose the quantisation condition $\Omega_\beta^T=-\Omega_\beta$, which then gives all integrals of motion fixed  as a (multi-valued) function of $\xi$
and also allows us to determine the gluing matrix $\Gamma$ (which is parametrised by one constant $\gamma$ from \eq{eq:glue}, fixed by \eq{eq:symom}).
This procedure was already well established in the previous works~\cite{Gromov:2017cja,Cavaglia:2020hdb,Gromov:2019jfh} and is a simplified version of the similar method~\cite{Gromov:2015wca} developed for the spectrum of $\mathcal{N}=4$ SYM.

Next, in order to evaluate the bilinear
combinations of the Q-functions we start from the expression \eq{asSum}. Consider the first term for example:
\beq\la{thesum}
\sum_{n=1+m}^\infty
\left.\omega_{\beta,ac}^B \; (q^B)^{c\downarrow}(u)
\;u^k\; q_b^{A\downarrow}(u+im)\right|_{u=\theta_\beta+i n}\;.
\eeq
In principle we know already all ingredients such as $\omega_\beta$
and $q$ which we can compute at any point. However, as we discussed previously, the sum \eq{thesum} does not necessarily converge in the usual sense and a $\zeta$-type of regularisation is needed.

What we do in practice is fix some cut-off $N$ (in practice around $100$) and split the sum into two parts. For $n<N$ we compute the sum without any further approximation, for $n>N$
we replace $q's$ by their asymptotics \eq{eq:asyQB} and re-expand it for large $n$, so that it takes the form
\beq
\sum_{n=N}^\infty \Lambda^n n^{-\alpha}\(1+\frac{C_1}{n}+\frac{C_2}{n^2}+\dots\)\;.
\eeq
For each of the above terms we can use the analytic result for the $\zeta$-regularised sum\footnote{$\Phi$ is  \verb+HurwitzLerchPhi+ in Mathematica.}
\beq
\sum_{n=N}^\infty \Lambda^n n^{-\alpha}=\Lambda ^N \Phi (\Lambda ,\alpha ,N)\;.
\eeq
We found that this procedure is very accurate and allows one to compute the Q-bilinear forms with $20$ digit precision quite easily. We will give some examples in section~\ref{sec:ntest}.

In addition to the infinite sums described above, in some cases we might need to evaluate the contributions of the double 
poles \eq{extra}. These contributions may potentially require evaluation of derivatives of Q-functions. In the examples we considered, the second order pole cancels and becomes a single pole, so the term \eq{extra} can be evaluated on equal footing with others.

\subsection{Completeness of the basis of the Q-bilinear forms} \la{sec:num2}
Here we discuss why the basis of periodic functions~\eq{eq:defmu} is complete in the sense described below.
We will assume that the double poles do not appear.

First, we notice that the following periodic function will result in a zero 
\beq
\mu_0(u) \equiv \prod_{\alpha =1}^J \left( {1 - e^{2 \pi (u - \theta_{\alpha} )} }\right)
\eeq
for the integral of the type \eq{eq:Ka} as 
$\mu_0(u)$ will cancel all residues inside the integration contour. Then we notice that any polynomial in $e^{2\pi u}$ can be represented as $\tilde P(e^{2\pi u})\mu_0(u)+\sum A_\beta\;\mu_{\beta}(u)$, for some polynomial $\tilde P(u)$. The first term, divisible by $\mu_0(u)$, gives zero and can be removed. 
Similarly, any $i$-periodic
function $F(u)$ with no poles inside the integration contour can be written as
\beq
F(u)=
\sum_{\beta=1}^J
F(\theta_\beta) \mu_\beta(u)
+\mu_0(u)R(u)
\eeq
where $R(u)$ is an $i$-periodic function, also analytic inside the  contour.
The last term does not give any contribution and thus this case also reduces to our basis. 

More complicated is the case when the periodic function $F(u)$ has poles inside the integration contour. To discuss this case, let us assume that the twists are phases $\lambda_a=e^{i\phi_a}$ such that
$\phi_a^B > \phi_b^A$ meaning that
the integrand, which behaves as $e^{u(\phi_b^A-\phi_a^B)}$, is decaying to the right of the contour. In this case the function 
\beq
\mu(u) = \frac{e^{2\pi u}}{e^{2\pi u}-e^{2\pi u_0}}\;,
\eeq
which has a pole at $u=u_0+i{\mathbb Z}$, 
will give zero under the integral \eq{eq:Ka}. This is because the integrand decays exponentially for large positive ${\rm Re}\;u$, due to the asymptotics of $q$ and $p$, and also
at large negative ${\rm Re}\;u$, due to the exponential decay of the function $\mu(u)$ itself -- hence the contour can be pushed to infinity giving a zero result. 
For the opposite $\phi_a^B < \phi_b^A$ case,  we rewrite
\beq
\mu(u)=\frac{e^{2\pi u_0}}{e^{2\pi u}-e^{2\pi u_0}}+
1\;,
\eeq
where now the first term will be zero under the integral and the second term is regular. In conclusion, we see that one can subtract simple poles in the measure factor $F(u)$ in terms of  such functions which do not contribute to the Q-bilinear forms, bringing us back to the case with no poles.

This shows that the basis $K^{AB}_{ab,\alpha}$ of Q-bilinear forms is the most general. In the next section we will show how to use it to compute nontrivial correlation functions in the fishnet theory.

Note that the above consideration also allows us to relate the basis of forms described above, to the one defined with opposite
arrow combinations
$\tilde K_{ab,\alpha}\equiv \langle p^{\downarrow}_a\hat O q^{\uparrow}_a \rangle_{\mu_\alpha}$. For that, one can use the  matrix $\Omega$ to reverse the direction of the arrows~\eq{qOq}. Since  $\Omega$ is $i$-periodic, at the level of the integrals it can be replaced by a combination of the elementary measures $\mu_{\beta}$ (the coefficients are state-dependent in this case). 
For some observables, a basis obtained with a certain pattern of the arrows might be more natural, and lead to simplifications. One such example will be the $g$-function, which we consider in section \ref{sec:gsec}.

\section{Functional SoV for a class of correlation functions}\label{sec:orthosec}
In this section we show how the Q-bilinear forms, introduced in the previous section, can be combined into physical observables of the QFT. 

In particular, we focus on a certain class of observables which can be obtained as a result of variation of the numerous parameters (such as inhomogeneities, twist angles and coupling). Those observables can be studied within the functional SoV approach introduced in~\cite{Cavaglia:2019pow} and do not require further input from the more conventional operatorial SoV method. 
Following~\cite{Cavaglia:2019pow}, the variations of the integrals of motion are found as solutions of a linear system of equations.

Like in the case of spin chains~\cite{Gromov:2020fwh}, these results are expected to provide deep structural insights on the form of the result for more general observables, including the off-diagonal cases. In particular, they should fix explicitly the measure factor in the SoV basis, and lead to the determinant form for an  even wider class of correlators.

Among these observables, we will discuss in detail the explicit  expression for the diagonal 3-point functions involving the Lagrangian. We also describe the form factor of the variation of the Hamiltonian w.r.t. local weights which amounts to a nontrivial local insertion at the level of diagrams.

\subsection{Conjugation property of Baxter equations}

The key starting point for the functional SoV consideration is the Baxter equation satisfied by the Q-functions $q_a$
and also the dual Baxter equation, satisfied by $q^a$ or $p_a$
as defined in the previous section in \eq{eq:pdef}.

 \subsection{Baxter TQ-relation as a finite difference operator}
 Following \cite{Cavaglia:2019pow}, we introduce a useful notation, and represent the Baxter equation in terms of a finite difference operator $\mathcal{B}$. To define it we use the shift operators $D \equiv e^{\frac{i}{2} \partial_u }$ acting on a function of the spectral parameter as $D \circ f(u) = f(u + \frac{i}{2})$. 
 The Baxter finite-difference operator is given by
 \beqa\label{eq:BaxOpdef}
 \mathcal{B}  &\equiv& 
\cD_4  + \cD_{-4} -  \left( (u+\tfrac{i}{2})^J\chi_{\mathbf{4}}\;\cD_2- u^J\chi_{\mathbf{6}} \;\cD_0
  + (u-\tfrac{i}{2})^J\chi_{\bar{\mathbf{4}}}\;\cD_{-2} 
 \right)
 \\
 \nn&+& 
 \sum_{b\in \{-2,0,\bar 0 ,2\}} (-1)^{\tfrac b2}\sum_{\alpha=1}^{J} (u+\tfrac{ib}{2})^{\alpha-1} \, I_{(b,\alpha)} \, \cD_b 
\;, 
 \eeqa
 where the  state-dependent factors
 $I_{(b,\alpha)}$ are the $4J$ nontrivial eigenvalues of integrals of motion,
 $\cD_b$ are state-independent finite difference operators defined as
 \begin{align*}
 \cD_4 &\equiv 
 \hat{D}^2
 \mathcal{R}^{[-2]}_{\theta} \, 
 \mathcal{Q}_{\theta} \,  \hat{D}^2 , &
  \cD_0 &\equiv \frac{1}{{\mathcal{P}_{\theta}}} ,  
   & 
 \cD_{-4}  &\equiv
\frac{1}{ \hat{D}^2}\mathcal{R}_{\theta}^{[-2]}
 \mathcal{Q}_{\theta} \, \frac{1}{ \hat{D}^{2}}
 , \\
 \cD_2 &\equiv  \hat{D}\,\mathcal{R}^{-}_{\theta}\,  \hat{D} ,
 &\cD_{\bar 0} &\equiv \frac{u^J}{\mathcal{P}_{\theta}}  ,  &
 \cD_{-2}   &\equiv \frac{1}{ \hat{D}}\mathcal{R}_{\theta}^{-} \, \frac{1}{ \hat{D}}
 ,
 \end{align*}
 and $\mathcal{P}_{\theta}$, $\mathcal{Q}_{\theta}$, $\mathcal{R}_{\theta}$ are fixed polynomials:
 \beq\label{eq:defQR}
 \mathcal{R}_{\theta}(u) \equiv \prod_{\alpha=1}^J (u - \theta_{\alpha} )^{{\mathbf{I}}_{\alpha} }\;\;,\;\;\mathcal{P}_{\theta}(u) \equiv \prod_{\alpha=1}^J (u - \theta_{\alpha} )^{1 - {\mathbf{I}}_{\alpha}  }\;\;,\;\; \mathcal{Q}_{\theta}(u) \equiv \prod_{\alpha=1}^J (u - \theta_{\alpha} )\;.
 \eeq
 With these definitions the Baxter equation (\ref{eq:Bax}) is written as 
 \beq
 \mathcal{B} \circ q_a = 0\; .
 \eeq
  Similarly, the dual Baxter equation satisfied by the Q-functions with upper indices $q^a$ is
  given by 
\beq  \la{BP}
  \mathcal{B}^{\text{dual}} \circ q^a = \mathcal{B}^{\text{dual}} \circ p_a = 0\;,
\eeq
where
 \beqa\label{eq:defduBax}
 \mathcal{B}^{\rm dual}  &\equiv& 
\cD_{-4}  + \cD_{+4} -  \left( (u-\tfrac{i}{2})^J\chi_{\mathbf{4}}\;\cD_{-2}- u^J\chi_{\mathbf{6}} \;\cD_0
  + (u+\tfrac{i}{2})^J\chi_{\bar{\mathbf{4}}}\;\cD_{2} 
 \right)
 \\
 \nn&+& 
 \sum_{b\in \{-2,0,\bar 0 ,2\}} (-1)^{\tfrac{b}{2}}\sum_{\alpha=1}^{J} (u+\tfrac{ib}{2})^{\alpha-1} \, I_{(-b,\alpha)} \, \cD_b\; 
 . 
 \eeqa

\paragraph{The conjugation property.}
Let us show that the Baxter finite-difference operators $\mathcal{B}$ and its dual  $\mathcal{B}^{\text{dual}}$ defined in \eq{eq:BaxOpdef} and \eq{eq:defduBax} are conjugated to each other in the sense defined in section~\ref{sec:conjugationdef},
\beq\la{Bconj}
\mathcal{B}^{\text{dual}} = \mathcal{B}^{\dagger}\;. 
\eeq
To see this, it is sufficient to notice that $\cD_{-4}^\dagger= \cD_{+4}$ and $\cD_{-2}^\dagger= \cD_{+2}$.
The property \eq{Bconj} is similar to what was observed in \cite{Cavaglia:2019pow} and will be used intensively in the next several sections.

\subsection{Variation of the Baxter equation}\label{sec:diagM}
We now consider the variation of the integrals of motion of a physical state with respect to a tunable parameter $p$. A natural application is when this parameter is the coupling constant, but we can consider also varying the twists or inhomogeneities which enter the definition of the Hamiltonian. The most general variation is a combination of all  these. 

Any type of  variation induces a change in the Q-functions $ q_a \rightarrow q_a + \delta q_a$, and simultaneously in the constants appearing in the Baxter equation. In general we have
 \beqa\label{eq:delB}
 \delta\mathcal{B}  &\equiv& 
\delta\cD_4  + \delta\cD_{-4} -  \delta\left( (u+\tfrac{i}{2})^J\chi_{\mathbf{4}}\;\cD_2- u^J\chi_{\mathbf{6}} \;\cD_0
  + (u-\tfrac{i}{2})^J\chi_{\bar{\mathbf{4}}}\;\cD_{-2} 
 \right)
 \\
 \nn&+& 
 \sum_{\mathbf{b}\in \{-2,0,\bar 0 ,2\}} (-1)^{\frac{ \mathbf{b}}{2}}\sum_{\alpha=1}^{J} (u+\tfrac{i\mathbf{b}}{2})^{\alpha-1} \, (\delta I_{(\mathbf{b},\alpha)} \, \cD_{\mathbf{b}}+I_{(\mathbf{b},\alpha)} \, \delta\cD_{\mathbf{b}}) 
\; , 
 \eeqa
so that the Baxter equation to linear order in the variation reads
\beq\label{eq:varBax}
(\mathcal{B} + \delta \mathcal{B} ) \circ ( q_a + \delta q_a ) = 0 \;.
\eeq
Inserting this condition inside the brackets with the dual Q-function $p_a$ (to ensure the bracket is finite we should make sure the indices coincide) we find  $0=\langle p_{a}^{\uparrow} (\mathcal{B} + \delta \mathcal{B} ) \circ ( q_{a}^{\downarrow} + \delta q_{a}^{\downarrow} ) \rangle_{\mu_\alpha}$, and expanding we find
\beq
\cancel{ \langle p_{a}^{\uparrow} \mathcal{B}  \circ ( q_{a}^{\downarrow} + \delta q_{a}^{\downarrow} ) \rangle_{\mu_\alpha} } + \langle p_{a}^{\uparrow}  \delta \mathcal{B}   \circ q_{a}^{\downarrow}   \rangle_{\mu_\alpha} \simeq \langle p_{a}^{\uparrow}  \delta \mathcal{B}   \circ  q_{a}^{\downarrow}   \rangle_{\mu_\alpha} = 0\; ,
\eeq
where the first term on the lhs vanishes due to (\ref{BP}) and \eq{Bconj}, and we keep only the first order in the variation. Using the  basis of measures $\mu_{\alpha}$ discussed in the previous section, these are precisely $4 J$ linear equations 
as $a=1,\dots,4$ and $\alpha=1,\dots,J$, 
for  $4 J$ variables $\partial_p I_{(b, \beta)}$. 
It is convenient to rewrite the linear system as
\beq\label{eq:varsys}
\mathcal{M} \cdot \delta \vec I = \delta \vec V\;,
\eeq
 where $\vec I \equiv I_{(a,\alpha)}$ is a $4J$
 dimensional vector built out of the integrals of motion. We use the conventions where they are ordered as 
 \beq
 \vec{I} = \left( I_{(-2,1)},I_{(-2,2)},\dots |  I_{(2,1)},I_{(2,2)},\dots |  I_{(0,1)},I_{(0,2)},\dots| I_{(\bar 0,1)},I_{(\bar 0,2)},\dots\right),
 \eeq
 the matrix of coefficients is defined by blocks as
\beqa\label{eq:diagM}
\left(\mathcal{M} \right)_{(a,\alpha)}^{\;(b,\beta)} &\equiv&  (-1)^{\frac{b}{2}} \langle   \; p_{a}^{\uparrow} \times  \;  (u+\tfrac{ib}{2})^{\beta-1}\cD_b \, \circ q_{a}^{\downarrow} 
\rangle_{\mu_\alpha}
\;, 
\eeqa
for $1 \leq a\leq 4$, $b \in\left\{ 2,0,\bar{0},-2\right\}$ and  $1 \leq \alpha,\beta \leq J$.
The r.h.s. $\delta\vec V$ contains the part of the variation for fixed value of $\vec I$:
\beq
(\delta\vec V )_{(a, \alpha)} \equiv  - \langle p_a^{\uparrow}  \(\left.\delta\mathcal{B}\right|_{\delta I\to 0} \) \circ q_a^{\downarrow} \rangle_{\mu_\alpha} \;.  
\eeq
In the next sections we consider some particular cases revealing additional features in comparison with the HW spin chain cases. 

\subsection{Zero mode of the variation matrix}

An important observation, which we will further explore here, is that the matrix ${\cal M}$ is in fact degenerate, and in the generic situation should have one null vector.

As we will see, the zero mode is related to a  physically very important case of the variation in parameters -- where we vary the coupling constant for fixed values of the twists and inhomogeneities. 
This is particularly interesting because even in the non-twisted theory the quantity $\partial_{\xi^2} \Delta$ gives a nontrivial structure constant~\cite{Costa:2010rz}. 

To understand how that is related to the existence of the null vector, we recall that both $\Delta$ and $\xi$ are two conserved charges. Thus, in order to compute $\d_{\xi^2}I_{a,\alpha}$ 
we are not actually changing any parameters of the system, we just follow the one dimensional space of solutions of the same spin chain.
Due to this, we have the inhomogeneous part set to zero $\delta_{\xi^2} \cD_b = \delta_{\xi^2} \vec V = 0$, therefore  (\ref{eq:varsys}) reduces to a homogeneous equation:
\beq\label{eq:varsys0}
\mathcal{M} \cdot \d_{\xi^2} \vec I = 0\;.
\eeq
Equation \eq{eq:varsys0} implies that there is a one-dimensional null space. The null vector $\d_{\xi^2}\vec I$ has to be normalised so that its $(0,1)$ component is fixed according to \eq{eq:I31} to
be
\beq\label{eq:normI}
\left(\partial_{\xi^{2}} \vec I \right)_{(0,1)} = J\, \xi^{2 J - 2}\;.
\eeq
After that, the other components are fixed (according to our numerical experiments) and give indeed the variation of all other integrals of motion (which also include $\Delta$) with respect to the coupling constant $\xi$.

The observation that the matrix ${\cal M}$
has rank $4J-1$ is based on a large variety of numerical tests, but the analytic proof seems to be complicated.
We leave this problem to future studies.
We also note that $\Delta(\xi)$ is a multi-valued function and has branch cuts at various nontrivial values of $\xi$. Of course, at the branch points  the derivative $\Delta'(\xi)$
diverges, which should manifest in the  vanishing of the $(0,1)$ component of the null eigenvector of $\mathcal{M}$.

\subsubsection{Explicit result for $\d_{\xi^2}\Delta$}
For SoV applications it could be useful to express the result for $\d_{\xi^2}\Delta$ explicitly as a ratio of determinants -- something which the operatorial SoV is expected to give like in~\cite{Gromov:2020fwh}.

To make the solution explicit, we turn the homogeneous linear system into an inhomogeneous one by moving the column ``$(0,1)$'' to the right hand side of the equation. Using the normalisation (\ref{eq:normI}), we find:
\beq\label{eq:inhomog}
\sum_{(b,\beta) \neq (0,1)}\mathcal{M}_{(a,\alpha)}^{(b,\beta)} \, \partial_{\xi^2} \vec I _{(b,\beta)} = -  J \xi^{2 J -2} \, \langle \,\frac{p_a^{\uparrow } q_a^{\downarrow}  }{\prod_{\alpha=M+1}^J (u-\theta_{\alpha} ) }  \,\rangle_\alpha \equiv \mathcal{V}_{(a,\alpha)}\;,
\eeq
which is a system of $4 J$ equations in $4 J - 1$ unknowns. One equation out of $4J$
is linearly dependent so there is a unique solution. Furthermore, some of the 
variables $(\partial_{\xi^2} \vec I )_{(b,\beta)}$ are related to each other in a simple way, since due to \eq{III} we have
\beqa\label{eq:dIs}
\partial_{\xi^2} I_{(2,J)} = \frac{\lambda_{++--}\partial_{\xi^2}\Delta }{-2 i}
\;,\;
\partial_{\xi^2}I_{(\bar 0,J)} =     \frac{ \lambda_3 \lambda_4-\lambda_1 \lambda_2  }{i} \partial_{\xi^2} \Delta 
\;
, \;
\partial_{\xi^2} I_{(-2,J)} = \frac{\bar\lambda_{++--}\partial_{\xi^2}\Delta}{2 i}\;.\nn
\eeqa
Using these relations and selecting $4J-3$ equations out of $4J$
we can get a non-degenerate non-homogeneous linear system. The  solution can be written as a ratio of determinants
\beq\label{eq:minorformula}
\partial_{\xi^2} \Delta = -\frac{ \texttt{N} }{\texttt{D}}\;,
\eeq
where $\texttt{N}$, $\texttt{D}$ are determinants of $(4J-3)\times (4J-3)$ dimensional matrices, where each element is a Q-bilinear form. 

In the particular case $J=1$, $M=0$
the determinants become one-dimensional and we reproduce the result we obtained with Amit Sever in~\cite{withAmitupcoming}:
\beq
\partial_{\xi^2}\Delta =  \frac{2 i\int_{|} du \, \frac{ p_a^{\uparrow}  q_a^{\downarrow}  }{(u-{\theta} )} }{ \int_{|} du \,  p_a^{\uparrow} \left(\mathcal{L}_1  q_a^{\downarrow ++} +  \mathcal{L}_2   q_a^{\downarrow --} + \mathcal{L}_4 \frac{ u \, q_a^{\downarrow} }{(u-\theta) } \right)  }\;\;, \;\; \text{  for any } a \in \left\{1,\dots,4\right\}\;.
\eeq
where $\mathcal{L}_a$ are $\mathcal{L}_1 \equiv \frac{i}{2} \left( \lambda_1 + \lambda_2 - \lambda_3 - \lambda_4 \right) $, $\mathcal{L}_2 \equiv \frac{i}{2} \left( \bar\lambda_3 + \bar\lambda_4 - \bar\lambda_1 - \bar\lambda_2 \right) $, $\mathcal{L}_4 \equiv i \left( \lambda_1 \lambda_2 - \lambda_3 \lambda_4 \right)$. 

\subsubsection{Variations in other parameters}
\label{sec:varoth}
As discussed above in section \ref{sec:diagM}, we can also consider more general variations, with respect to an external parameter such as  the twist angles or inhomogeneities. In this case, we obtain an inhomogeneous linear system \eq{eq:varsys} with a nonvanishing r.h.s. and the same matrix of coefficients $\mathcal{M}$ as for the $\xi$ variations. Since ${\cal M}$ has rank $4J-1$, such a linear system has a one-parameter family of solutions, which differ by the null eigenvector of $\mathcal{M}$. As we saw in the previous section, the null vector represents the variation with respect to the coupling constant. Depending on which observable we are computing, this ambiguity should be fixed accordingly. For instance, in order to compute variation of the scaling dimension w.r.t. the twist, we would keep the value of the coupling fixed (as the graph building operator is well defined regardless of the twist value). This means that we should set the solution's component ``$(0,1)$'' to zero, which fixes the solution uniquely. Below we will also consider variation of the Hamiltonian w.r.t the weights $h_\alpha$, in which case as we will discuss the ambiguity is fixed by choosing appropriate values for the leading nontrivial integrals of motion \eq{III}.

By choosing a subset of $4J-1$ among the original $4J$ equations, we can write the solution as a ratio of two determinants of size $(4J-1)$, built from Q-bilinear forms, similarly to \eq{eq:minorformula}. 

\subsubsection{Numerical test}\la{sec:ntest}
We verified the above general formalism numerically. For instance, we studied the case $J=2$, $M=0$. For  inhomogeneities $\theta_1 = -\theta_2 = -\frac{1}{100}$, twists $\lambda_1 = e^{\frac{i}{3} }$, $\lambda_2 = e^{-\frac{i}{4} }$, $\lambda_3 = e^{\frac{i}{2} }$, $\lambda_4 = e^{i\frac{7}{12} }$, and coupling $\xi^2 \simeq 0.100992$, we considered a solution corresponding to a state with zero spins, and $\Delta \simeq 1.93034 $. We computed numerically the coefficients of the linear system,\footnote{For this numerical test we took a different basis of Q-bilinear forms, using combinations of the type $\langle p^{\downarrow} q^{\uparrow} \rangle$. }
\beq
\mathcal{M} \equiv  \underbrace{\left( \begin{array}{c|c}
\langle p^{\downarrow}_a   u^{\beta-1}\mathbb{D}_{2} \circ q^{\uparrow}_a  \rangle_{1} & \langle p^{\downarrow}_a u^{\beta-1} \mathbb{D}_{0} \circ q^{\uparrow}_a \rangle_1 \\
\hline 
\langle p^{\downarrow}_a u^{\beta-1} \mathbb{D}_{2} \circ q^{\uparrow}_a  \rangle_{2} & \langle p^{\downarrow}_a u^{\beta-1} \mathbb{D}_{0} \circ q^{\uparrow}_a \rangle_2 
\end{array} \right) }_{\texttt{M}_1} \oplus \underbrace{ \left( \begin{array}{c|c}
\langle p^{\downarrow}_a u^{\beta-1} \mathbb{D}_{\bar{0}} \circ q^{\uparrow}_a  \rangle_{1} & \langle p^{\downarrow}_a u^{\beta-1} \mathbb{D}_{-2} \circ q^{\uparrow}_a \rangle_1 \\
\hline 
\langle p^{\downarrow}_a u^{\beta-1} \mathbb{D}_{\bar{0}} \circ q^{\uparrow}_a  \rangle_{2} & \langle p^{\downarrow}_a u^{\beta-1} \mathbb{D}_{-2} \circ q^{\uparrow}_a \rangle_2 
\end{array} \right)
}_{\texttt{M}_2} 
\eeq
where  the indices range over
$1 \leq a \leq 4 , 1\leq \beta\leq 2$. We find
\beqa
\texttt{M}_1 &\simeq& {\tiny \left(
\begin{array}{cc|cc}
 4.31901\, -11.4414 i & 22.6063\, -49.4069 i & 0.0046265\,
   -5.98084 i & -4.80953+2.51189 i 
   \\
 0.809298\, +7.85372 i & -3.27609-11.5111 i & 0.401649\, +5.97 i &
   4.44227\, +1.97328 i  \\
 1.00063\, +0.946839 i & -3.3413-4.13463 i & -0.18902-0.620784 i &
   -0.0600478+0.740181 i  \\
 9.14976\, -16.4014 i & 54.6825\, -118.479 i & 0.126059\, +3.50622
   i & 1.4338\, +6.18232 i
   \\
   \hline
 -4.3639+11.574 i & -22.9599+50.229 i & -0.0046265+5.97614 i &
   4.80953\, -2.53922 i  \\
 -0.807428-7.76595 i & 3.22652\, +11.225 i & -0.400613-5.97466 i &
   -4.44744-1.95001 i \\
 -1.02467-0.96949 i & 3.39688\, +4.20952 i & 0.19209\, +0.603913 i
   & 0.0863731\, -0.757081 i 
   \\
 -8.94188+16.0332 i & -53.6573+116.191 i & -0.111173-3.60207 i &
   -1.26448-6.03864 i 
\end{array}
\right)}\; , \\
\texttt{M}_2 &\simeq& {\tiny \left(
\begin{array}{cc|cc}
 -0.414138+11.9493 i &
   0.435697\, +54.2825 i & -3.43629-11.3103 i & -23.5189-49.5152 i
   \\
 1.145\, -7.55161 i & -3.24604+11.4394 i
   & -2.93838+6.92369 i & 9.2625\, -8.897 i \\
 -0.29778-1.3448 i & 1.14678\, +5.15391
   i & -0.502725+1.26785 i & 1.2753\, -5.16149 i \\
 0.00878648\, +18.7776 i &
   -0.00274115+130.198 i & -9.11569-16.3378 i & -54.6917-118.479 i
   \\
   \hline
  0.414138\, -12.0894 i & -0.435697-55.175
   i & 3.48119\, +11.4428 i & 23.8725\, +50.3373 i \\
  -1.12592+7.46583 i & 3.1836\, -11.1587 i &
   2.89948\, -6.84499 i & -9.09632+8.65886 i \\
 0.304503\, +1.37583 i &
   -1.16684-5.24384 i & 0.514594\, -1.2967 i & -1.2941+5.25293 i
   \\
 -0.0077489-18.3607 i & 0.00241745\,
   -127.698 i & 8.91184\, +15.9772 i & 53.6653\, +116.191 i 
\end{array}
\right) }\; ,
\eeqa
and indeed $\mathcal{M}$  has a null eigenvector \beq
V_0 \simeq \left( \begin{array}{c}
1.5601\, +0.0101706 i \\0.544237\, -0.725972 i \\1\\-0.152213\\
5.35905\\1.19863\, \\1.5601\, -0.0101706 i\\0.544237\, +0.725972 i
\end{array}\right)\propto \partial_{\xi^4} \left( \begin{array}{c}I_{(2,1)}\\I_{(2,2)} \\ I_{(0,1)}\\I_{(0,2)}\\I_{(\bar{0},1)}\\I_{(\bar{0},2)}\\I_{(-2,1)}\\I_{(-2,1)}\end{array}\right)\; .
\eeq
At the same time we computed the derivative of the dimension $\Delta$ as a finite difference to obtain
\beq
\frac{\partial\Delta}{\partial\xi^2} \simeq -1.45431\; , 
\eeq
which, using (\ref{eq:dIs}), implies $\partial_{\xi^4} I_{(2,2)} = \partial_{\xi^4} I_{(-2,2)}\simeq 0.544237 -0.725972 i$, $\partial_{\xi^4} I_{(0,2)} \simeq 1.19863 $, consistently with the form of the null eigenvector. 

\subsection{Variations as spin chain form factors}
At the operator level variations of the  integrals of motion w.r.t. a parameter correspond to nontrivial spin chain expectation values, as was understood in \cite{Cavaglia:2019pow,Gromov:2019wmz,Gromov:2020fwh} for rational $gl(N)$ spin chains. Let us briefly recall this argument here. Consider a left and a right eigenstate of the transfer matrix $\Psi_L$ and $\Psi_R$ corresponding to the same eigenvalue. When we change the values of the parameter $p\to p+\delta p$, these wavefunctions will also change, as will the transfer matrix eigenvalues, so we have 
 \beq
\langle \langle \Psi_L \;,\; ( \hat{\mathbb{T}}^{\mathbf{r}} + \delta \hat{\mathbb{T}}^{\mathbf{r}} )\circ (\Psi_R + \delta \Psi_R ) \rangle\rangle = (\mathbb{T}^{\mathbf{r}} + \delta \mathbb{T}^{\mathbf{r}} ) \;  \langle \langle \Psi_L  \;,\;(\Psi_R  + \delta \Psi_R  \rangle\rangle \ .
 \eeq
Expanding it to the first order in the variation and using that $\Psi_L$ is a left eigenvector of $\hat{\mathbb T}^{\mathbf{r}}$, we see that several terms cancel and we are left with
\beq
\langle \langle \Psi_L \;,\;   \delta \hat{\mathbb{T}}^{\mathbf{r}}  \circ \Psi_R \rangle\rangle= \delta \mathbb{T}^{\mathbf{r}} \langle \langle \Psi_L \;,\;  \Psi_R \rangle\rangle \ .
\eeq
Thus we find
\beq\label{eq:finalresv}
  \frac{\langle \langle \Psi_L \;,\;  \frac{\delta\hat{\mathbb{T}}^{\mathbf{r}}}{\delta p} \circ \Psi_R \rangle \rangle}{\langle \langle \Psi_L \;,\; \Psi_R \rangle \rangle} =\frac{\delta \mathbb{T}^{\mathbf{r}}}{\delta p}\ .
\eeq

So we see that the diagonal form factor of the nontrivial operator $\frac{\delta \hat{\mathbb{T}}^{\mathbf{r}}}{\delta p}$ is written in terms of the variation of the eigenvalue, or, equivalently, of the integrals of motion. As a result, like discussed in section \ref{sec:varoth}, this form factor can be written in terms of determinants of the type \eq{eq:minorformula} built from Q-functions. 

The operator whose form factor we are computing can be quite nontrivial. In the next section we study as an example the variation of the Hamiltonian eigenvalue with respect to the local weight.

To link the spin chain picture with the fishnet CFT, we can notice that the right spin chain eigenstates become the CFT wavefunctions (i.e. correlators of the type \eq{eq:wfdef}) once we set the inhomogeneities to the corresponding values \eq{eq:shiftedtheta}. Moreover, as shown in appendix~\ref{app:subsecRL}, the left state $\Psi_L$ also can be identified with an appropriately chosen CFT wavefunctions that involve a conjugated set of fields, with the conjugated operator sitting at the other fixed point $x_{\bar{0}}$ and twisted by the inverse map $G^{-1}$. 
The form factors act as operators on the chain in between the two wave functions.
In general, the result  can be interpreted as a certain off-shell observable given by diagrams with nontrivial modification or insertion of some propagators and vertices, whose precise form is dictated by the operator we consider. In other words, this way we can compute a class of rather nontrivial Feynman diagrams similar to those appearing in multipoint correlators.

\subsection{Variation with respect to local weights $h_\alpha$}

The variations in the twist angles $\lambda_a$ and the inhomogeneities $\theta_\alpha$ were studied already in the spin chain context in~\cite{Gromov:2020fwh}.
In this section we consider another type of variation corresponding to changing the 
 weights $h_\alpha$ that define the representation of the conformal group at each site $\alpha$ of our spin chain. It would be interesting also to further explore this type of variation for the usual $gl(n)$ spin chains.
 
 In the case of fishnet CFT, we fix $h_\alpha$ to the values $1$ or $2$, however by taking a variation around these points we will be able to compute rather nontrivial quantities. Below we discuss several subtleties related to these variations and then present a particular nontrivial example -- the variation of the Hamiltonian.

\subsubsection{Variation with respect to weights and scalar product}

As the weights $h_\alpha$ explicitly enter the scalar product \eq{eq:scalar}, it is not immediately clear that the argument leading to the form factor result \eq{eq:finalresv} still goes through. Let us show that the result is still valid. A small variation of one of the weights $h_\alpha \rightarrow h_\alpha + \delta h_\alpha$ induces a change in  the transfer matrix eigenstate  $\Psi \rightarrow \Psi + \delta \Psi$, and  in the integrals of motion ${\hat I} \rightarrow {\hat I} + \delta {\hat I}$. At the same time, the conformally invariant scalar product (\ref{eq:scalar}) itself will change infinitesimally, which can be represented by the insertion of an operator
 \beq
 \left. \frac{\partial }{\partial \delta h_\alpha }\langle \langle f , g \rangle \rangle_{h_\alpha +  \delta h_\alpha } \right|_{{ \delta h_\alpha } = 0 } = \langle \langle f \; , \; \hat{A}_\alpha \circ g \rangle \rangle_{h} , \;\;\;\; \hat{A}_\alpha \equiv - \log \Box_{\alpha}  \ .
 \eeq
where we assumed that the wave functions $f$, $g$ have no dependence on the weights.

Let us then consider a left and right eigenstates of the transfer matrices, $\Psi_L$ and $\Psi_R$, respectively, corresponding to the same transfer matrix eigenvalue. 
 Under a small variation of the weights, the eigenvalue equation becomes
 \beq
\langle \langle \Psi_L \;,\; ( \hat{\mathbb{T}}^{\mathbf{r}} + \delta \hat{\mathbb{T}}^{\mathbf{r}} )\circ (\Psi_R + \delta \Psi_R ) \rangle\rangle_{h_\alpha + \delta h_\alpha } = (\mathbb{T}^{\mathbf{r}} + \delta \mathbb{T}^{\mathbf{r}} ) \;  \langle \langle \Psi_L   \;,\; (\Psi_R  + \delta \Psi_R ) \rangle\rangle_{h_\alpha + \delta h_\alpha } \ .
 \eeq
Expanding it at the first order in the variation and using the same logic as before, we see that there will be extra terms in the r.h.s. and l.h.s. involving the insertion of $\hat A_\alpha$,
\beqa &&
\langle \langle \Psi_L \;,\;   \delta \hat{\mathbb{T}}^{\mathbf{r}}  \circ \Psi_R \rangle\rangle_{h} +  \langle \langle \Psi_L \;,\;  ({\hat{A}}_\alpha ) \circ  \hat{\mathbb{T}}^{\mathbf{r}}  \circ \Psi_R \rangle\rangle_{h}\;\delta h_\alpha 
\\ \nn && = \delta \mathbb{T}^{\mathbf{r}} \langle \langle \Psi_L \;,\;  \Psi_R \rangle\rangle_{h} + \mathbb{T}^{\mathbf{r}} \;  \langle \langle \Psi_L \;,\;  ({\hat{A}}_\alpha )   \circ \Psi_R \rangle\rangle_{h} \;\delta h_\alpha
\eeqa
(with no summation over $\alpha$).
We notice that the extra terms involving $\hat A_\alpha$ cancel against each other. Thus  the expression \eq{eq:finalresv} is still valid. 

\subsubsection{Fixing the zero mode}

Let us also clarify how to fix the ambiguity in solving the linear system \eq{eq:varsys0} (coming from the Baxter equation) for the variations of integrals of motion when we vary $h_\alpha$. Keeping the value of $I_{(0,1)}$ fixed in this case is not what we would like to do, since the relation between $\hat I_{(0,1)}$ and coupling only exists at the specific values of $h_\alpha$ corresponding to the fishnet CFT.
Instead what we are trying to achieve is to find the diagonal matrix element of the derivative w.r.t. $h_\alpha$ of the integrals of motion, which are differential operators with coefficients dependent on the weights $h_\alpha$.

We notice that the $h$-dependence of the first subleading integrals of motion \eq{III} arises only due to the linear in $h$ term in the dilatation operator \eq{eq:diffoperators} (these integrals of motion are simply linear combinations of the Cartan charges) which appears in \eq{III} via $\Delta$. Thus we have 
\beq
     \d_{h_\alpha} I_{(2,J)}=\frac{1}{2i}\lambda_{++--} \ , \ \ \d_{h_\alpha} I_{(-2,J)}=-\frac{1}{2i}\bar\lambda_{++--} \ , \ \ 
     \d_{h_\alpha} I_{(0,J)}=-i(\lambda_1\lambda_2-\lambda_3\lambda_4) \ .
 \eeq
 These are the conditions we impose in order to fix the zero mode ambiguity in solving the linear system, which as a result allows us to compute the variations of all other integrals of motion in terms of the Q-functions.

\subsubsection{Example: variation of the inverse graph-building operator}

A particular nontrivial operator whose variation we can consider is the Hamiltonian $\hat H$, i.e. inverse of the graph-building operator, described in section \ref{sec:wfspinchain} in \eq{eq:invB}, since it is one of the coefficients of the transfer matrix ${\mathbb T}^{\bf 6}$, namely the integral of motion $\hat I_{(0,1)}=\left.{\mathbb T}^{\bf 6}\right|_{u=0}$. Let us compute its variation w.r.t the weight on one site $h_\alpha$. We will restrict to the case of the state without magnons, i.e. after taking the variation we set all $h_\alpha=1$ and $\theta_\alpha=0$. We find
\beq
\label{dH}
    \delta \hat H=\delta h_\alpha\; {\rm Tr}\(\hat{\mathbb L}^{\bf 6}_J(0)\dots \hat{\mathbb L}^{\bf 6}_{\alpha+1}(0)\hat V_\alpha \hat{\mathbb L}^{\bf 6}_{\alpha-1}(0)\dots \hat{\mathbb L}^{\bf 6}_1(0)G^{\bf 6}\)
\eeq
where the operator $\hat V$ is read off from  the Lax operator \eq{eq:Lax6},
\beq
\label{defV}
    \hat V\equiv\left.{\d_h \hat{\mathbb L}^{\bf 6}}\right|_{u=0}=-i\d_h\hat  q+\frac{1}{2}\hat q \cdot\d_h \hat q+\frac{1}{2}\d_h\hat q\cdot  \hat q-\frac{1}{2}
\eeq
in terms of the conformal generators $\hat q^{MN}$ defined in section \ref{sec:conformal0}. The variation $\d_h\hat q$ comes about due to the explicit $h$-dependence in the realisation of the generators in \eq{4Dgen} which leads to the nonzero components of $\d_h\hat q$ being
\beq
    \d_h \hat q^{-1,0}=-\d_h \hat q^{0,-1}=-i \ , \ \ \d_h \hat q^{0,\mu}=-\d_h \hat q^{\mu,0}=ix^\mu \ , \ \ \d_h \hat q^{-1,\mu}=-\d_h \hat q^{\mu,-1}=-ix^\mu\;.
\eeq

In order to write the operator \eq{defV} more explicitly, it is very useful to employ the 6D realisation of the conformal group we outlined in section \ref{sec:conformal0}. In  this formalism, the $\hat q^{MN}$ operators do not depend on $h$ explicitly, rather the $h$-dependence is contained in the 6D function \eq{f6f4} on which they act. After a somewhat lengthy calculation (similar to that done in \cite{Gromov:2019bsj} to compute explicitly the Hamiltonian), we find many cancellations and the result for the variation of the Lax operator in $h$ is surprisingly simple, namely
\beq
    \hat V^{MN}=\frac{1}{2}\(X^M\d^N+X^N\d^M\)
\eeq
in 6D notation. Translating it into 4D and combining it with the remaining operators inside the trace \eq{dH}, we find\footnote{For simplicity we restrict to the case where none of the sites $\alpha-1,\alpha,\alpha+1$ are at position $J$, otherwise one finds a slightly different result.} 
\beqa\label{eq:dinvB}
\frac{\d \hat H}{\d h_\alpha} &=& \frac{1}{(-4)^{J-2}} 
\left|\frac{\partial G(x_J) }{\partial x_J} \right|^{-\frac{1}{4} } \prod_{\beta<\alpha \;,\;\beta>\alpha+1}  x_{\beta, \beta-1}^2 
\, \\ \nn
&\times&\frac{1}{2}
 \[-\frac{x_{\alpha,\alpha-1}^2+x_{\alpha,\alpha+1}^2}{2}\(1+x^\mu_\alpha\frac{\d}{\d x_\alpha^\mu} \)+(x_{\alpha,\alpha-1}^2x_{\alpha+1}^\mu  +x_{\alpha,\alpha+1}^2x_{\alpha-1}^\mu) \frac{\d}{\d x_\alpha^\mu} \]
 \prod_{\beta\neq\alpha}^J \Box_{\beta}\; .
\eeqa
We see that the change from the original Hamiltonian  \eq{eq:invB} amounts to removing the Laplacian operator at the site $\alpha$ and replacing the two inverse propagators involving that site with a combination of $x$'s and derivatives. Thus we can compute the form factor of this operator according to \eq{eq:finalresv}. We can further simplify the result by considering the operator $\frac{\d \hat H}{\d h_\alpha}\hat H^{-1}$ whose form factor is trivially related to that of $\frac{\d \hat H}{\d h_\alpha}$ (via multiplication by $\xi^{2J}$) since the states we consider diagonalise $\hat H$. Nicely, the result is a `local' operator that acts only on the three neighbouring sites $\alpha-1$, $\alpha$ and $\alpha+1$,
\beqa \nn
\frac{\d \hat H}{\d h_\alpha}\hat H^{-1} &=& -8
 \[-\frac{x_{\alpha,\alpha-1}^2+x_{\alpha,\alpha+1}^2}{2}\(1+x^\mu_\alpha\frac{\d}{\d x_\alpha^\mu} \)+(x_{\alpha,\alpha-1}^2x_{\alpha+1}^\mu  +x_{\alpha,\alpha+1}^2x_{\alpha-1}^\mu) \frac{\d}{\d x_\alpha^\mu} \]
 \\ 
 &\times& \Box_\alpha^{-1}\frac{1}{x_{\alpha,\alpha-1}^2}\frac{1}{x_{\alpha,\alpha+1}^2} \ .
\eeqa
The expectation value of this operator corresponds to a class of nontrivially modified Feynman diagrams that are therefore computable within our functional SoV approach. It would be also interesting to try and link it with concrete OPE coefficients and other observables. In general, using variations in the local spin chain parameters such as the weights and the inhomogeneities opens the way to computing many nontrivial observables, and it would be important to explore them further.

\section{Scalar Products in Separated Variables}\label{sec:sovscalar}
In the spin chains with more regular HW representations, it was shown in \cite{Gromov:2019wmz,Gromov:2020fwh} that
one can obtain a lot of structural information
from the functional SoV approach~\cite{Cavaglia:2019pow}. In particular one can deduce the scalar product in SoV basis in determinant form.
In the current case, when none of the Q-functions are polynomial, there are additional subtleties which can be properly addressed with extra additional input from the operatorial SoV approach, like in~\cite{Gromov:2020fwh}. 
The manifestations of these additional complications can be seen for example in 
the properties of the scalar product in the coordinate space (see section~\ref{sec:div}). A rather new feature, in comparison to the quasi-periodic systems studied with SoV methods so far, is that the scalar product has a log-divergence 
when both states have the same dimension $\Delta$ (and the twists are opposite); in this case, the meaningful quantity is the coefficient in front of the logarithm. At the same time, for two generic states we expect the properly regularised scalar product to be finite and thus we may be able to follow the usual path of~\cite{Gromov:2019wmz,Gromov:2020fwh} and establish the link between the SoV and coordinate representations of the scalar product. 

In this section we will introduce the SoV-like scalar product, following the functional SoV approach. Then we discuss its key properties such as orthogonality and also show that the log-divergence presented in the coordinate representation emerges naturally in the functional SoV formalism. 

\subsection{General philosophy}
First let us schematically repeat the general logic.
As in the case of the simple spin chains we expect that there is an SoV basis $\langle x|$ such that the eigenvectors of the transfer matrix (as well as many other states) $|\Psi\rangle$
factorise. For convenience in this section we will use the bra and ket notations, but the scalar product in the case of the conformal spin chain should coincide with the one introduced in section~\ref{sec:scaladef}. In general we can write
\beq
\langle x|\Psi^A\rangle = \prod_{\alpha=1}^J Q^A(x_{\alpha})
\eeq
where $Q(u)$ are some simple combinations of the Q-functions $q_a(u)$, corresponding 
to the state $A$, with some shifts. Similarly, the conjugate states (as defined in section~\ref{sec:quantwf}) are factorised in general in the dual SoV basis $|y\rangle$
\beq
\langle \Psi^B|y\rangle = \prod_{\alpha=1}^J Q^B(y_{\alpha})\;.
\eeq
Then, using the expected completeness of the SoV basis, we get
\beq\la{scalinsov}
\langle \Psi^B |\Psi^A\rangle =
\sum_{x,y}\langle \Psi^B
|y\rangle
M_{x,y}
\langle x|\Psi^A\rangle = 
\sum_{x,y}\prod_{\alpha=1}^J Q^B(y_{\alpha})
M_{x,y}
\prod_{\alpha=1}^J Q^A(x_{\alpha})
\eeq
where the SoV measure $M_{x,y}\equiv \langle x|y\rangle^{-1}$,
can be deduced  most efficiently from the functional SoV approach, like in~\cite{Gromov:2020fwh}.
The main point of the expression \eq{scalinsov}
is that it allows one to concentrate all state-dependent information in the Q-functions, and combines it together with some universal measure factor $M_{x,y}$ into the scalar product.
Furthermore the r.h.s. of \eq{scalinsov} can be usually written as a determinant, which makes it very useful in practice.

The idea of~\cite{Cavaglia:2019pow} was that the r.h.s.
of \eq{scalinsov} can be deduced directly from the Baxter TQ-relations and is greatly constrained by the requirement that for two different eigenfunctions of the transfer matrices the combination of Q-functions should obey orthogonality, i.e. vanish for any pair of states with distinct eigenvalues. 

In this section we will follow~\cite{Cavaglia:2019pow} to establish the possible form of the r.h.s. of \eq{scalinsov} and discuss the main properties of the resulting expression. In order to conclude with certainty about the relation to the l.h.s. of \eq{scalinsov}, i.e. to the conformally  invariant scalar product \eq{eq:scalar}, further studies, in particular at the operatorial SoV side, are required and will be reported elsewhere.

\subsection{Orthogonality relation for generic states}\label{sec:scalardisc}
Here we repeat the argument of~\cite{Cavaglia:2019pow} for the orthogonality relation. The starting point is the Baxter operator \eq{eq:BaxOpdef}. We will need two copies of that for two different states $A$ and $B$ such that
\beq
{\cal B}^A q_a^A=0\;\;,\;\;{\cal B}^{{\rm dual}B} p_a^B=0\;.
\eeq
Using the conjugation property of the Q-bilinear form we have again
\beq
 \langle p_{a}^{B}  ({\cal B}^A-{\cal B}^B)  \circ  q_{c}^{A}   \rangle_\alpha = 0\;\;,\;\;a,b=1,\dots,4\;\;,\;\;\alpha=1,\dots,J\;.
\eeq
Writing the difference of the Baxter operators more explicitly we get
 \beqa\label{eq:delB2}
 {\cal B}^A-{\cal B}^B  &=& 
 \sum_{{\bf b}\in \{-2,0,\bar 0 ,2\}} (-1)^{\tfrac {\bf b}2}\sum_{\alpha=1}^{J} (u+\tfrac{i{\bf b}}{2})^{\beta-1} \, ( I^A_{({\bf b},\beta)}- I^B_{({\bf b},\beta)}) \, \cD_{\mathbf b }
 \;.
 \eeqa
 Combining the two we get 
\beq\label{eq:varsys2}
\sum_{\beta=1}^J\sum_{{\bf b}\in \{-2,0,\bar 0 ,2\}}\left(\mathcal{M}^{AB} \right)_{(a,c,\alpha)}^{\;({\bf b},\beta)} \cdot ( I^A_{({\bf b},\beta)}- I^B_{({\bf b},\beta)}) =0\;,
\eeq
where like in \eq{eq:diagM} we have
\beqa\label{eq:diagM2}
\left(\mathcal{M}^{AB} \right)_{(a,c,\alpha)}^{\;({\bf b},\beta)} &\equiv&  (-1)^{\frac{{\bf b}}{2}} \langle   \; p_{a}^{A}   \;  (u+\tfrac{i{\bf b}}{2})^{\beta-1}\cD_{\bf b} \, \circ q_{c}^{B} 
\rangle_{\mu_\alpha}\;.
\eeqa
The main difference with the section~\ref{sec:diagM},
where we were interested in the continuous variation of the conserved charges w.r.t. the internal (e.g. $\xi$) or external (e.g. $\theta_\alpha$) parameters, is that for the convergence of the Q-bilinear form constituting $\mathcal{M}^{AB}$ we are no longer required to assume $a$ and $c$ to be related to each other. So in total we have $16$ possible combinations of the indices $a$ and $c$. In the case of spin chains with HW representation the natural choice for $a$ and $c$ is given by the polynomiality constraint of $p_a$ and $q_c$. In that case the set of possible values of $a$ and $c$ is such that the analogue of the matrix 
$\mathcal{M}^{AB}$ is a square matrix.
In our case this would correspond to a selection of $4$ combinations of
$(a,c)$
out of $16$. Imagine this was done, then the system 
\eq{eq:varsys2} becomes a homogeneous  $4J\times 4J$
linear system on $4J$ unknowns $I^A- I^B$. Following the logic of~\cite{Cavaglia:2019pow}, this is only possible if
\beq\la{orth}
\det \mathcal{M}^{AB} = 0\;\;,\;\;A\neq B\;.
\eeq
The expression \eq{orth} represents the so-called orthogonality relation.
Even though the above expression is derived for two different eigenstates of the same transfer matrix, the interpretation \eq{orth} goes beyond that case. In the cases of the HW spin chains, where the operatorial formalism was worked out explicitly, it was shown~\cite{Gromov:2020fwh} that it gives the realisation of the expression \eq{scalinsov} for the scalar product of two states
\beq\la{scalinsov2}
\langle \langle \Psi^B ,\Psi^A\rangle\rangle =
\bar{\cal N}_B{\cal N}_A\det \mathcal{M}^{AB}\;,
\eeq
where $\bar{\cal N}_A,\;{\cal N}_B$ are the normalisation coefficients, dependent on one state only. Of course, when $|\Psi^A\rangle$ and $\langle \Psi^B |$ are two different eigenstates of the same transfer matrix then \eq{orth}
should hold indeed. Furthermore, in~\cite{Gromov:2020fwh}
it was shown that the relation stays correct in a number of less trivial
situations, for example when the states correspond to the transfer matrices with different twist eigenvalues $\lambda_a$, 
or even when the states are not eigenstates of any transfer matrix, but are factorisable in the same SoV bases $\langle x|$ and $|y\rangle$. In all these cases \eq{scalinsov2} gives a nontrivial r.h.s. and allows to bring together the SoV representation and the coordinate representations of the various overlaps.

Whereas the proper proof of the proposal \eq{scalinsov2}
requires further insights from the operatorial SoV approach, like in~\cite{Gromov:2020fwh}, we can analyse some common features of the l.h.s. of \eq{scalinsov2}, which is represented by a $4J$-dimensional integral of two CFT wave-functions, and the r.h.s. which is given by the determinant of one dimensional integrals of bilinears of Q-functions.
One key property, discussed in section~\ref{sec:div}, is the logarithmic divergence in the case $\Delta^A=\Delta^B$. This can be reproduced in the r.h.s. of \eq{scalinsov2} as some of the Q-bilinear forms are also divergent when $\Delta^A=\Delta^B$, due to a non-regularisable $1/n$-divergence in \eq{general}.
In order to mimic the cutoff in the coordinate space, we can introduce a slight difference in the twist eigenvalues $\lambda_a$, then the $1/n$-divergence will get replaced in \eq{general} by
\beq
\sum_{n=1}^\infty\frac{(\lambda^B/\lambda^A)^n}{n}=\log\(
\frac{\lambda^A}{\lambda^A-\lambda^B}\)\;.
\eeq
Analysing the finite part of the determinant $\mathcal{M}^{AB}$
under such regularisation, one can also deduce the SoV representation 
for the particularly important case $\Delta^A=\Delta^B$.

We will leave all these very intriguing questions for future investigation. In particular, this includes the question of how to pick correctly the $4$ combinations of $a,c$ out of $16$ possibilities.
Let us point out, however, that a similar ambiguity/freedom can be observed in the
spin chains in finite dimensional representations.  In this case, all Q-functions are (twisted) polynomials and multiple combinations of indices $a$ and $c$ would produce the correct expression for the norm in SoV. 
Different possibilities correspond to using different reference  states in the algebraic Bethe ansatz.

\section{On the structure of the g-function in separated variables}\label{sec:gsec}
The g-function is an important object appearing in studies of 
 integrable systems with boundaries \cite{Affleck:1991tk}, and recently connected to interesting observables in $\mathcal{N}$=4 SYM  \cite{deLeeuw:2015hxa,Komatsu:2020sup,WLToappear,Jiang:2019xdz}. In our context it can be interpreted as an overlap between a CFT wave-function and a fixed boundary state. Whereas the exact form of the overlap would depend on the details of the boundary state, it also contains a universal part, which is, however, very hard to calculate. 
This universal part satisfies  special selection rules, and in this section we propose a construction based on the Q-bilinear forms which nontrivially obeys these properties and is a suitable candidate for the universal part of the g-function.

The construction here is inspired by \cite{Buhl-Mortensen:2015gfd,Caetano:2020dyp,Gombor:2021uxz}, where it was observed that the universal part has a very suggestive structure, which is related in a simple way to the expression for the norm -- both in the case of spin chains and for some field theories.

\subsection{Review: the g-functions and their parity selection rules}
We start with some introductory remarks where we will review the definition of these observables, and describe the parity selection rules that they have to satisfy.

\begin{figure}[t!]
\begin{center}
\includegraphics[scale=0.4]{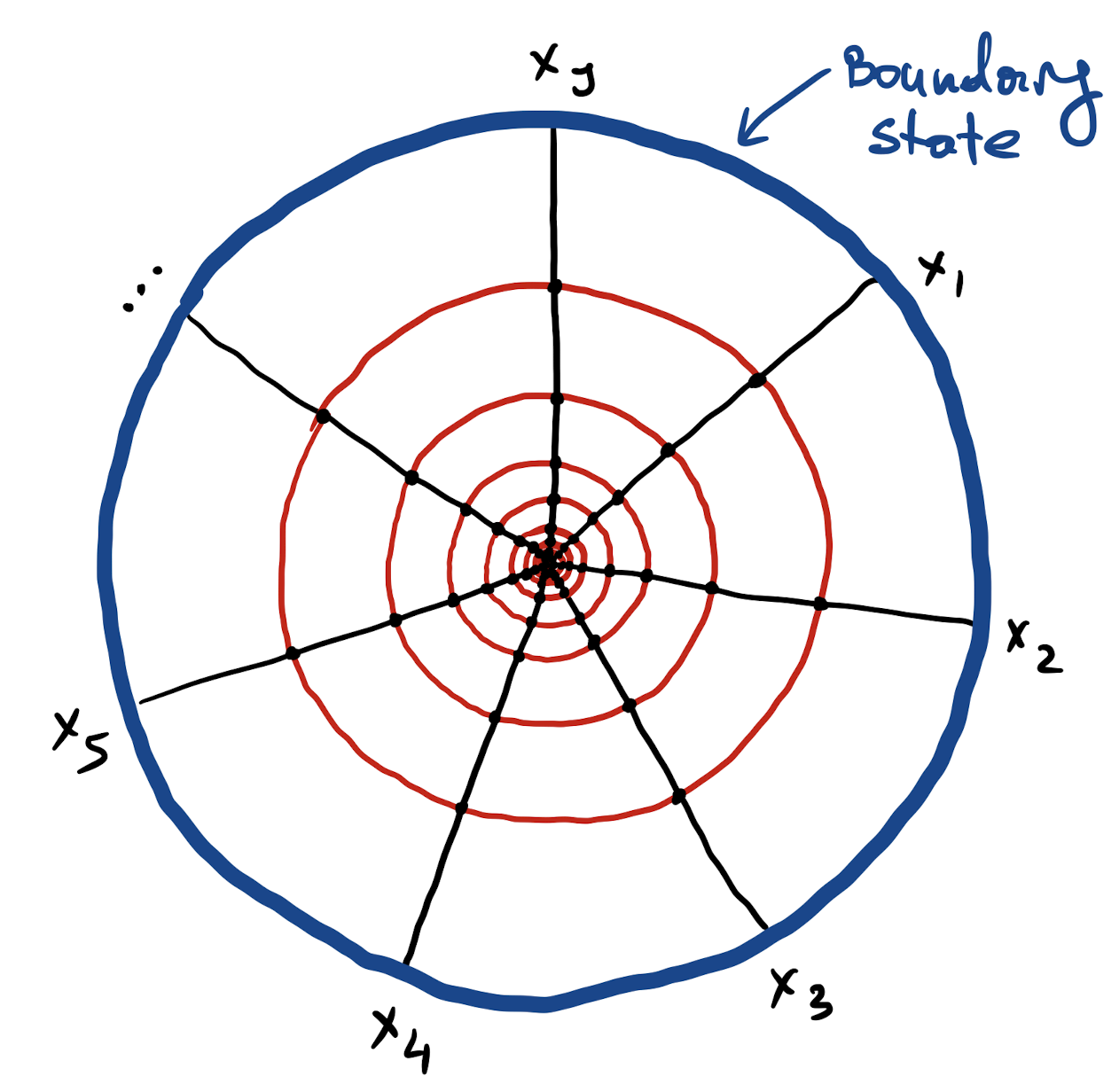}
\end{center}
\caption{An example of the overlap of CFT wave function with a boundary state in the fishnet model is a correlation function between a local operator and a circular Wilson-Maldacena loop. \label{fig:boundary}}
\end{figure}

The g-functions measure the overlap between a generic state $|\Psi\rangle$
and an integrable boundary state $\langle B |$. Usually the boundary state $\langle B |$
can be obtained as a ``Wick rotation" of an integrable boundary condition, which satisfies a suitable boundary-Yang-Baxter equation~\cite{Sklyanin:1988yz}. For example, a version of the Wilson line 
in the fishnet theory
was shown to be an example of an integrable boundary condition~\cite{Gromov:2021ahm}
and the corresponding boundary state can be written explicitly as (see figure \ref{fig:boundary})
\beq
\langle B_{WL}| =
\sum_{n=1}^J
\int_{0}^{2\pi}\frac{d\phi_1}{2\pi}
\int_{\phi_1}^{2\pi }\frac{d\phi_2}{2\pi  
}
\dots
\int_{\phi_{J-1}}^{2\pi}\frac{d\phi_J}{2\pi }
\prod_{i=1}^J\frac{1}{(x_{i+n,1}-R\sin\phi_i)^2+
(x_{i+n,2}-R\cos\phi_i)^2
}\;,
\eeq
where for simplicity we assume that the contour is in the $(1,2)$ plane.\footnote{More explicitly for $J=1$ and $J=2$:
$\langle B_{WL}| =\prod_{i=1}^J\frac{1}{x_{i,1}^2+x_{i,2}^2-R^2}$ and becomes more complicated for $J>2$.
}
Other examples from $\mathcal{N}$=4 SYM could arise in various situations: for instance, when we consider the expectation value of a single-trace operator in presence of a domain-wall defect \cite{Komatsu:2020sup,Gombor:2020kgu}, or a Wilson loop \cite{WLToappear}, or its contraction with two determinant operators \cite{Jiang:2019xdz}, \cite{VescoviToappear}. In $\mathcal{N}$=4 SYM, all these quantities have been related to a g-function with an appropriate boundary state. 

A simple characterisation of integrable boundary states was proposed in \cite{Ghoshal:1993tm} for integrable 1+1 dimensional QFT and generalised in \cite{Piroli:2017sei} for lattice systems. This condition states that quasi-cyclic invariant integrable boundary states are  annihilated by all the conserved charges of the spin chain that are odd under a certain ``chain-reflection" operation $\Pi$. Assuming that $\Pi$ also maps
the integrals of motion into linear combinations of the integrals of motion, they can be organised in two families, denoted as $H_+$ and $H_-$, which are respectively even or odd with respect to $\Pi$:
\beq
\Pi \cdot \hat H_{\pm} \cdot \Pi = \pm \hat H_{\pm}\ .
\eeq
Then the nontrivial integrability condition of boundary states proposed in \cite{Piroli:2017sei} reads:
 \beq\label{eq:intH}
 \hat H_{-} | B \rangle = 0\;,
 \eeq
namely the boundary states has to be annihilated by all odd charges.
Usually \eq{eq:intH} is a nontrivial condition, which we verified for $\ket{B_{WL}}$ explicitly at $J=2$ to be true for trivial twists $\lambda_a=1$.

In our case, a possible choice of the space-reflection is the transformation of a CFT wave function
\beq\la{eq:spinreflection}
\Pi\circ \Psi( x_1,\dots, x_J)\to 
\Psi(\tilde x_J,\dots,\tilde x_1)\;,
\eeq
where in addition we perform a reflection
in the space-time $\tilde x_\alpha =  (-x_{\alpha,1},x_{\alpha,2},-x_{\alpha,3},x_{\alpha,4})$.
Note that the boundary state $\langle B_{WL}|$ 
is invariant under this transformation, as reordering changes the orientation of the WL, which is then restored with the spatial reflection.

The g-function is defined as the normalised overlap between an integrable boundary state and an eigenstate of the integrals of motion $|\Psi \rangle $:
\beq\label{eq:gfunc0}
g \equiv \sqrt{\frac{\langle B | \Psi \rangle\, \langle  \Psi | B  \rangle }{   \langle \Psi | \Psi \rangle } }\;.
\eeq
Note that \eq{eq:intH} implies that there is a selection rule for g-function
\beq\label{eq:selection}
\langle B | \Psi \rangle = 0 , \text{   if   } \vec{\hat H}_{-} \cdot | \Psi \rangle \neq 0\;,
\eeq
namely, the overlap is non-vanishing only for symmetric states.

The g-function was studied quite intensively recently.
One can encode the information on the boundary state $| B\rangle$ 
through the boundary reflection operator~\cite{BoundaryYB}, which in many interesting cases
reduces to a reflection phase $\Theta(u)$.
The first proposal for a method to compute the g-function in integrable systems was made in \cite{LeClair:1995uf}, as a convolution involving the reflection phase and the Y-functions. However,  it was first pointed out in \cite{Dorey:1999cj} that the g-function must also contain an additional \emph{universal factor}, that does not depend on the boundary state, but only on the  state $|\Psi \rangle$.  
A correct proposal for this universal factor in rank-1 system was made for the first time in \cite{Dorey:2004xk}. Later, path integral arguments  to prove this result were presented in \cite{Woynarovich:2004gc} (reproducing part of the result), and \cite{Pozsgay:2010tv} (reproducing the whole result). Recently a rigorous derivation of the general form of the g-function, extended to generic rank, was given in \cite{Kostov:2018dmi}, see also \cite{Jiang:2019xdz} for a different argument.

Schematically, the structure of the g-function is the following:
\beq
g = \underbrace{ \text{exp}\left( \int_{0}^{\infty} \Theta(u) \log(1 + Y(u) ) du \right) }_{\text{boundary-dependent, simple}} \times \underbrace{\sqrt{\frac{\text{det}\left[1- \hat G_-\right]}{\text{det}\left[1- \hat G_+\right]} }}_{\text{universal factor, hard}} \;,\label{eq:gstruct}
\eeq
where we are omitting sums over different types  of Y-functions for simplicity. The universal factor is a ratio of Fredholm determinants of integral operators $\hat G_{\pm}$, which are defined in terms of the Y-functions (see for example \cite{Jiang:2019xdz} for details).

Here, we want to discuss the structure of the universal factor in terms of the Q-functions, following similar arguments to \cite{Caetano:2020dyp} in the case of the Sinh-Gordon model. In that work, it was suggested that the Fredholm determinants can be related to factors of definite parity of the determinant defining the norm of the state in separated variables.\footnote{ This  association with the norm emerges quite clearly in the asymptotic large-volume limit, where the Fredholm determinants give rise to Gaudin-type determinants, see \cite{Jiang:2019xdz}. }
 An important guiding principle in formulating such a proposal is that any g-function has to satisfy the selection rule \eq{eq:selection},   which should be a property of the universal factor. 
Below we demonstrate that there is such combination of Q-functions in the  fishnet theory, which we propose as a natural candidate for the universal factor.

\subsection{Chain-reflection symmetry in the fishchain}
We start by discussing the chain-reflection symmetry $\Pi$ in the fishnet spin chain. 
For simplicity we consider the case with no magnons $M=0$.
In fact the presence of magnons explicitly violates parity, as in the fishnet theory the magnons spiral in one direction around the ``worldsheet" of the Feynman diagram. 
The chain-reflection, acting on the CFT wave-function,
is defined in \eq{eq:spinreflection}. 

Now we need 
to see how this symmetry acts on the integrals of motion.
In order to map the integrals of motion to each other we will additionally require that the twist eigenvalues satisfy
$\lambda_1 = 1/\lambda_2$, $\lambda_3 =1 /\lambda_4$
and that the inhomogeneities are such that $\theta_{\alpha}  =  - \theta_{J+1-\alpha}$.
A reversal of the order of spin chain sites is equivalent to transposition in the auxiliary space in the definition of the transfer matrices. 
From the definition of the Lax matrices in appendix \ref{app:Ts}, we see that 
\beq
(\hat{ \mathbb{L}}^{\mathbf{6}} (u) )^{MN} = (\hat{ \mathbb{L}}^{\mathbf{6}} (-u) )^{NM}\;\;,\;\;
(\hat{ \mathbb{L}}^{\mathbf{4}} (u) )^{a}_{\;\;\; b} = -(\hat{ \mathbb{L}}^{\bar{\mathbf{4}}}(-u) )_b^{\;\;\;a }
\;,
\eeq
namely, transposition in auxiliary space sends $u \leftrightarrow -u$ and $\mathbf{4} \leftrightarrow \bar{\mathbf{4}}$. 
The reflection $x_\alpha\to \tilde x_\alpha$ is a conformal transformation which in the $\mathbf{4}$ representation interchanges $1\leftrightarrow 2$ and 
$3\leftrightarrow 4$, and thus transforms the diagonal twist matrix 
$\Lambda^{\bf 4}=\text{diag}\left(\lambda_1,\lambda_2,\lambda_3,\lambda_4\right)\to
\text{diag}\left(\lambda_2,\lambda_1,\lambda_4,\lambda_3 \right)=\Lambda^{\bar{\bf 4}}$.
At the same time the $\Lambda^{\bf 6}$ is not diagonal but the transposition is exactly compensated by the twist interchange.
Thus we conclude that
\beq
\mathbb{T}^{\bf 6}(u)\to \mathbb{T}^{\bf 6}(-u)\;\;,\;\;
\mathbb{T}^{\bf 4}(u)\to (-1)^J \mathbb{T}^{\bar{\bf 4}}(-u)\;,
\eeq
which then results in the following transformation of the polynomials $P^{\mathbf{r}}(u)$:
\beq\label{eq:Pipar}
\Pi: \;\;\;\;\; \left\{P^{\mathbf{4}}_J(u), \;P^{\bar{\mathbf{4}}}_J(u), \; P^{\mathbf{6}}_{2 J}(u) \right\} \rightarrow \left\{   (-1)^J \, P^{\bar{\mathbf{4}}}_{J}(-u), \; (-1)^J \,  P^{\mathbf{4}}_{J}(-u), \; P^{\mathbf{6}}_{2 J}(-u) \right\}\;.
\eeq
As the parity changes the eigenvalues of the integrals of motions, it also acts nontrivially on the  eigenstates
and consequently on the Q-functions.
If the initial $q(u)$ was solving the TQ-relation
${\cal B}\circ q(u)=0$, then the Baxter equation with transformed coefficients $\tilde{\mathcal{B}}\circ \tilde q(u)=0$ is solved simply by $\tilde q(u)=q(-u)$.

More precisely, taking into account the conventions for the asymptotics  (\ref{eq:Qasy}) and analyticity we get
\beq\label{eq:Qpar}
\tilde q_a^{\downarrow } (u)  \propto  q_{\sigma(a)}^{\uparrow}(-u) , \;\;\;\;\;  
\tilde q_{a}^{ \uparrow }(u)  \propto   q_{\sigma(a)}^{\downarrow}(-u) \ ,  
\eeq
where $\sigma$ is a permutation acting as
 $\left\{1,2,3,4\right\} \rightarrow \left\{2,1,4,3\right\}$. 

\paragraph{Parity-even and parity-odd integrals of motion.}
Finally, from (\ref{eq:Pipar}) we introduce even and odd combinations of the integrals of motion:
\beqa
&& H_{(+ , \alpha)} \equiv I_{(2,\alpha)} + (-1)^{J + \alpha-1} I_{(-2,\alpha)} \;,\label{eq:newbasis40} \\  && H_{(-, \alpha)} \equiv I_{(2,\alpha)} - (-1)^{J + \alpha-1} I_{(-2,\alpha)}\; , 
\label{eq:newbasis1}\\ \nn
&& H_{(+ , \alpha)}'\equiv \left( I_{(0,1)}, \dots I_{(0, J)} , I_{(\bar{0},1)}, \dots, I_{(\bar{0}, J)} \right)_{2 \alpha -1}\; , \\ 
&& H_{(- , \alpha)}'\equiv \left( I_{(0,1)}, \dots I_{(0, J)} , I_{(\bar{0},1)}, \dots, I_{(\bar{0}, J)} \right)_{2 \alpha }\; ,\label{eq:newbasis4}
\eeqa
where $1 \leq \alpha \leq J$. 
Under a parity transformation, (\ref{eq:Pipar}) implies that \beq
\Pi: \;\;\;\;\;
\left\{ H_{(\pm, \alpha)} , H_{(\pm, \alpha)}' \right\} \rightarrow \pm \left\{ H_{(\pm, \alpha)} , H_{(\pm, \alpha)}' \right\}\; .
\eeq
Parity-symmetric states are the ones for which $ H_{(-, \alpha)}  =  H_{(-, \alpha)}'  = 0$, 
 which implies in particular that they must have zero spins $S_i = 0$, $i=1,2$, whereas $\Delta$ is parity even and does not transform.

\subsection{Expression in terms of the Q-bilinear forms}
In section \ref{sec:sovscalar} we introduced a linear system of equations associated to any pair of states $A$, $B$ (see \eq{eq:varsys2}), the solution of which is the difference of their integrals of motion. 
Now let us apply the same argument for a pair of states  chosen as follows: a generic state $A$, and the state $\tilde{A}$ which is the image of $A$ under the parity transformation $\Pi$. Note that in this case the values of $\Delta$ for both states coincide and we have to take $a=c$ in \eq{eq:varsys2} in order to keep the matrix elements ${\cal M}^{A\tilde A}$ finite.

These two states must have the same value for all the even charges $H_+$, $H_+'$, while the odd charges differ by a sign $H_-^A = - H_-^{\tilde{A}}$, $H_-^{'\,A} = - H_-^{' \, \tilde{A} }$. 
In this case the null eigenvector 
$I^A-I^B$ in \eq{eq:varsys2}
has only $2J$ nontrivial components,
since when written in the basis of $H_\pm$ the components corresponding to $H_+^A-H_+^{\tilde A}$ should be zero by construction.
After this change of basis of the integrals of motion, we get
\beq\label{eq:blockMtxt}
\tilde{\mathcal{M}}^{A \tilde{A}} = \left(\begin{array}{c|c|c|c}  m_{-}^{(1)} & \; m_{-}^{'(1)} & \; m_{+}^{ (1) } & \; m_{+}^{' (1) }\\
  \hline m_{-}^{(2)} & \; m_{-}^{'(2)} & \; m_{+}^{ (2) } & \; m_{+}^{' (2) }\\
\hline   m_{-}^{(3)} & \; m_{-}^{'(3)} & \; m_{+}^{ (3) } & \; m_{+}^{' (3) }\\
    \hline  m_{-}^{(4)} & \; m_{-}^{'(4)} & \; m_{+}^{ (4) } & \; m_{+}^{' (4) }
 \end{array}\right)\;\;,\;\;
\tilde{\mathcal{M}}^{A \tilde{A}} 
\cdot \left( \begin{array}{c} 2\vec{H}_{(-)} \\\hline
 2\vec{H}_{(-)}' 
 \\\hline
 0\vec{H}_{(+)} 
 \\\hline
 0\vec{H}_{(+)}' 
 \end{array} 
 \right)=0
 \;,
\eeq
where the blocks $m_\pm$ are constructed explicitly in terms of Q-bilinear forms in appendix \ref{app:gfunc} (in particular, see (\ref{eq:smallblocks})).

Now let us consider a non-symmetric state, where $\tilde{A}  \neq A$. In this case the system of equations (\ref{eq:blockMtxt}) has a nonzero solution, which implies  that any $2 J \times 2 J$ minor extracted from the left half of the linear system has to vanish. In particular, for a non-symmetric state we have:
\beq\label{eq:selection20}
| M _- | = 0  , \;\;\;\;\;\text{ iff } A \neq \tilde{A}
\eeq
where
\beq\label{eq:defMpm0}
M_{-} = \left(\begin{array}{c|c}  m_{-}^{(1)}  & \; m_{-}^{'(1) } \\
  \hline m_{-}^{(2)}  & \; m_{-}^{ '(2) } \end{array} \right)\;.
\eeq
The reason we pick this particular minor is because, as we show in appendix \ref{app:gfunc}, for the partity symmetric 
states $A = \tilde{A}$ the matrix $\tilde{\cal M}^{A\tilde A}$
becomes block diagonal:
\beq\la{tM}
\tilde{\mathcal{M}}^{A \tilde{A}} = \left(\begin{array}{c|c}  M_- &  0 \\ \hline 
0 &  M_+ 
\end{array}\right)\;,
\eeq
and the only minor which is non-zero in this case is $|M_-|$. 
For example, we computed these determinants in the case $J=2$, $M=0$, for twists $\lambda_1 = 1/\lambda_2 = e^{\frac{i}{3} }$,  $\lambda_3 = 1/\lambda_4 = e^{\frac{i}{2} }$, impurities $\theta_1 = -\theta_2 = -\frac{1}{100}$, coupling $\xi^2 = 0.100992$, and for the state with $\Delta \simeq 1.92864 $ (and with  zero spins). This state is symmetric under chain reflection. We found
\beq
M_- \simeq {\tiny 
\left(
\begin{array}{cccc}
 2.01167\, -0.0171415 i & -0.295886-24.8005 i & -2.48553+1.16921 i & 0.283543\, +27.2419 i \\
 -2.03801+0.0171415 i & 0.295886\, +25.2569 i & 2.48553\, -1.18821 i & -0.283543-27.7381 i \\
 1.90268\, -0.425068 i & -5.51714-24.1991 i & 0.00626019\, +1.40562 i & 6.061\, +26.5768 i \\
 -1.94376+0.433256 i & 5.61628\, +24.6345 i & 0.0280348\, -1.43976 i & -6.16937-27.0517 i \\
\end{array}
\right)
} \;,
\eeq
(which has a non-vanishing determinant $|M_-| \simeq -0.00338656 $), and
\beq
M_+ \simeq {\tiny 
\left(
\begin{array}{cccc}
 0.261583\, -5.84906 i & 11.627\, +0.0235689 i & 0.444272\, -2.76135 i & -0.242355+6.14948 i \\
 -0.261583+5.92727 i & -11.8254-0.0235689 i & -0.444272+2.75742 i & 0.242355\, -6.23208 i \\
 0.78032\, +3.40643 i & -11.132+2.53844 i & -0.243301-0.976073 i & -0.890288-3.91207 i \\
 -0.797245-3.47857 i & 11.3255\, -2.58256 i & 0.241469\, +0.95587 i & 0.908886\, +3.99468 i \\
\end{array}
\right)
}\;  ,
\eeq
which has $|M_+|\simeq 0$, while its null vector is  proportional to $\partial_{\xi^2} (\vec{H}_+ \, , \, \vec{H}_+' )$. 
\paragraph{The structure of the $g$-function. }
We see that the determinant $|M_-|$ vanishes if and only if the state $A$ is parity-symmetric.  This is the same selection rule (\ref{eq:selection20}) exhibited by the overlaps with integrable boundary states. That makes it a natural candidate for the universal part of the overlap appearing in the g-function, 
\beq\label{eq:overlapconj}
\sqrt{\langle B | \Psi_A \rangle\,\langle \Psi_A | B \rangle } \propto | M_-^{(A)} |\;.
\eeq
The meaning of this proposal is that we expect the r.h.s. to appear naturally in the SoV construction of the overlap, in  the same spirit as the discussion of the scalar product in section \ref{sec:sovscalar}. Crucially, the relations  (\ref{eq:selection20}) imply that the r.h.s  of (\ref{eq:overlapconj}) satisfies the selection rules, which provide an infinite number of consistency conditions. 
However, one would still have to figure out explicitly how to build  the non-universal part of this overlap, in terms of the boundary reflection operator corresponding to a given boundary state.

Finally, let us point out that for the parity symmetric states, due to the block diagonal structure of~\eq{tM}, we also find that 
the determinant corresponding to the finite part of the norm has a factorised form:
\beq\label{eq:Mfactorised}
\text{det}( \tilde{\mathcal{M}}^{A A} ) \propto | M_-^{(A)} | \, |M_+^{(A)}| \;,
\eeq
the same way as was found in  \cite{Buhl-Mortensen:2015gfd,Caetano:2020dyp,Gombor:2021uxz}.
 As we discussed in sections \ref{sec:orthosec} and \ref{sec:sovscalar}, for the standard choice of Q-bilinear forms the determinant (\ref{eq:Mfactorised}) vanishes, which in this case is due to $|M_+| = 0$. To find a connection with the (finite part of) the norm of the state, a regularisation prescription is required, which would be clarified in the operatorial SoV construction. Here, we point out that a natural possibility is to consider $|M_+|_{\ast}$, defined as the product of the $2 J -1$ non-vanishing eigenvalues of $|M_+|$. This leads to a proposal for the structure of the universal factor of the $g$-functions, similar to \cite{Caetano:2020dyp}:
 \beq
 (g_{\text{universal}} )^2 \propto \frac{|M_-|}{|M_+|_{\ast} }\; ,
 \eeq
 which is also very similar to the structure found rigorously in \cite{Gombor:2021uxz}  in the case of the Heisenberg spin chain.

\section{$\mathcal{N}=4$ super Yang-Mills: generalisations and speculations}\label{sec:secYM}

In the previous sections we demonstrated that the functional SoV approach to spin chains can be also applied in such integrable field theories as the fishnet CFT.
Here we speculate on how this method can be further extended to more complicated cases such as $\mathcal{N}$=4 SYM. 
In this section we assume the most general deformed version of the theory, which also includes the twist parameters $\lambda_a$ in $AdS_5$.

In this case, the theory is described by a more complicated $PSU(2,2|4)$ Q-system. Four of the Q-functions, usually denoted as $\bQ_i$, are directly connected with the Q-functions of the fishnet model in the double scaling limit of \cite{Gurdogan:2015csr}. They represent $AdS_5$ degrees of freedom in the dual string picture. In addition, in this case we also have the $\bP_a$ functions, which represent the $S^5$ degrees of freedom. In the fishnet limit they simplify drastically and become rational functions with a pole at the origin, and the coefficients in the numerator of these rational functions encode the integrals of motions $I_{({\bf b},\alpha)}$.

\begin{figure}
    \centering
    \includegraphics[scale=0.45]{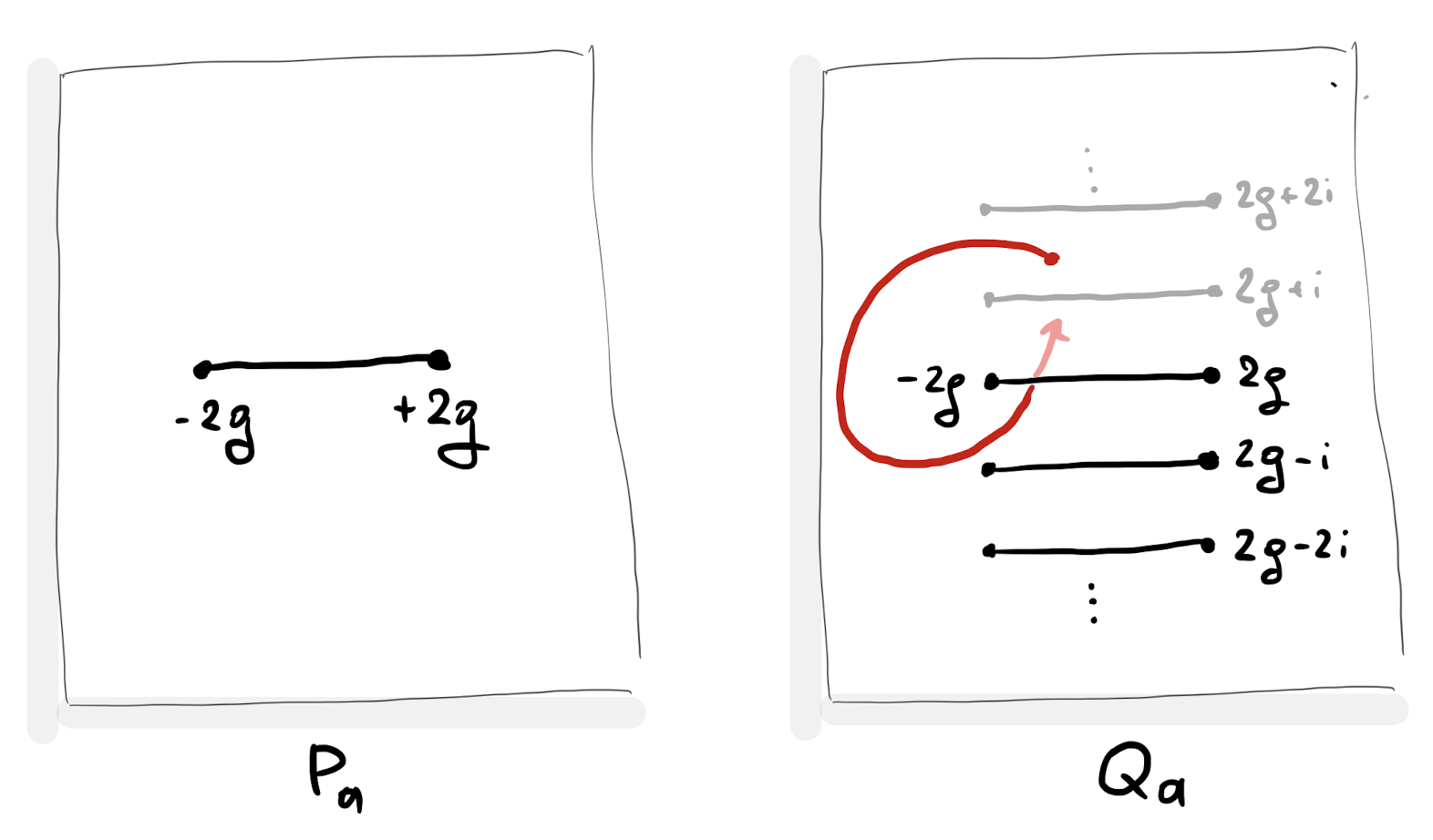}
    \caption{Analytic structure of ${\bf P}$ and ${\bf Q}$ functions. In the fishnet limit $g\to 0$ and the cuts collapse into poles.}
    \label{fig:cuts}
\end{figure}

The $\bP$ and $\bQ$  functions have a characteristic analytic structure  illustrated in figure~\ref{fig:cuts}. The size of the cuts is given by the `t Hooft coupling $g=\frac{\sqrt{\lambda}}{4\pi}$ and tends to zero in the fishnet limit, with cuts of ${\bf Q}_a$ becoming the already familiar poles of $q_a$, whereas ${\bf P}_a$ become rational functions with a pole at the origin.

The Baxter equation satisfied by $q_a$
generalises to a fourth order finite-difference equation~\cite{Alfimov:2014bwa}:
\beq\la{baxsym}
\bQ^{[+4]}_i \mathbb{A}_{+2} + \bQ^{[+2]}_i \mathbb{A}_{+1} + \bQ^{}_i \mathbb{A}_{0} + \bQ^{[-2]}_i \mathbb{A}_{-1} + \bQ^{[-4]}_i \mathbb{A}_{-2} = 0\; ,
\eeq
where the coefficients are ${\bf P}_a$ dependent combinations defined as
\beqa\label{eq:defA}
&&\mathbb{A}_{2} \equiv \frac{1}{\bar{\mathcal{D}}_0 } ,\;\;\;\; \mathbb{A}_{-2} \equiv -\frac{1}{{\mathcal{D}}_0} ,\;\;\; \mathbb{A}_0 \equiv \frac{\mathcal{D}_2 - \mathbf{P}_a \mathbf{P
}^{a[+2]}\mathcal{D}_1 + \mathbf{P}_a \mathbf{P
}^{a[+4]}\mathcal{D}_0 }{\mathcal{D}_0 \bar{\mathcal{D}}_0 },\\
&& \mathbb{A}_1 \equiv -\frac{
{\cal D}_1-\bP_a^{[+2]}\bP^{a[+4]}{\cal D}_0 }{\mathcal{D}_0 \bar{\mathcal{D}}_0 },\;\;\;\; \mathbb{A}_{-1} \equiv \frac{\bar {\mathcal{D}}_1 +\bP_a^{[-2]}\bP^{a[-4]}\bar {\mathcal{D}}_0 }{\mathcal{D}_0 \bar{\mathcal{D}}_0 }\;,
\eeqa
and ${\mathcal{D}}_k$ are some combinations of ${\bf P}_a$'s defined in appendix~\ref{app:baxdet}. 

In the fishnet model, the transfer matrix eigenvalues appearing in the TQ equations were polynomials, leading to a natural basis for the integrals of motion. The coefficients $\mathbb{A}_k(u)$ have a more complicated analytic structure, see appendix~\ref{app:baxdet}.  Crucially, they are asymptotically constant at infinity, and
are analytic outside a certain radius $R_{\ast}$ in the complex plane. 
This means that they admit a convergent Taylor expansion around infinity: 
\beq\label{eq:Aexp}
\mathbb{A}_k(u) = \sum_{n=0}^{\infty} \frac{\mathcal{I}_{(k,n)}}{u^n} , \;\;\;\; |u|>R_{\ast} .
\eeq
The coefficients  $\mathcal{I}_{(k,n)}$  should contain a basis of integrals of motion,
generalising  $I_{({\bf b},\alpha)}$ of the fishnet theory. At any fixed order in the weak coupling expansion, only a finite number of integrals of motion are linearly independent: 
as it is known from the scaling of 
the coefficients of ${\bf P}_a$, at order $O(g^{2l})$ 
 such functions truncate 
to finite Laurent polynomials in $u$ with $\sim l$ coefficients, see e.g.~\cite{Marboe:2014gma}. However, at finite coupling, we have an infinite number of integrals of motion, as expected since the theory  corresponds to a sigma model with infinite number of degrees of freedom (unlike the fishnet theory, which is dual to a system of $J$ particles with $4J$ degrees of freedom~\cite{Gromov:2019aku}\footnote{See also \cite{Basso:2019xay,Basso:2018agi} for a different approach to formulating the dual model.}). 
Finding the most convenient basis of the coefficients $\mathcal{I}_{(k,n)}$, which behaves nicely at weak coupling and takes into account the cut structure of the coefficients $\mathbb{A}_k(u)$, is an important 
problem for future studies. One possibility is to 
use a spectral representation of $\mathbb{A}_k(u)$, which has a finite number of cuts, with the discontinuities expanded into Zhukovsky variables $x^n$ -- this basis will behave well at weak coupling, but could still contain some linear dependencies.

In order to advance further in the analogy with the fishnet theory, we have to introduce Q-bilinear forms
and make sure that the Baxter equation~\eq{baxsym},
written as a finite difference operator
\beq
{\cal B} = \sum_{n=-2}^{2} \mathbb{A}_{n} D^{2 n}
\;\;,\;\;
{\cal B}\circ \bQ_i = 0
\eeq
has a simple conjugation property under this bilinear form. Fortunately, this step can be realised in  $\mathcal{N}=4$ SYM too. We introduce the dual Q-function $\mathbb{Q}^i$ by 
\beq
    \begin{vmatrix}
    \bQ_i(u+i) & \bQ_i(u) & \bQ_i(u-i) \\
    \bQ_j(u+i) & \bQ_j(u) & \bQ_j(u-i) \\
    \bQ_k(u+i) & \bQ_k(u) & \bQ_k(u-i) 
    \end{vmatrix}  \equiv \frac{1}{6}\epsilon_{hijk} \mathbb{Q}^h(u)\;,
\eeq
which in the QSC Q-system notation correspond to the Q-functions $Q_{\emptyset|ijk}$. 
This dual Q-function does indeed satisfy the dual
TQ-relation
as we demonstrate in appendix \ref{app:dualBproof},
\beq
    {\cal B}^{\text{dual}} \circ \mathbb{Q}^i=0\;,
\eeq
where ${\cal B}^{\text{dual}}$ is the conjugate operator to ${\cal B}$ in the same sense as in the fishnet CFT case, i.e. as defined in section~\ref{sec:conjugationdef}. 
This means, for example, that we can repeat the 
derivation of the orthogonality relations from section~\ref{sec:sovscalar},
giving the SoV form of the scalar product starting from the equation
\beq\label{eq:dpSYM}
    \langle \mathbb{Q}^{B\;i} \;   ({\cal B}^B-{\cal B}^A)\circ \bQ^A_j\rangle_{\mu(u)}=0\;,
\eeq
where as before we
define the bracket $\langle f\rangle_{\mu}$ as a
formal integral over the contour $\int_{|}   \equiv  \int_{c -i\infty}^{c + i\infty} - \int_{-c -i\infty}^{-c + i\infty}  $ , where we assume that $c>2g$. Such an integral will encircle the infinite ladder of cuts of the Q-functions. In practice, say for numerical calculations, one would write this integration as a sum over the infinite set of contours encircling the branch-cuts of ${\bf Q}_i$ and $\mathbb{Q}_j$. Then for each cut one can use a strategy very similar to  section~\ref{sec:numericsreview} -- one can flip the location of poles by using the $i$-periodic, anti-symmetric matrix $\omega^{ij}(u)$, which then should allow us to have full analytic control over the tail of the sum over cuts 
and enable efficient $\zeta$-regularisation just like in the fishnet case. A convenient set of periodic functions $\mu(u)$ is
\beq
\mu_n(u) \equiv   \delta_{n,0} + \oint \frac{dv}{2 \pi i} \coth{\pi (u-v)} \left( x^n(v) - x^{-n}(v) \right)\;, \nn
\eeq
which generalises naturally the corresponding basis  in the fishnet CFT~\eq{eq:defmu}.

Furthermore, like in the fishnet case,
there are exactly $4$ combinations of the indices $(i,j)$
in the Q-bilinear forms \eq{eq:dpSYM} which give a finite result in the important case when $\Delta^A=\Delta^B$ and also
the twists $\lambda_a^A=\lambda_a^B$.
Exactly  the same combinations of the Q-function indices $(i,j) \in\left\{ (1,2),(2,1),(3,4),(4,3)\right\}$
lead to a finite result for the integral \eq{eq:dpSYM}
when the Q-functions are on-shell i.e. all QSC analyticity constraints are satisfied including the gluing condition~\cite{Gromov:2015vua,Alfimov:2018cms}.  
For the more general deformed setup of two states with different twists, we expect again that the integrals are convergent for generic pairings the same way as was discussed in section~\ref{sec:sovscalar}.

Of course the plan outlined in this section deserves very intensive further studies, but we believe the main ingredients presented here will become instrumental in the future.
The main difference with the fishnet case is that the number of integrals of motion is now infinite, leading to infinite size determinant for the SoV scalar product. In order to overcome this additional complication, one should develop a good truncation at perturbative level, where the number of independent integrals of motions becomes effectively finite. Then this perturbative method can be extended to finite coupling by an appropriate extrapolation procedure, similarly to how that is done for the spectrum~\cite{Gromov:2015wca}.

One of the applications of this approach would be the calculation of the overlaps between local operators and integrable boundary states, given by
the g-function~\cite{Caetano:2020dyp,Jiang:2019xdz}. 
For that one should follow the steps done in section~\ref{sec:gsec} for the fishnet model.
The main message of this section is that for
the applications of the SoV methods to models like ${\cal N}=4$ SYM one can essentially follow the same steps as for the fishnet model. This makes the development of the SoV for spin chains in general paramount for the progress with the non-perturbative description of  correlators in AdS/CFT.

\section{Summary}

In this paper we initiated the program of studying the multiple wrapped diagrams in fishnet theory by means of Separation of Variables methods. This approach promises to extend the power of the Quantum Spectral Curve and alike methods, initially developed for the spectrum, to more general classes of observables. In contrast with other integrability based methods for correlation functions the presented approach resums all wrapping diagrams and the calculations are valid non-perturbatively for finite length operators\footnote{The SoV has been previously successfully constructed for certain types of fishnet graphs with open boundary conditions, which enter the computation of Basso-Dixon diagrams \cite{Basso:2017jwq}, \cite{Derkachov:2019tzo,Derkachov:2020zvv,Derkachov:2018rot}. This construction by its ideology is closer to the Hexagon formalism, and may lead to its rigorous derivation.
At the same time it inherits the same weaknesses from Hexagon approach when it comes to the wrapping corrections and further resummations at finite coupling.}. 

In this paper we develop the so-called functional SoV approach introduced in \cite{Cavaglia:2019pow}. The next obvious step would be to join our results with the insights coming from the operatorial SoV approach~\cite{Sklyanin:1995bm,Gromov:2016itr,Gromov:2019wmz}, which would allow us to construct more general overlaps and form factors in fishnet theory like in ~\cite{Gromov:2019wmz}. 
 Moreover, an explicit construction of the CFT wave functions (which should be obtainable with the operatorial approach) would be a very significant step towards the full solution of the planar theory. This is because the wave functions contain all the nontrivial coupling dependence of generic correlators. 

Within the functional SoV approach, we made a proposal for the expression for the scalar product in the SoV basis. Furthermore, we derived a closed expression for the derivative of the dimension w.r.t. to the coupling constant $\xi$, which is written in terms of the Q-functions evaluated at one fixed value of $\xi$. This result, which is related to a $4J\times 4J$ determinant of one dimensional integrals of Q-functions, gives a prototype for the type of structures which should appear in more general correlators.
We also developed general numerical methods for efficient evaluation of such integrals of Q-functions. This method is based on the core properties of the Q-functions and the QQ-relations.

Another application of our method is the computation of the g-function, which is closely related to the scalar product from the SoV point of view \cite{Caetano:2020dyp,Gombor:2021uxz}. We discuss this in detail in section~\ref{sec:gsec}.

One unexpected observation we made is that the quantisation condition, which comes as an extra condition on the Q-functions, is tightly related with the finiteness of the norm in SoV representation.
 
In section~\ref{sec:secYM} we speculated on how our methods can be extended and applied for the more complicated theories like ${\cal N}=4$ SYM. There, we argued that all main ingredients of our construction find their counterparts in this more general case, even though more detailed future studies are required. 
 
Let us also mention that in a particular (Gaudin) limit the spin chain we considered here should be linked closely to the computation of multi-point conformal blocks in any dimension, in the light of the results from \cite{Buric:2020dyz}. It would be important to explore the implications of SoV techniques
developed here in this context, where further simplification is expected.

Another important direction is to apply the methods of this paper to the boundary problems like ~\cite{Gromov:2021ahm}. This could lead to generalisation of the initial observation of~\cite{Cusp} for the case of the simple cusp 3-point correlator.

\section*{Acknowledgements}
We are very grateful to A. Sever for many discussions and collaboration on related topics, including SoV for the single-site chain, and E. Vescovi for discussions and explanations on the g-function. We thank D.~Anninos, N.~Drukker, G.~Ferrando, T.~Fleury, T.~Gombor, J.~Julius, V.~Kazakov,  E.~Olivucci, B.~Pozsgay, N.~Primi, P.~Ryan, D.~Serban, E.~Sobko, D.~Volin and G.~Watts for interesting discussions. The work of N.G. and A.C. was supported by European Research Council (ERC) under the European Union’s Horizon 2020 research and innovation programme (grant agreement No. 865075)  EXACTC.

\appendix
\addtocontents{toc}{\protect\setcounter{tocdepth}{1}}
\section{Sigma matrices}
\label{app:sigmas}
The Sigma matrices $\Sigma_{MN}$ in our conventions read
\beq
\begin{array}{ccc}
    \Sigma _{1,2}=\left(
\begin{array}{cccc}
 \frac{i}{2} & 0 & 0 & 0 \\
 0 & \frac{i}{2} & 0 & 0 \\
 0 & 0 & -\frac{i}{2} & 0 \\
 0 & 0 & 0 & -\frac{i}{2} \\
\end{array}
\right) & 
\Sigma _{1,3}=\left(
\begin{array}{cccc}
 0 & 0 & 0 & -\frac{1}{2} \\
 0 & 0 & \frac{1}{2} & 0 \\
 0 & -\frac{1}{2} & 0 & 0 \\
 \frac{1}{2} & 0 & 0 & 0 \\
\end{array}
\right)
& 
\Sigma _{1,4}=\left(
\begin{array}{cccc}
 0 & 0 & 0 & -\frac{i}{2} \\
 0 & 0 & -\frac{i}{2} & 0 \\
 0 & -\frac{i}{2} & 0 & 0 \\
 -\frac{i}{2} & 0 & 0 & 0 \\
\end{array}
\right) 
\\
   \Sigma _{1,5}=\left(
\begin{array}{cccc}
 0 & 0 & \frac{i}{2} & 0 \\
 0 & 0 & 0 & -\frac{i}{2} \\
 \frac{i}{2} & 0 & 0 & 0 \\
 0 & -\frac{i}{2} & 0 & 0 \\
\end{array}
\right)
& 
\Sigma _{1,6}=\left(
\begin{array}{cccc}
 0 & 0 & -\frac{1}{2} & 0 \\
 0 & 0 & 0 & -\frac{1}{2} \\
 \frac{1}{2} & 0 & 0 & 0 \\
 0 & \frac{1}{2} & 0 & 0 \\
\end{array}
\right)
& \Sigma _{2,3}=\left(
\begin{array}{cccc}
 0 & 0 & 0 & -\frac{1}{2} \\
 0 & 0 & \frac{1}{2} & 0 \\
 0 & \frac{1}{2} & 0 & 0 \\
 -\frac{1}{2} & 0 & 0 & 0 \\
\end{array}
\right)
\\
\Sigma _{2,4}=\left(
\begin{array}{cccc}
 0 & 0 & 0 & -\frac{i}{2} \\
 0 & 0 & -\frac{i}{2} & 0 \\
 0 & \frac{i}{2} & 0 & 0 \\
 \frac{i}{2} & 0 & 0 & 0 \\
\end{array}
\right)
&
\Sigma _{2,5}=\left(
\begin{array}{cccc}
 0 & 0 & \frac{i}{2} & 0 \\
 0 & 0 & 0 & -\frac{i}{2} \\
 -\frac{i}{2} & 0 & 0 & 0 \\
 0 & \frac{i}{2} & 0 & 0 \\
\end{array}
\right)
&
\Sigma _{2,6}=\left(
\begin{array}{cccc}
 0 & 0 & -\frac{1}{2} & 0 \\
 0 & 0 & 0 & -\frac{1}{2} \\
 -\frac{1}{2} & 0 & 0 & 0 \\
 0 & -\frac{1}{2} & 0 & 0 \\
\end{array}
\right)
\\
\Sigma _{3,4}=\left(
\begin{array}{cccc}
 \frac{1}{2} & 0 & 0 & 0 \\
 0 & -\frac{1}{2} & 0 & 0 \\
 0 & 0 & \frac{1}{2} & 0 \\
 0 & 0 & 0 & -\frac{1}{2} \\
\end{array}
\right)
&
\Sigma _{3,5}=\left(
\begin{array}{cccc}
 0 & \frac{1}{2} & 0 & 0 \\
 \frac{1}{2} & 0 & 0 & 0 \\
 0 & 0 & 0 & \frac{1}{2} \\
 0 & 0 & \frac{1}{2} & 0 \\
\end{array}
\right)
&
\Sigma _{3,6}=\left(
\begin{array}{cccc}
 0 & \frac{i}{2} & 0 & 0 \\
 -\frac{i}{2} & 0 & 0 & 0 \\
 0 & 0 & 0 & -\frac{i}{2} \\
 0 & 0 & \frac{i}{2} & 0 \\
\end{array}
\right)
\\ 
\Sigma _{4,5}=\left(
\begin{array}{cccc}
 0 & \frac{i}{2} & 0 & 0 \\
 -\frac{i}{2} & 0 & 0 & 0 \\
 0 & 0 & 0 & \frac{i}{2} \\
 0 & 0 & -\frac{i}{2} & 0 \\
\end{array}
\right)
&
\Sigma _{4,6}=\left(
\begin{array}{cccc}
 0 & -\frac{1}{2} & 0 & 0 \\
 -\frac{1}{2} & 0 & 0 & 0 \\
 0 & 0 & 0 & \frac{1}{2} \\
 0 & 0 & \frac{1}{2} & 0 \\
\end{array}
\right)
&
\Sigma _{5,6}=\left(
\begin{array}{cccc}
 \frac{1}{2} & 0 & 0 & 0 \\
 0 & -\frac{1}{2} & 0 & 0 \\
 0 & 0 & -\frac{1}{2} & 0 \\
 0 & 0 & 0 & \frac{1}{2} \\
\end{array}
\right)
\end{array}
\eeq

\section{Derivation of the form of the scalar product}\label{app:scalar}
The scalar product is a bilinear functional. Without loss of generality, we can represent it in the form
\beq
\langle \langle f , g \rangle \rangle \equiv \int \prod_{i=1}^J d^dx_i\, f(x_1, \dots, x_J ) \hat{K}\circ g(x_1, \dots, x_J ),
\eeq
where $\hat{K}$ is a linear operator, which can be written in terms of an integration kernel (which might be a distribution):
\beq
( \hat{K} \circ f )(x_1, \dots, x_J)\equiv \int \prod_{i=1}^J d^d y_i \, \mathcal{K}(x_1, \dots, x_J |  y_1\dots, y_J ) \,f(y_1, \dots, y_J ).
\eeq
Let us now impose the covariance of the scalar product under a generic conformal transformation $C^{-1}$, acting on the site $n$ as 
\beq C^{-1}_{(n)} \circ f (x_1, \dots, x_J) \equiv f(x_1,\dots,x_{n-1} , C(x_n), x_{n+1},\dots, x_J) \ .
\eeq 
The covariance condition reads:
\beq\label{eq:covarianceapp}
(\hat{K} \circ C^{-1}_{(n)} \circ f )(x_1,\dots, x_J) =   \left| \frac{\partial C(x) }{\partial x} \right|^{\frac{D-h_n}{d} }_{x=x_n} \,  (\hat{K} \circ  f )(x_1,\dots,x_{n-1} , C(x_n), x_{n+1},\dots, x_J) ,
\eeq
where $D$ is the spacetime dimension, i.e. $D=4$ in our case. 
Since this must be satisfied for all sites $n$, we can look for a solution in terms of a factorised kernel:
\beq
\mathcal{K}(x_1, \dots, x_J | y_1, \dots , y_J ) \equiv \prod_{i=1}^J \mathcal{K}_{h_i}(x_i | y_i) .
\eeq
Imposing the conditions (\ref{eq:covarianceapp})  for all sites $n$ is equivalent to:
\beq
\mathcal{K}_h(x|y) = \left| \frac{\partial C(x) }{\partial x} \right|^{\frac{D-h}{D} }\,\mathcal{K}_h(C(x)|C(y) )   \left| \frac{\partial C(y) }{\partial y} \right|^{\frac{D-h}{d} }.
\eeq
This is the same transformation rule as for a two-point function with scaling weight $(D-h)$. The solution is therefore fixed to:
\beq
\mathcal{K}_h(x|y) \propto  \frac{1}{|x-y|^{{ 2 D - 2 h  }} }.
\eeq
Multiplying by an appropriate  normalisation factor, this integral kernel can be related to the fractional operator $\Box^{\frac{D}{2} - h }$ defined in (\ref{eq:boxfrac}).

\section{Dual Baxter equation}\label{app:dualBproof}

Here we describe in general the form of the dual Baxter equation satisfied by $3\times 3$ determinants of Q-functions.

Consider a finite difference equation of the form
\beq\la{BQ}
A_{-2}(u)Q_a(u-2i)
+
A_{-1}(u)Q_a(u-i)
+
A_{0}(u)Q_a(u)
+
A_{+1}(u)Q_a(u+i)
+
A_{+2}(u)Q_a(u+2i)=0
\eeq
where $a=1,\dots,4$ labels $4$ independent solutions of this equation. Define dual Q-functions by
\beq\la{Pa}
P^a(u)=F(u)\epsilon^{a a_1 a_2 a_3}Q_{a_1}(u+i)Q_{a_2}(u)Q_{a_3}(u-i)\;.
\eeq
Then the $4$ functions $P^a(u)$ 
satisfy the dual finite difference equation
\beq\la{baxP}
B_{-2}(u)P^a(u-2i)
+
B_{-1}(u)P^a(u-i)
+
B_{0}(u)P^a(u)
+
B_{+1}(u)P^a(u+i)
+
B_{+2}(u)P^a(u+2i)=0
\eeq
where
\beqa
B_{-1}(u)&=&\frac{A_1(u-i)  F(u-2 i)}{A_{-2}(u-i) F(u-i)}B_{-2}(u),\\
B_{0}(u)&=&\frac{A_0(u) A_2(u-i)  F(u-2 i)}{A_{-2}(u) A_{-2}(u-i) F(u)}B_{-2}(u),\\
B_{+1}(u)&=&\frac{A_{-1}(u+i) A_2(u) A_2(u-i)  F(u-2 i)}{A_{-2}(u) A_{-2}(u-i)
   A_{-2}(u+i) F(u+i)}B_{-2}(u),\\
B_{+1}(u)&=&\frac{A_{-1}(u+i) A_2(u) A_2(u-i)  F(u-2 i)}{A_{-2}(u) A_{-2}(u-i)
   A_{-2}(u+i) F(u+i)}B_{-2}(u),\\
B_{+2}(u)&=&\frac{A_2(u) A_2(u-i) A_2(u+i) F(u-2 i)}{A_{-2}(u) A_{-2}(u-i) A_{-2}(u+i)
   F(u+2 i)} B_{-2}(u)\;.
\eeqa
To show this we plug the definition \eq{Pa}
into \eq{baxP} and then use \eq{BQ}
to exclude $Q_a(u+3i),\;Q_a(u+2i)$ and $Q_a(u+i)$ 
from the resulting equation. 

To make the result simpler, we can make use of the arbitrary overall multiplier $B_{-2}(u)$ which we set to be $B_{-2}(u)=A_{+2}(u-2i)$, and also adjust the factor $F(u)$ in the definition of $P^a(u)$ so that it satisfies
\beqa
\frac{F(u+\tfrac{i}{2})}{F(u-\tfrac{i}{2})}=
\frac{A_{+2}(u-\tfrac{i}{2})}{A_{-2}(u+\tfrac{i}{2})} \ .
\eeqa
Then we see that the dual Baxter equation \eq{baxP} takes a very simple form,
\beq
A_{+2}^{[-4]}P^{a[-4]}+A_{+1}^{[-2]}P^{a[-2]}+A_0P^a+A_{-1}^{[+2]}P^{a[+2]}+A_{-2}^{[+4]}P^{a[+4]}=0\;.
\eeq

\section{Additional properties of the Q-functions on-shell}\label{app:Qplus}
In this appendix we review a reformulation of the quantisation introduced in \cite{GrabnerToAppear}, and already used in \cite{Cavaglia:2020hdb}. We will consider vanishing rapidities $\theta_i \rightarrow 0$ in this section. The most important aspect of this alternative method, is that it provides a way to compute the eigenvalue of the graph-building operator, which also applies in the  more complicated case of quantum numbers $|M| = J$.

Consider
\beq
T_n^{[+n]}\equiv 2 q_a^\downarrow(u) q^\uparrow_b(u+i n)\Gamma^{ab} ,
\eeq
where we assume that the Q-functions are on-shell, namely satisfy the quantisation condition, and $\Gamma^{ab}$ is the gluing matrix. 
We want to show that, as a consequence of the quantisation conditions, this function has   an analyticity strip $[-in/2,in/2]$. To see this, we rewrite
\beq
T_n^{[+n]} = 
2\; q_a^\downarrow(u) \;\Gamma^{ab}\Omega_{b}^c\; q^\downarrow_c(u+i n) = 2\; q_a^\downarrow(u) \;\omega^{ac}\; q^\downarrow_c(u+i n),
\eeq
and using the antisymmetry of $\omega^{ab} = (\Gamma \cdot \Omega)^{ab}$, we find
\beq
-2\; q_a^\downarrow(u) \;\Omega_{b}^a\Gamma^{bc}\; q^\downarrow_c(u+i n) =
-2\; q_a^\uparrow(u) \;\Gamma^{ab}\; q^\downarrow_b(u+i n) ,
\eeq
where the r.h.s. is explicitly analytic at $u=0,\dots,-in$. Furthermore one has
\beq
T_n=\Gamma^{ab} \(
{q_a^\downarrow}^{[-n]}(u)
{q_b^\uparrow}^{[+n]}(u)-
{q_a^\uparrow}^{[-n]}(u)
{q_b^\downarrow}^{[+n]}(u)
\)=-T_{-n} .
\eeq
Among these quantities, a particularly important one is $Q_+(u) \equiv T_1^{[-1]}(u)$, which has a pole of order $J$ at $u \sim  \frac{3 i}{2}$. From the coefficient of this pole, one can extract the eigenvalue of the Hamiltonian (identified with the coupling constant)~\cite{GrabnerToAppear}:
\beq
\xi^{2 J} = \lim_{\epsilon \rightarrow 0} \epsilon^{J} \frac{Q_+(\frac{3 i}{2} - i \epsilon)}{Q_+(\frac{i}{2})} .
\eeq
This equation applies to all states of the theory, including those with $|M| = J$ for which the simple
 relation $I_{(0,1)} = \xi^{2 J}$ is not valid. 
 
\section{Covariance of twisted correlators}\label{app:maps}
In this appendix we review the covariance transformation rules for colour-twisted correlation functions in a large-$N$ CFT, and the insertion points $x_i$ are fixed points of these maps, see \cite{Cavaglia:2020hdb}. 
Consider a correlator of colour-twisted operators, $\langle \prod_{i=1}^n \mathcal{O}_{G_i}(x_i) \rangle $, where $G_1, \dots, G_n$ are conformal transformations used in the definition of the twisted operators. 
Under any conformal transformation $C$, the correlator is covariant in the sense that
\beq\label{eq:covariance}
\langle \prod_{i=1}^n \mathcal{O}_{G_i}(x_i) \rangle = \langle \prod_{i=1}^n\mathcal{O}_{\widetilde G_i}(\tilde x_i) \rangle \, \prod_{i=1}^n \left| \frac{\partial \tilde x_i }{\partial x_i }\right|^{\frac{\Delta_i}{4} } ,  
\eeq
where $\tilde x_i \equiv C(x_i)$, and the twist maps are also conjugated as $\tilde G_i \equiv C G_i C^{-1}$. We considered scalar operators for simplicity. 
\paragraph{Two-point function at opposite twists. } 
A particularly important case, already studied in  \cite{Cavaglia:2020hdb}, is the 2-point function of two operators with opposite twists, $G_2 = G_1^{-1}$. In this case, the 2-point function has the same kinematical dependence on the points as in the standard CFT without twists. Let us recall how this works. We consider twist maps with fixed points $x_0$, $x_{\bar{0}}$.
Such maps always admit the following decomposition (we use the notations explained in the main text):
\beq\label{eq:Gsapp}
G_1 \equiv G= K \, r \, \Lambda \, r^{-1} \, K^{-1} \;\;\;\; G_2 \equiv G^{-1} = K \, r \, \Lambda^{-1} \, r^{-1} \, K^{-1} ,
\eeq
where $\Lambda$ is the diagonal part, $r$ is a 4D rotation, and $K$ is the special conformal transformation defined in (\ref{eq:Kdef}), satisfying $K ( 0 )= x_0$, $K (\infty) = x_{\bar{0}}$. 

It is simple to show that the  2-point function $\langle \mathcal{O}_{G}(x_0) \mathcal{O}_{G^{-1} }'(x_{\bar{0}}) \rangle$ does not depend on the form of the rotation $r$. In fact, consider the covariance rule   (\ref{eq:covariance}),  for the map $C = K \circ r' \circ K^{-1} $, where $r'$ is any  different $SO(4)$ rotation. Since such map $C$ has unit Jacobian at the fixed points, we conclude that the 2-point function is the same with $r \rightarrow r' \circ r$. 

Therefore, the only relevant parameters are the fixed points and the diagonal part of $G$: we can write $\langle \mathcal{O}_{G}(x_0) \mathcal{O}_{G^{-1} }'(x_{\bar{0}}) \rangle \equiv \mathcal{F}( x_0, x_{\bar{0}} ; \vec{\lambda} )$.  

Now, (\ref{eq:covariance}) just becomes the familiar
\beq
\mathcal{F}( x_0, x_{\bar{0}} ; \vec{\lambda} ) = \mathcal{F}( \tilde{y}_1, \tilde{y}_2 ; \vec{\lambda} )\left| \frac{\partial \tilde x_0 }{\partial x_0 }\right|^{\frac{\Delta_{\mathcal{O} } }{4} } \, \left| \frac{\partial \tilde x_{\bar{0}} }{\partial x_{\bar{0}} }\right|^{\frac{\Delta_{\mathcal{O}' } }{4} } ,  
\eeq
so we can proceed as in the standard CFT case.  Considering the effect of dilatations, rotations around an arbitrary point, and translations, we conclude that 
\beq
\mathcal{F}( x_0, x_{\bar{0}} ; \vec{\lambda}) = \frac{\mathcal{N}_{\mathcal{O} , \mathcal{O}' } }{|x_0 - x_{\bar{0}} |^{ \Delta_{\mathcal{O}} + \Delta_{\mathcal{O}' } } }.
\eeq
Finally, consider the covariant transformation property for the map $C \equiv K \circ e^{\rho \mathbb{D} } \circ K^{-1}$, which leaves $G$ invariant. Then, (\ref{eq:covariance}) implies that the 2-point function has to vanish unless $\Delta_{\mathcal{O}} = \Delta_{\mathcal{O}'}$.

\paragraph{Transformation of the CFT wave function. }
For the CFT wave function, which is defined in terms of the twisted operator $\mathcal{O}_G$ and a twisted trace (see (\ref{eq:wfdef})), the covariance property (\ref{eq:covarianceapp}) gives
\beq\label{eq:covariancewf}
\varphi_{\mathcal{O}_G}( x_1, \dots, x_J ) = \varphi_{\mathcal{O}_{\tilde{G}} }(  \tilde x_1, \dots, \tilde x_J ) \times \left| \frac{\partial \tilde x_0 }{\partial x_0 }\right|^{\frac{\Delta}{4} } \times \prod_{i=1}^J \left| \frac{\partial \tilde x_i }{\partial x_i }\right|^{\frac{I_i + 1}{4} } ,
\eeq
where $\Delta$ is the scaling dimension of the operator $\mathcal{O}$. Notice that we took into account that, as a correlator, the wave function depends on $J+1$ insertion points, one of them being the fixed point of the twist $x_0$ where the operator $\mathcal{O}$  sits.

\section{Properties of the transfer matrices under transposition}\label{app:Ts}
In this appendix we derive some   properties of the adjoints (or transposes) of the transfer matrices with respect to the conformally invariant scalar product, which are denoted as $(\mathbb{T}^{\mathbf{r} })^T$:
\beq\label{eq:deftranspose2}
\langle \langle f , \, \mathbb{T}^{\mathbf{r} }\circ g \rangle\rangle = \langle \langle ( \mathbb{T}^{\mathbf{r} } )^T\circ f , \, g \rangle\rangle.
\eeq
We will derive a convenient explicit equation for the transpose, as a combination of reversal of the order of sites in the chain, and conjugation with a special conformal map. This representation  gives a simple map between right and left eigenvectors in the case of homogeneous chain, and allows us to prove directly that the transfer matrices and their transposes have the same spectrum. 
At the end of this appendix, we also use this relation to show that, in the fishnet theory, the left eigenvectors can be interpreted as conjugate CFT wave functions.  

\subsection{An explicit formula for the transpose}
\paragraph{Statement of the result. }
Recall the definition of the transfer matrices\footnote{For clarity, in this appendix we occasionally denote the Lax matrices and conformal generators as $\hat{\mathbb{L}}^{{\bf r}}_{x_k, h_k}$, $\hat{q}^{MN}_{x, h}$, to mark the space-time variable on which they act, and the weight of the representation. }
\beq
\hat{\mathbb{T}}^{{\bf r}}\equiv \text{Tr}_{{\bf r}}\left[  \hat{\mathbb{L}}^{{\bf r}}_{x_J, h_J}\left(u -\vartheta_J \right)\dots  \hat{\mathbb{L}}^{{\bf r}}_{x_1, h_1}\left(u -\vartheta_1 \right)\cdot G  \right].
\eeq
In this section we will obtain the formula: 
\beq\label{eq:resultTadjapp}
(\mathbb{T}^{{\bf r} })^{T} 
= F \circ \hat{\mathbb{T}}^{{\bf r}}_{\texttt{rev}} \circ 
F ,
\eeq
where $\hat{\mathbb{T}}^{{\bf r}}_{\texttt{rev}}$ denotes the transfer matrix with reversed order of sites inside the trace:
\beq
\hat{\mathbb{T}}^{{\bf r}}_{\texttt{rev}}\equiv \text{Tr}_{{\bf r}}\left[  \hat{\mathbb{L}}^{{\bf r}}_{x_1, h_1}\left(u -\vartheta_1 \right)\dots  \hat{\mathbb{L}}^{{\bf r}}_{x_J, h_J}\left(u -\vartheta_J \right)\cdot G \right] ,
\eeq
and $F = K \circ \tilde{\mathcal{I}} \circ K^{-1}$ is the conformal map defined in (\ref{eq:defF2}), with  $\tilde{\mathcal{I}}$  the holomorphic inversion (\ref{eq:holoI2}).
\paragraph{Properties of $F$.}
Before presenting the proof, let us list some important properties of the conformal map $F$, which can be easily verified. \begin{enumerate}[-]
\item
It is equal to its inverse, $F  = F^{-1}$ ; 
\item
In the frame reached by this transformation, the twist is inverted:
\beq
F \circ G \circ F^{-1} = G^{-1},
\eeq
where we used $\tilde{\mathcal{I}} \circ \Lambda \circ \tilde{\mathcal{I}} = \Lambda^{-1}$ for the diagonal part of the twist;
\item
The map changes the sign of the Cartan generators:
\beq\label{eq:CartanF}
F \circ \mathbb{Q}_a \circ F^{-1} = -\mathbb{Q}_a , \;\;\;\;\; a = 0,1,2,
\eeq
which follows from $\tilde{\mathcal{I}} \circ \left\{ \mathbb{D}, \mathbb{S}_{1,2}, \mathbb{S}_{3,4} \right\} \circ \tilde{\mathcal{I}} = - \left\{ \mathbb{D}, \mathbb{S}_{1,2}, \mathbb{S}_{3,4} \right\} $.
\item
The map swaps the two fixed points of the twist map $x_0 = F(x_{\bar{0}})$, $x_{\bar{0}} = F(x_0)$. 
\end{enumerate}

\subsubsection{Derivation}
\paragraph{Transposition of Lax matrices.}
From the very definition of the  conformally invariant scalar product, the adjoints of the conformal generators satisfy:
\beq\label{eq:moving}
(\hat{q}^{MN}  )^T = - {\hat{q}^{MN} } =\, {\hat{q}^{NM} },
\eeq
which is valid locally at every site. 
In virtue of this relation, the Lax operators satisfy
\beqa
&& \left( {\left[\mathbb{L}^{\mathbf{6} }(u) \right]^{MN}} \right)^T  =   {\left[\mathbb{L}^{\mathbf{6} }(u) \right]^{NM} } \label{eq:conjuL6}\\
&& \left({\left[\mathbb{L}^{\mathbf{4} }(u) \right]_{a}^{\; b}} \right)^T =  -{\left[\mathbb{L}^{\mathbf{4} }(-u) \right]_{a}^{\; b}} , \label{eq:conjuL4}
\eeqa
where we kept explicit the indices in auxiliary space, and we remind that $O^T$ denotes the transpose of an operator  w.r.t. the scalar product, i.e. the adjoint. 
\paragraph{Intertwining with $F$.}
There are two key identities that we will use in the following:
\beqa
 \tilde{\mathcal{I}}  \circ \hat{q}^{MN} \circ   \tilde{\mathcal{I}}  &=& (\tilde{\mathcal{I}^{\bf 6}})_{M'}^{\;M} \, \hat{q}^{M'N'}\; (\tilde{\mathcal{I}^{\bf 6}})_{N'}^{\;N},\label{eq:ide1appspec}\\
( \tilde{\mathcal{I}^{\bf 6}} )_M^{M'} \cdot (\Sigma_{M'N'})_a^{\,b} \cdot ( \tilde{\mathcal{I}^{\bf 6}} )_{N}^{\, N'} &=& -   (\Sigma_{MN})_{b}^{\,a} ,\label{eq:conjP}
\eeqa
where we kept explicit the  representation indices $ 1\leq a,b \leq 4$, and the holomorphic inversion acts in 6D as $\tilde{\mathcal{I}^{\bf 6}}={\rm diag}\{1, -1, 1, -1, -1, 1\}$. Combined with (\ref{eq:conjuL6}),(\ref{eq:conjuL4}) , these identities give
\beqa
\left({\left[\mathbb{L}^{\mathbf{6} }(u) \right]^{MN}} \right)^T&=&  \tilde{\mathcal{I}}  \circ \left( ( \tilde{\mathcal{I}^{\bf 6}} )_{N'}^{\,N}{\left[\mathbb{L}^{\mathbf{6} }(u) \right]^{N' M'}} \, ( \tilde{\mathcal{I}^{\bf 6}} )_{M'}^{\,M} \right) \circ  \tilde{\mathcal{I}}  ,\label{eq:conjuI6} \\
 \left({\left[\mathbb{L}^{\mathbf{4} }(u) \right]_{a}^{\; b}} \right)^T &=&  \tilde{\mathcal{I}}  \circ \left( {\left[\mathbb{L}^{\mathbf{4} }(u) \right]_b^{\, a}} \, \right) \circ  \tilde{\mathcal{I}}  .\label{eq:conjuI4}
\eeqa
Notice that the terms in round brackets on the right hand side of \eq{eq:conjuI6},\eq{eq:conjuI4} involve a transposition of the auxiliary space indices. 

\paragraph{Transfer matrices.}

We will work in the frame where the twist map is the diagonal transformation $G \equiv \Lambda$. 

Consider the case of the vector representation, defined as
\beq
\hat{\mathbb{T}}^{\mathbf{6}} = \text{Tr}_{\mathbf{6}}\left[  \left({\hat{\mathbb{L}}^{\mathbf{6}} }_{X_J}(u - \vartheta_{J}  ) \right) \dots \left({\hat{\mathbb{L}}^{\mathbf{6}} }_{X_1}(u - \vartheta_{1} ) \right)\,\Lambda^{\mathbf{6}}  \right] ,
\eeq
where the Lax matrices without explicit mention of the indices are assumed to have the index structure  $(\hat{\mathbb{L}}^{\mathbf{6}} )^M_{\; N} \equiv  (\hat{\mathbb{L}}^{\mathbf{6}})^{ M M'} \, \eta_{M' N} $. Using \eq{eq:conjuL6}, for its transpose  we find
\beq
(\hat{\mathbb{T}}^{\mathbf{6} })^{T} = \text{Tr}_{\mathbf{6}}\left[ \hat{\mathbb{L}}^{\mathbf{6}}_{X_1, h_1}\left(u -\vartheta_1 \right)\dots \hat{\mathbb{L}}^{\mathbf{6}}_{X_J, h_J}\left(u -\vartheta_J \right) \left( \eta \cdot \Lambda^{\mathbf{6}} \cdot \eta^{-1} \right)^{T_{\text{aux}}} \right],\label{eq:difftwist}
\eeq
which is a similar to the definition of the transfer matrix, but with the order of sites reversed, and a different twist. Above, by $A^{T_{\text{aux}} }$ we denote transposition in the auxiliary space indices.  
The twist matrix appearing in (\ref{eq:difftwist}) can be rewritten as
\beq\label{eq:ideapp}
\left( \eta \cdot \Lambda^{\mathbf{6}} \cdot \eta^{-1} \right)^{T_{\text{aux}}}  = (\tilde{\mathcal{I}^{\bf 6}} )^{ T_{\text{aux}} } \cdot \Lambda^{\mathbf{6}} \cdot \tilde{\mathcal{I}^{\bf 6}}.
\eeq
Using the fact that $(\tilde{\mathcal{I}^{\bf 6}} \cdot\tilde{\mathcal{I}^{\bf 6}})_M^{\;N}= \delta_M^{\;N}$ and $\tilde{\mathcal{I}^{\bf 6}} =(\tilde{\mathcal{I}^{\bf 6}} )^{ T_{\text{aux}} }$, we can use (\ref{eq:ideapp}) to write
\beqa
(\hat{\mathbb{T}}^{\mathbf{6} })^{ T} &=& \text{Tr}_{\mathbf{6}}\left[ \left(\tilde{\mathcal{I}^{\bf 6}} \cdot \hat{\mathbb{L}}^{\mathbf{6}}_{X_1, h_1}\left(u -\vartheta_1 \right) \cdot \tilde{\mathcal{I}^{\bf 6}} \right)\dots \left( \tilde{\mathcal{I}^{\bf 6}}\cdot  \hat{\mathbb{L}}^{\mathbf{6}}_{X_J, h_J}\left(u -\vartheta_J \right)\cdot \tilde{\mathcal{I}^{\bf 6}} \right)\cdot \Lambda^{\mathbf{6}}  \right],\label{eq:result6}\\
&=&  \tilde{\mathcal{I}}  \circ \text{Tr}_{\mathbf{6}}\left[  \hat{\mathbb{L}}^{\mathbf{6}}_{X_1, h_1}\left(u -\vartheta_1 \right)\dots  \hat{\mathbb{L}}^{\mathbf{6}}_{X_J, h_J}\left(u -\vartheta_J \right)\cdot \Lambda^{\mathbf{6}}  \right] \circ \tilde{\mathcal{I}}  ,\nn
\eeqa
where we used (\ref{eq:ide1appspec}) in the last line. This matches the announced result (\ref{eq:resultTadjapp}). 

Now we consider $\hat{\mathbb{T}}_{\mathbf{4}}$.  Identity (\ref{eq:conjuI4}) tells us that, upon integration by parts, $\mathbb{L}^{\mathbf{4} }(u) $ gets transposed in auxiliary space, and conjugated with the holomorphic inversion in quantum space. 
The twist matrix $\Lambda_{\mathbf{4}}$ is not affected by such transposition, since it is diagonal in this representation. Therefore, repeating the steps described for the previous case, we find immediately
\beq
(\hat{\mathbb{T}}^{\mathbf{4} })^{ T} 
=  \tilde{\mathcal{I}}  \circ \text{Tr}_{\mathbf{4}}\left[  \hat{\mathbb{L}}^{\mathbf{4}}_{X_1, h_1}\left(u -\vartheta_1 \right)\dots  \hat{\mathbb{L}}^{\mathbf{4}}_{X_J, h_J}\left(u -\vartheta_J \right)\cdot \Lambda_{\mathbf{4}}  \right] \circ 
\tilde{\mathcal{I}}  .\label{eq:result4}
\eeq
The derivation for the case of $\bar{\mathbf{4}}$ is a simple generalisation. 

We have presented the above results (\ref{eq:result6}),(\ref{eq:result4}) for the cases where the twist matrix is $G \equiv \Lambda$. The general case is related to this by a conformal transformation denoted as $K$ in the main text. Using the covariance of twisted transfer matrices (\ref{eq:brokenconf}) under this transformation, we see that this has simply the effect of replacing the twist $\Lambda \rightarrow G$, as well as changing the holomorphic inversion to the map $F$. This leads to the announced result (\ref{eq:resultTadjapp}).

\subsection{Spectral properties}\label{app:TsSpectral}
Now we establish that the spectrum of $\hat{\mathbb{T}}^{{\bf r}}$ on the subspace of wave functions with fixed conformal charges $Q_a^R = \left\{ i \Delta, S_1, S_2 \right\}$, is the same as the spectrum of $(\mathbb{T}^{{\bf r}})^{T}$, on the subspace of wave functions with opposite conformal charges $Q_a^L = -\left\{ i \Delta, S_1, S_2 \right\}$. 

We have seen that the adjoint of the transfer matrix is given by the chain of Lax matrices applied on the sites in reverse order, and conjugated by  the map $F$ defined in (\ref{eq:defF2}):
\beq\label{eq:Tadjide}
(\hat{\mathbb{T}}^{{\bf r}})^{T} = F \circ \hat{\mathbb{T}}^{{\bf r}}_{\texttt{rev}} \circ F,
\eeq
with
\beq
\mathbb{T}^{{\bf r}}_{\texttt{rev}}\equiv \text{Tr}_{{\bf r}}\left[  \hat{\mathbb{L}}^{{\bf r}}_{x_1, h_1}(u - \vartheta_{1}  ) \cdot \dots \cdot \hat{\mathbb{L}}^{{\bf r}}_{x_J, h_J}(u - \vartheta_{J}  ) \cdot G \right].
\eeq
First, we notice that the Baxter equation is invariant under synchronised permutation of the inhomogeneities and the order of magnons. Therefore, the spectrum is invariant under this operation. This means that, for every right eigenstate satisfying $\hat{\mathbb{T}}^{{\bf r}} \circ \Psi^R_A = \mathbb{T}^{{\bf r} R}_A \, \Psi^R_{A} $, there is an associated wave function $\Psi_{A, \texttt{rev}}$ satisfying $\hat{\mathbb{T}}^{{\bf r}}_{\texttt{rev}} \circ \Psi_{A, \texttt{rev}} = \mathbb{T}^{{\bf r} R}_A \, \Psi_{A, \texttt{rev}} $, with the same eigenvalue. 
For fixed conformal charges, the map between states $\Psi^R_{A}$ and $\Psi_{A, \texttt{rev}}$ is one-to-one since one expects the spectrum is non-degenerate. This wave function has the same Cartan charges $Q_a^R $ as $\Psi^R_{A}$, since they can be read from the integrals of motion and are unaffected by permutations of the sites. 

Now, we can construct $\Psi^L_A \equiv F \circ \Psi_{A, \texttt{rev}}$, which, by the identity (\ref{eq:Tadjide}), will be an eigenstate of $(\hat{\mathbb{T}}^{{\bf r}})^{T}$ with the same eigenvalue. 
Moreover, if $\Psi^R_{A} $ (and therefore $\Psi_{A, \texttt{rev} }$) has Cartan charges $Q_a^R $, due to property (\ref{eq:CartanF}) we find that the left eigenvector $\Psi^L_A $ has exactly opposite Cartan charges.

\subsection{Left eigenvectors in the fishnet case}\label{app:subsecRL}
In this section we consider the fishnet theory, specifying to zero inhomogeneities. 

As we discussed in section \ref{sec:wavefs}, the CFT wave functions, which are right eigenvectors of the integrals of motion, are defined as certain correlators:
\beq\label{eq:wfdefapp}
\varphi_{\mathcal{O}}(x_1, x_2, \dots, x_J ) 
=\langle \mathcal{O}_G(x_0) \; \text{Tr}\left( \chi_{{\bf I}_1}(x_1) \chi_{{\bf I}_2}(x_2)  \dots \chi_{{\bf I}_J}(x_J) \mathcal{T}_{G^{-1} } \right) \rangle\; ,\;\;\;\;\;
\hat{\mathbb{T}}^{\mathbf{r}} \circ \varphi_{\mathcal{O}} = \mathbb{T}^{\mathbf{r}} \, \varphi_{\mathcal{O}}.
\eeq
The associated left eigenvectors,  $\varphi_{\mathcal{O}}^{\texttt{Left}}$ satisfying  $(\hat{\mathbb{T}}^{\mathbf{r}} )^T \circ \varphi_{\mathcal{O}}^{\texttt{Left}} = \mathbb{T}^{\mathbf{r}} \, \varphi_{\mathcal{O}}^{\texttt{Left}}$, also have a very simple interpretation. We will use the representation found in the previous subsection, 
\beq\label{eq:applyF}
\varphi_{\mathcal{O}}^{\texttt{Left}} \equiv F \circ (\varphi_{\mathcal{O}} )_{\texttt{rev}} ,
\eeq
where $ (\varphi_{\mathcal{O}} )_{\texttt{rev}} $ is the eigenstate of the transfer matrices with reversed order of sites  (\ref{eq:Tadjide}), with the same eigenvalue. In the fishnet theory, such wave function can also be obtained as a correlator:
\beq
\varphi_{\mathcal{O}}^{ \texttt{rev}}(x_1, x_2, \dots, x_J ) 
=\langle \mathcal{O}_G(x_0) \; \text{Tr}\left( \chi_{{\bf I}_J}(x_J) \chi_{{\bf I}_{J-1}}(x_{J-1})  \dots \chi_{{\bf I}_1}(x_1) \mathcal{T}_{G^{-1} } \right) \rangle .
\eeq
Using (\ref{eq:applyF}), we then find
\beq
\varphi_{\mathcal{O}}^{ \texttt{Left}}(x_1, x_2, \dots, x_J ) 
= F \circ \varphi_{\mathcal{O}}^{ \texttt{rev}}(x_1, x_2, \dots, x_J )  = \varphi_{\mathcal{O}}^{ \texttt{rev}}( \tilde{x}_1, \tilde{x}_2, \dots, \tilde{x}_J )   \times \prod_{i=1}^J \left| \frac{\partial \tilde x_i }{\partial x_i }\right|^{\frac{\mathbf{I}_i + 1}{4} } ,
\eeq
with $\tilde{x}_i \equiv F\circ x_i$, and thanks to (\ref{eq:covarianceapp}), (\ref{eq:covariancewf}),  this can be interpreted as the correlator
\beq\label{eq:phiLeft}
\varphi_{\mathcal{O}}^{ \texttt{Left}}(x_1, x_2, \dots, x_J )  =\langle \mathcal{O}_{G^{-1}}(x_{\bar{0} }) \; \text{Tr}\left( \chi_{{\bf I}_J}(x_J) \chi_{{\bf I}_{J-1}}(x_{J-1})  \dots \chi_{{\bf I}_1}(x_1) \mathcal{T}_{G} \right) \rangle.
\eeq
Notice that going from right to left eigenvectors has the effect of inverting the twist of the operator, and changes the insertion point from $x_0$ to the second fixed point  $x_{\bar{0}} = F(x_0)$. 

Using the discrete symmetry of the fishnet model under conjugation of the elementary fields,  $\phi_i \leftrightarrow \phi_i^{\dagger}$, the left eigenfunction can also be written as
\beq\label{eq:conjleft}
\varphi_{\mathcal{O}}^{ \texttt{Left}}(x_1, x_2, \dots, x_J )  =\langle \mathcal{O}_{G^{-1}}^{\dagger}(x_{\bar{0} }) \; \text{Tr}\left( \chi_{{\bf I}_J}^{\dagger}(x_J) \chi_{{\bf I}_{J-1}}^{\dagger}(x_{J-1})  \dots \chi_{{\bf I}_1}^{\dagger}(x_1) \mathcal{T}_{G} \right) \rangle.
\eeq
This relation is useful to derive the interpretation of the norm in CFT, presented in section \ref{sec:normmeaning}. In particular, in Figure \ref{fig:scalar}, the sum of Feynman diagrams on the left of the green cut can be identified with the correlator on the r.h.s. of (\ref{eq:conjleft}); while the diagrams on the right side of the cut give the CFT wave function (\ref{eq:wfdefapp}).

\section{Details on the $\d_\xi \Delta$ calculation and relation between scalar product and 2-pt function}\label{app:proofdDelta}

In this appendix we consider the eigenfunctions for zero inhomogeneities, corresponding to CFT wave functions of the fishnet theory. We show how they are normalised according to the scalar product. This argument was already implicit in \cite{Gromov:2019jfh} and we thank A. Sever for discussing it with us. We  report it here for completeness. 

For simplicity, we will consider the  state for lowest scaling dimension at any $J$, with $M=0$. 

We will start from the perturbative theory, and then resum the diagrams. 
 We take the propagator to be 
\beq
    \langle\phi(x)\phi(y)\rangle=\frac{1}{4\pi^2}\frac{1}{(x-y)^2} ,
\eeq
which in 4D inverts the kinematical term as $
    -\Box \frac{1}{4\pi^2x^2}=\delta^4(x) $. 
    
    We want to study operators of the schematic form $\mathcal{O}(x_0)\propto\text{Tr}( \phi_1^J(x_0) \mathcal{T}_{G} )$, $\tilde{\mathcal{O}}(x_{\bar{0}})\propto\text{Tr}( \phi_1^{\dagger J }(x_{\bar{0}}) \mathcal{T}_{G^{-1}} )$ at weak coupling. We define them using an explicit point-splitting regularisation scheme. 
    To regularise the two-point function $\langle \mathcal{O}(x_0) \tilde{\mathcal{O}}(x_{\bar{0}}) \rangle$, we consider 
\beq
W(x_1, \dots, x_J | y_1, \dots, y_J) = \sum_{n=0}^{\infty} \xi^{2 J n} W^{(n)}(\mathbf{x}, \mathbf{y} ) ,
\eeq
where $W^{(n)}$ represents the fishnet diagram with $n$ wheels\footnote{We extract a prefactor with the coupling as ${\cal B}$ already contains it} \footnote{The $J$ vertices in each wheel  contribute a $(16\pi^2)^J$ factor and the propagators linking them give $1/(4\pi^2)^J$, leaving $4^J$. It is partially cancelled by the corresponding ladder (spokes) propagators that give $1/(4\pi^2)^J$ which leaves only the $1/(\pi^2)^J$ factor that we have included in ${\cal B}$.}:
\beq
W^{(n)} =\frac{1}{\xi^{2Jn}}\underbrace{\hat{ \mathcal{B} }\circ \dots \circ \hat{ \mathcal{B} } }_{\text{ n times } } \circ \frac{1}{\prod_i4\pi^2(x_i-y_i)^2 } . \eeq
In the  coincident points limit,  $(x_i - x_0)^2\sim (y_i-y_{\bar{0}})^2 \sim \epsilon^2 $, this quantity will be denoted as $W_{\epsilon, \epsilon}(x_0|x_{\bar{0}})$. It exhibits a power-like divergence, which we can remove by multiplicative renormalisation of the operators. Choosing a certain normalisation, this gives\footnote{
In the untwisted theory, it would be more natural to add an additional factor $J$ on the l.h.s. of \eq{OOW}, due to the  ambiguity in contracting $x_i$ with $y_i$. With a nontrivial twist, there is a distinct  class of diagrams for each choice of twist-cut. Here, by convention, we consider the diagrams with cut between the coordinates labelled $J$ and $1$, as in the main text. Different classes are related as explained in \cite{Cavaglia:2020hdb}. 
}
\beq
\label{OOW}
   W_{\epsilon, \epsilon}(x_0|x_{\bar{0}})\simeq \epsilon^{2 \Delta-2 J}  \; \langle {\mathcal O}(x_0) \widetilde{\mathcal{O}}(x_{\bar{0}} )\rangle = \epsilon^{2 \Delta - 2 J} \, \frac{\mathcal{N} }{x_{0 \bar{0}}^{2 \Delta}  } ,
  \eeq
  where the constant $\mathcal{N}$ is defined implicitly by the l.h.s. 
Similarly, we denote as 
$W_{\epsilon}(x_0| \mathbf{y} )$ the quantity $W(\mathbf{x}, \mathbf{y} )$ with $(x_i - x_0)^2 \sim \epsilon^2$. It is a regularisation of the wave function
\beq
\label{phiW}
 W_{\epsilon}(x_0, \mathbf{y} ) \simeq \epsilon^{ \Delta-J} \; \varphi_{\mathcal{O}}(x_0| \mathbf{y} )\ ,
\eeq
\beq
\label{phiW2}
  W_{\epsilon}(\mathbf{x}, x_{\bar 0} ) \simeq \epsilon^{ \Delta-J} \; \tilde\varphi_{\mathcal{O}}(x_{\bar 0}| \mathbf{x} )\ .
\eeq
Now let us consider the integral defining the norm, applied to the regularised wave functions:
\beq
a =\langle \langle\;  W_{\epsilon}(\dots | x_{\bar{0}}),\ W_{\epsilon}(x_0 | \dots ) \;\rangle \rangle .
\eeq
From \eq{phiW}, \eq{phiW2} we see this differs from the scalar product of two wave functions by an infinite normalisation: 
\beq\label{eq:normaliseapp}
a =\langle \langle \tilde\varphi, \varphi \rangle \rangle \; \epsilon^{2 \Delta-2 J}. 
\eeq
Looking at the explicit expression for the scalar product, we notice that the $\Box$ factors first remove one layer of propagators,  and after that the ladders are glued  together. This produces some combinatorial factors, so  that the result is given by the sum of diagrams: 
\beq
a = \sum_{n=0}^{\infty} (n+1) \,\xi^{2 J n} W_{\epsilon, \epsilon}^{(n)}(x_0|x_{\bar{0}}) .
\eeq
Notice that this has the form of a derivative, in fact we can obtain the same sum as
\beq
\label{acomp1}
a =\left[\frac{1}{J} \xi^2 \partial_{\xi^2} + 1 \right] ( W_{\epsilon, \epsilon}(x_0 | x_{\bar{0}} )  )  \sim \frac{2}{J } (\xi^2 \partial_{\xi^2} \Delta ) 
\log\left(\frac{\epsilon}{x_{0\bar{0}} } \right)\; 
\epsilon^{2 \Delta-2 J} \, \frac{\mathcal{N} }{x_{0 \bar{0}}^{2 \Delta}  } + \dots ,
\eeq
where the second term contains subleading singularities (without the log for example). 
Comparing with (\ref{eq:normaliseapp}), we conclude that 
\beq
\label{phi2O2}
\langle \langle \tilde{\varphi}, \varphi \rangle \rangle \simeq \frac{2}{{J} } \;  (\xi^2 \partial_{\xi^2} \Delta ) \log(\epsilon_{\text{UV}} ) \; \frac{\mathcal{N} }{x_{0 \bar{0}}^{2 \Delta}  }= \frac{2}{{J} } \;  (\xi^2 \partial_{\xi^2} \Delta ) \log(\epsilon_{\text{UV}} )\langle \mathcal{O}(x_0) \widetilde{\mathcal{O}}(x_{\bar{0}} )\rangle ,
\eeq
which is the result stated in \eq{eq:dDelta}.

\section{Parity structure of integral orthogonality relations}\label{app:gfunc}
In this appendix we follow the  approach of section \ref{sec:sovscalar} to derive a linear system of equations associated to  a generic state $A$, and its $\Pi$-transformed state denoted by $\tilde{A}$, where $\Pi$ is the chain-reflection symmetry discussed in section \ref{sec:gsec}.

 The general functional method described in section \ref{sec:sovscalar} gives us the relations
\beq
\langle p^{-\bullet_a A}_a \left(  \mathcal{B}_A - \mathcal{B}_{\tilde{A}} \right) \circ q^{\bullet_b \tilde{A}}_a \rangle_{{\mu}} = 0 ,
\eeq
which we can expand to obtain a linear system for the difference of the integrals of motion. To highlight the symmetries of the problem, it will be convenient to take a special basis of Q-bilinear forms: 
 we will pick a particular combination of functions analytic in the lower/upper half planes, 
\beq
\langle p^{-\bullet_a A} \hat{O} \circ q^{\bullet_a \tilde{A}} \rangle_{{\mu}} , \;\;\; {\bullet}_a \equiv \left( \uparrow , \downarrow, \uparrow,\downarrow \right)_a,\;\;\;-{\bullet}_a \equiv \left( \downarrow, \uparrow, \downarrow , \uparrow \right)_a, \;\; a=1,\dots,4 .
\eeq

 Using the basis of integrals of motion organised according to parity (see \eq{eq:newbasis40},\eq{eq:newbasis4})
\beq
\vec{H} \equiv \left( \vec H_{(-,\alpha)}  | \vec H_{(-, \alpha )}' ,  | \vec H_{(+, \alpha )}  | \vec H_{(+,\alpha )}'  \right)_{1\leq \alpha \leq J} ,
\eeq
 the linear system takes the form
\beq\label{eq:orthoPi}
{ {\mathcal{M}} }^{A \tilde{A}} \cdot \left( \vec H_A - \vec H_{\tilde{A}} \right) = 0  ,
\eeq
with 
\beq\label{eq:blockM}
{\mathcal{M}}^{A \tilde{A}} = \left(\begin{array}{c|c|c|c}  l_{-}^{(1)} & \; l_{-}^{'(1)} & \; l_{+}^{ (1) } & \; l_{+}^{' (1) }\\
  \hline l_{-}^{(2)} & \; l_{-}^{'(2)} & \; l_{+}^{ (2) } & \; l_{+}^{' (2) }\\
\hline   l_{-}^{(3)} & \; l_{-}^{'(3)} & \; l_{+}^{ (3) } & \; l_{+}^{' (3) }\\
    \hline  l_{-}^{(4)} & \; l_{-}^{'(4)} & \; l_{+}^{ (4) } & \; l_{+}^{' (4) }
 \end{array}\right) ,
 \eeq
 where the blocks are defined as
 \beqa\label{eq:smallblocks}
 l_{\pm }^{(a)} &\equiv & -\frac{1}{2}\,\left[\;\langle   \;   p_{a}^{-\bullet_a} \; \left( (u + i )^{\beta-1}\,\mathbb{D}_2 \pm (-1)^{J } \, (-u + i )^{\beta-1}\, \mathbb{D}_{-2} \right) \circ q_{a}^{\bullet_a } \; \rangle_{\alpha} \; \right]_{1 \leq \alpha,\beta \leq J } , \\
l_{\pm }^{'(a)} &=&  \left[ \; \langle \;   u^{2  \beta - 1 - \frac{1 \pm 1}{2}   } \, p_{a }^{-\bullet_a} \, \mathbb{D}_0 \circ q_{a}^{\bullet_a } \; \rangle_{\alpha} \;  \right]_{1 \leq \alpha, \beta \leq J } \ .
\eeqa

\paragraph{Factorisation for parity-symmetric states.}
Now let us show that, 
 for a parity-symmetric state $A = \tilde{A}$, the above matrix can be brought to a block structure. 
 
We will use the symmetry of the basis of Q-functions in the case $A = \tilde{A}$: indeed,  \eq{eq:Qpar} implies that in this case,
\beq 
q_a^{\downarrow } (u)  = \kappa_a\,  q_{\sigma(a)}^{\uparrow}(-u), \;\;\; p_a^{\downarrow } (u)  = \rho_a\,  p_{\sigma(a)}^{\uparrow}(-u) ,
\eeq
where $\sigma$ is the permutation acting as  $\sigma : \;\; (1,2,3,4) \rightarrow  (2,1,4,3) $, and $\kappa_a$, $\rho_a$ are simple constants, fixed by the large-$u$ asymptotics of Q-functions. Their value is not important in the following. 

The symmetry implies immediately, for instance,
\beq
\underset{u=\theta_\alpha - i n}{\rm res} \left( p^{\uparrow}_1(u) \,u^k \, q^{\downarrow}_1(u) \right) = (-1)^{k+1} \, \frac{\kappa_1}{\rho_1} \;\times \underset{u=-\theta_\alpha + i n}{\rm res} \left( p^{\downarrow}_2(u) \,u^k \, q^{\uparrow}_2(u) \right) ,
\eeq
and from the sum over poles,
\beq
\langle  p^{\uparrow}_1 \,u^k \, q^{\downarrow}_1  \rangle_{\alpha} = (-1)^{k+1} \, \frac{\kappa_1}{\rho_1} \;\times \langle  p^{\downarrow}_2 \,u^k \, q^{\uparrow}_2  \rangle_{J-\alpha+1} .
\eeq
With similar arguments we find in general,
\beq
\left(l_{\pm}^{(a)} \right)_{\alpha}^{\;\;\beta}= \mp A_a \, \left( l_{\pm}^{(\,\sigma(a) \,)} \right)_{J-\alpha+1}^{\;\;\beta}, \;\;\;\;\; \left(l_{\pm}^{'(a)} \right)_{\alpha}^{\;\;\beta}= \mp A_a \, \left( l_{\pm}^{'(\,\sigma(a) \,)} \right)_{J-\alpha+1}^{\;\;\beta},
\eeq
 where $A_a$ are simple (non-vanishing)  constants. Therefore, multiplying ${{ \mathcal{M}^{A \tilde{A} } }}$ on the left by the constant matrix
\beq
(L )_{(a,\alpha)}^{\,\,(b,\beta)} \equiv \frac{1}{2}\,\left( \begin{array}{cccc}
1 & A_1 & 0 & 0 \\
0 & 0 & 1 & +A_3\\
1 & -A_1 & 0 & 0 \\
0 & 0 & 1 & -A_3
\end{array} \right)_a^{\, b} \,\times \, \delta_{\alpha}^{\,J-\beta + 1} ,\;\;\;\; 
 {\tilde{\mathcal{M}}}^{A \tilde{A}}  \equiv \left( L \cdot {{\mathcal{M}}}^{A \tilde{A}}  \right) 
\eeq
we reach the block structure
\beq
\left.  {\tilde{\mathcal{M}}}^{A \tilde{A}}  \right|_{A = \tilde{A}} = \left(\begin{array}{cc} M_- & 0 \\
0 & M_+ \end{array}\right) ,
\eeq
in analogy with the case studied  in \cite{Caetano:2020dyp}. 
\paragraph{Comment on the selection rule. }
As proved in the main text,  $|M_-|=0$ whenever the state is non-symmetric, $A \neq \tilde{A}$. Let us show that, instead, when  $A =\tilde{A}$, we have $|M_-| \neq 0$ while $|M_+| = 0$. 

This is a consequence of the variation equation (\ref{eq:varsys0}). Since for a symmetric state the odd integrals of motion have the vanish, the latter reduces to a square system of $2J$ equations
\beq
M_+ \cdot \partial_{\xi^2} \left( \begin{array}{c} \vec{H}_{+} \\\hline \vec{H}_{+}'\end{array} \right) = 0 ,
\eeq
which implies (since the system has a nontrivial solution) that
\beq\label{eq:Mplus0}
| M_+ | = 0 .
\eeq
As discussed in section \ref{sec:orthosec}, the kernel of $\tilde{\mathcal{M}}$ is  expected to be one-dimensional in generic points of parameter space. This, together with \eq{eq:Mplus0}, implies that, for symmetric states, $|M_-|\neq 0$.   This determinant therefore gives a precise characterisation of the symmetry of the states: it vanishes precisely for states which are not symmetric.

\section{Determinants in the N=4 SYM Baxter equation}
\label{app:baxdet}

Here we present the determinants appearing in the coefficients of the $\cN=4$ SYM Baxter equation we presented in \eq{eq:defA}:
\beqa
{\mathcal{D}}_0&=&{\rm det}
\(
\bea{llll}
\bP^{1[+2]}&\bP^{2[+2]}&\bP^{3[+2]}&\bP^{4[+2]}\\
\bP^{1}&\bP^{2}&\bP^{3}&\bP^{4}\\
\bP^{1[-2]}&\bP^{2[-2]}&\bP^{3[-2]}&\bP^{4[-2]}\\
\bP^{1[-4]}&\bP^{2[-4]}&\bP^{3[-4]}&\bP^{4[-4]}
\eea
\)\;,\\
{\mathcal{D}}_1&=&{\rm det}
\(
\bea{llll}
\bP^{1[+4]}&\bP^{2[+4]}&\bP^{3[+4]}&\bP^{4[+4]}\\
\bP^{1}&\bP^{2}&\bP^{3}&\bP^{4}\\
\bP^{1[-2]}&\bP^{2[-2]}&\bP^{3[-2]}&\bP^{4[-2]}\\
\bP^{1[-4]}&\bP^{2[-4]}&\bP^{3[-4]}&\bP^{4[-4]}
\eea
\)\;,\\
{\mathcal{D}}_2&=&{\rm det}
\(
\bea{llll}
\bP^{1[+4]}&\bP^{2[+4]}&\bP^{3[+4]}&\bP^{4[+4]}\\
\bP^{1[+2]}&\bP^{2[+2]}&\bP^{3[+2]}&\bP^{4[+2]}\\
\bP^{1[-2]}&\bP^{2[-2]}&\bP^{3[-2]}&\bP^{4[-2]}\\
\bP^{1[-4]}&\bP^{2[-4]}&\bP^{3[-4]}&\bP^{4[-4]}
\eea
\)\;,\\
\bar {\mathcal{D}}_1&=&{\rm det}
\(
\bea{llll}
\bP^{1[-4]}&\bP^{2[-4]}&\bP^{3[-4]}&\bP^{4[-4]}\\
\bP^{1}&\bP^{2}&\bP^{3}&\bP^{4}\\
\bP^{1[+2]}&\bP^{2[+2]}&\bP^{3[+2]}&\bP^{4[+2]}\\
\bP^{1[+4]}&\bP^{2[+4]}&\bP^{3[+4]}&\bP^{4[+4]}
\eea
\)\;,\\
\bar {\mathcal{D}}_0&=&{\rm det}
\(
\bea{llll}
\bP^{1[-2]}&\bP^{2[-2]}&\bP^{3[-2]}&\bP^{4[-2]}\\
\bP^{1}&\bP^{2}&\bP^{3}&\bP^{4}\\
\bP^{1[+2]}&\bP^{2[+2]}&\bP^{3[+2]}&\bP^{4[+2]}\\
\bP^{1[+4]}&\bP^{2[+4]}&\bP^{3[+4]}&\bP^{4[+4]}
\eea
\)\;.
\eeqa
Since the $\bP$ functions have a single short cut on the first Riemann sheet (see figure \ref{fig:cuts}), these determinants all have a finite number of cuts, and are analytic outside a certain radius in the complex plane. We also notice that, using the QQ relations, these coefficients can be  written as the same expressions, but where all the $\bP$ functions are replaced by $\bQ$ functions with the same index.
The expression in terms of $\bQ$ functions makes it easy to evaluate the determinants for large $u$ when the $\bQ$ functions have generic twists $\lambda_i$. In this case, the determinants are non-vanishing constants at infinity. These analytic properties justify the expansion (\ref{eq:Aexp}). 

The fact that we can alternatively write the $\mathcal{D}$ coefficients in terms of the $\bQ$ or $\bP$ functions,  implies that they are functions with a finite number of either short or long cuts, on two connected Riemann sections. This suggests that they might have a simple analytic structure. Understanding it more fully  might be relevant to find a good basis of independent integrals of motion in $\mathcal{N}$=4 SYM.

\end{document}